
%
%
%
%

%
%
\font\twelvrm=cmr12
\font\ninerm=cmr9
\font\twelvi=cmmi12
\font\ninei=cmmi9
\font\twelvex=cmex10 scaled\magstep1
\font\twelvbf=cmbx12
\font\ninebf=cmbx9
\font\twelvit=cmti12
\font\twelvsy=cmsy10 scaled\magstep1
\font\ninesy=cmsy9
\font\twelvtt=cmtt12

\font\twelvsl=cmsl12

\font\abstractfont=cmr10
\font\abstractitalfont=cmti10

\def\twelvepoint{\def\rm{\fam0\twelvrm}
  \textfont0=\twelvrm \scriptfont0=\ninerm \scriptscriptfont0=\sevenrm
  \textfont1=\twelvi \scriptfont1=\ninei \scriptscriptfont1=\seveni
  \textfont2=\twelvsy \scriptfont2=\ninesy \scriptscriptfont2=\sevensy
  \textfont3=\twelvex \scriptfont3=\tenex \scriptscriptfont3=\tenex
	\textfont\itfam=\twelvit \def\it{\fam\itfam\twelvit}
	\textfont\slfam=\twelvsl \def\sl{\fam\slfam\twelvsl}
	\textfont\ttfam=\twelvtt \def\tt{\fam\ttfam\twelvtt}
	\textfont\bffam=\twelvbf \def\bf{\fam\bffam\twelvbf}
	\scriptfont\bffam=\ninebf  \scriptscriptfont\bffam=\sevenbf
        \skewchar\ninei='177
        \skewchar\twelvi='177
        \skewchar\seveni='177
}

\newdimen\normalwidth
\newdimen\double
\newdimen\single
\newdimen\indentlength		\indentlength=.5in

\newif\ifdrafton

\def\galley{
 \draftonfalse
 \twelvepoint
 \rm
 \font\chapterfont=cmbx10 scaled\magstep1
 \font\sectionfont=cmbx12
 \font\subsectionfont=cmbx12
 \font\headingfont=cmr10 scaled\magstep2
 \font\titlefont=cmbx10 scaled\magstep2
 \normalwidth=5.7in
 \double=.34in
 \single=.17in
 \hsize=\normalwidth
 \vsize=8.7in
 \hoffset=0.48in
 \voffset=0.1in
 \hfuzz=0.5pt
 \baselineskip=\double plus 2pt minus 2pt }

\parindent=\indentlength
\clubpenalty=10000
\widowpenalty=10000
\displaywidowpenalty=500
\overfullrule=2pt
\tolerance=100

\newcount\chapterno	\chapterno=0
\newcount\sectionno	\sectionno=0
\newcount\appno		\appno=0
\newcount\subsectionno	\subsectionno=0
\newcount\eqnum	\eqnum=0
\newcount\refno \refno=0
\newcount\chap
\newcount\figno \figno=0
\newcount\tableno \tableno=0
\newcount\lettno \lettno=0

\def\bodypaging{
 \headline={\ifodd\chap \hfil \else \tenrm\hfil\twelvrm\folio \fi}
 \footline={\rm \ifodd\chap \global\chap=0 \tenrm\hfil\twelvrm\folio\hfil
 \else \hfil \fi}}

%
\def\preq{\mainid.\the\eqnum}
\def\Elabel#1{\xdef#1{\mainid.\the\eqnum}}
\def\eq{\global\advance\eqnum by1 \eqno(\mainid.\the\eqnum)}
\def\eqlabel#1{\eq {\xdef#1{\mainid.\the\eqnum}}}
\def\quieteqlabel#1{\advance\eqnum by1 {\xdef#1{\mainid.\the\eqnum}}}
\def\newlett{\global\lettno=1\global\advance\eqnum by1
      \eqno(\mainid.\the\eqnum a)}
\def\lett{\global\advance\lettno by 1
    \eqno(\mainid.\the\eqnum{\ifcase\lettno\or a\or b\or c\or d\or e\or
    f\or g\or h\or \fi})}
\def\newlettlabel#1{\newlett {\xdef#1{\mainid.\the\eqnum}}}
\def\lettlabel#1{\lett {\xdef#1{\mainid.\the\eqnum}}}
%
\newwrite\refs
\def\startrefs#1{\immediate\openout\refs=#1
\immediate\write\refs{\global\chap=1}
\immediate\write\refs{\vfill\noexpand\eject\noexpand\vglue.2in}
\immediate\write\refs{\noexpand\centerline{\headingfont REFERENCES}}
\immediate\write\refs{\noexpand\vglue.5in\baselineskip=\single}
\immediate\write\refs{\parindent=16pt \parskip=\single}}

\def\ref#1{\advance\refno by1 \the\refno \immediate\write\refs
{\noexpand\item{\the\refno.}#1 \hfill\par}}
\def\andref#1{\advance\refno by1\kern-.4em[\the\refno]\immediate\write\refs
{\noexpand\item{\the\refno.}#1 \hfill\par}}
\def\cite#1{[#1]}

\def\refname#1{ \xdef#1{\the\refno}}

\newif\iftoc \tocfalse
\newif\ifpart \partfalse
\newwrite\conts
\def\startcontents{\iftoc\immediate\openout\conts=contents
\immediate\write\conts{\noexpand\centerline{\headingfont CONTENTS}}
\immediate\write\conts{\noexpand\vglue.5in\baselineskip=\single}
\immediate\write\conts{\parskip=0pt \parindent=0pt}\else\relax\fi}

\def\chaptercont#1{\iftoc\immediate\write\conts{
  \vskip\single\the\chapterno.\ #1\hfill\folio\vskip0cm}\else\relax\fi}
\def\andchaptercont#1{\iftoc\immediate\write\conts{
  \hskip .5in \ #1\hfill}\else\relax\fi}
\def\sectioncont#1{\iftoc\immediate\write\conts{\hskip.5cm
  \the\chapterno.\the\sectionno\ #1\hfill\folio\vskip0cm}\else\relax\fi}
\def\subsectioncont#1{\iftoc\immediate\write\conts{\hskip1cm
  \the\chapterno.\the\sectionno.\the\subsectionno\
  #1\hfill\folio\vskip0cm}\else\relax\fi}
\def\appendixcont#1{\iftoc\immediate\write\conts{\vskip\single
  Appendix \mainid.\ #1\hfill\folio\vskip0cm}\else\relax\fi}

\def\partcont#1{\iftoc
 \ifpart\immediate\write\conts{\noexpand\vfill\noexpand\eject}\fi
 \immediate\write\conts{\vskip\single
 \noexpand\centerline{- PART #1 -}}
 \parttrue
 \else\relax\fi}


\newif\ifpskip

\def\chapter#1{
  \ifpskip\vfill\eject \else \bigskip\bigskip\bigskip \fi
 \global\advance\chapterno by1 \sectionno=0 \subsectionno=0 \eqnum=0
 \def\mainid{\the\chapterno}
 \vglue.8cm\centerline{\chapterfont
 \uppercase\expandafter{\romannumeral\the\chapterno.\ #1}}
 \nobreak\vskip.2cm\nobreak
 \ifpskip\global\chap=1\fi
 \chaptercont{#1}}

\def\andchapter#1{\centerline{\chapterfont\raise.1cm
 \hbox{\uppercase{#1}}}\vskip.2cm\nobreak\andchaptercont{#1}}

\def\section#1{
 \global\advance\sectionno by1 \subsectionno=0
 \vskip.8cm\centerline{\sectionfont \the\chapterno.\the\sectionno\ #1}
 \nobreak\vskip.2cm\nobreak
 \sectioncont{#1}}

\def\andsection#1{\centerline{\sectionfont\raise.1cm\hbox{#1}}
 \nobreak\vskip.2cm\nobreak}
\def\subsection#1{
 \global\advance\subsectionno by1
 \vskip.2cm\leftline{\subsectionfont
 \the\chapterno.\the\sectionno.\the\subsectionno\ #1}
 \subsectioncont{#1}}
\def\appendix#1{
 \vfil\eject
 \global\advance\appno by1 \subsectionno=0 \eqnum=0
 \def\mainid{\ifcase\appno\or A\or B\or C\or D\or E\or F\or G\or H\or \fi}
 \vglue.8cm\centerline{\sectionfont
 \uppercase{\mainid .\ \ #1}}
 \nobreak\vskip.2cm\nobreak
 \global\chap=1
 \appendixcont{#1}}

\def\loneappendix#1{
 \vfil\eject
 \global\advance\appno by1 \subsectionno=0 \eqnum=0
 \def\mainid{\ifcase\appno\or A\or B\or C\or D\or E\or F\or G\or H\or \fi}
 \vglue.8cm\centerline{\sectionfont
 \uppercase{APPENDIX: #1}}
 \nobreak\vskip.2cm\nobreak
 \global\chap=1
 \appendixcont{#1}}
\def\Slabel#1{\xdef#1{\mainid
              \ifnum\sectionno>0
                    { {.\the\sectionno}
                        \ifnum\subsectionno>0
                             { .\the\subsectionno}
                         \fi}
                \fi}}
%
\newif\iftable \tablefalse
\newwrite\tablelist
\def\starttablelist{\iftable
 \immediate\openout\tablelist=tablelist
 \immediate\write\tablelist{\noexpand\centerline{\headingfont LIST
      OF TABLES}}
 \immediate\write\tablelist{\vskip\double\vskip\single\baselineskip=\single}
 \immediate\write\tablelist{\parskip=0pt \parindent=0pt}\else\relax\fi}
\def\inserttable#1#2{\global\advance\tableno by 1 \vfill\vskip\double
 \centerline{Table \the\tableno:\ #2}\nobreak\vskip\double\nobreak
 \begingroup\input#1\endgroup \vskip\double
 \iftable\immediate\write\tablelist{\hskip1cm
 \the\tableno. #2 \hfill\folio\vskip0cm}\else\relax\fi}
\def\andinserttable#1#2{\vfill\vskip\double
 \centerline{Table \the\tableno\ (continued):\ #2}
 \nobreak\vskip\double\nobreak
 \begingroup\input#1\endgroup \vskip\double}

\def\noadvancetable#1#2{\vfill\vskip\double
 \centerline{Table \the\tableno:\ #2}\nobreak\vskip\double\nobreak
 \begingroup\input#1\endgroup \vskip\double
 \iftable\immediate\write\tablelist{\hskip1cm
 \the\tableno. #2 \hfill\folio\vskip0cm}\else\relax\fi}
\def\continuetable#1#2{\vfill\vskip\double
 \centerline{Table \the\tableno{ }(continued):\ #2}\nobreak
 \vskip\double\nobreak
 \begingroup\input#1\endgroup \vskip\double}
\def\Tlabel#1{\global\advance\tableno by 1
       \xdef#1{\the\tableno} \the\tableno}
%
\newif\iffig \figfalse
\newwrite\figlist
\newwrite\figs
\def\startfiglist{\iffig\immediate\openout\figlist=figlist
 \immediate\openout\figs=figs
 \immediate\write\figlist{\noexpand\centerline{\headingfont LIST OF
        FIGURES}}
 \immediate\write\figlist{\vskip\double\vskip\double\baselineskip=\single}
 \immediate\write\figlist{\parskip=0pt \parindent=0pt}\else\relax\fi}
\def\insertpagefig#1{\iffig\global\advance\figno by 1
 \immediate\write\figs{\vfil\noexpand\eject\bodypaging\pageno=\the\pageno
 \noexpand\vglue 20cm}
 \immediate\write\figs{\noexpand\centerline{Figure \the\figno. #1}}
 \immediate\write\figlist{\hskip1cm
 \the\figno. #1 \hfill\folio\vskip0cm}\advance\pageno by1 \else\relax\fi}
\def\insertfig#1#2{\iffig\global\advance\figno by 1
 \vglue #2
 \centerline{Figure \the\figno. #1}
 \immediate\write\figlist{\hskip1cm
 \the\figno. #1 \hfill\folio\vskip0cm} \else\relax\fi}
\def\insertfig#1#2{\iffig\global\advance\figno by 1
 \vglue #2
 \centerline{Figure \the\figno. #1}
 \immediate\write\figlist{\hskip1cm
 \the\figno. #1 \hfill\folio\vskip0cm} \else\relax\fi}

\def\Flabel#1{\global\advance\figno by 1 \xdef#1{\the\figno} \the\figno}
\def\noadvancefig#1#2{\iffig
 \vglue #2
 \centerline{Figure \the\figno. #1}
 \immediate\write\figlist{\hskip1cm
 \the\figno. #1 \hfill\folio\vskip0cm} \else\relax\fi}


\galley

%
%

%

\def \amp{{\cal M}}

\def \Jhat{\skew5\widehat{J}}
\def \Jms{{\cal{J}}}
\def \Jslash{\thinspace{\not{\negthinspace J}}}
\def \Jbar{\skew5\bar{\Jms}}

\def \Ims{{\cal{I}}}
\def \Ibar{\skew3\bar{\Ims}}
\def \Ihat{\skew5\widehat{I}}
\def \Xms{{\cal{X}}}

\def \psihat{\skew2\widehat{\psi}}
\def \psibarhat{\skew2\widehat{\skew2\bar{\psi}}}
\def \psibar{\skew3\bar{\psi}}

\def \GAMMA{{\mit\Gamma}}

%
%
\def \permsum#1#2{\sum_{{\cal{P}}(#1\ldots #2)}}

\def \braket#1#2{ \langle #1 \thinspace\thinspace #2 \rangle }
\def \bra#1{ \langle #1 | }
\def \ket#1{ | #1 \rangle }
\def \num{{\cal{N}}}

\def \eps{\epsilon}
\def \vareps{\varepsilon}

\def \zee{{\cal{Z}}}
\def \coef{c\thinspace}

\def \link#1#2#3{ {{\braket{#1}{#3}}\over{\bra{#1}#2\ket{#3}}} }
\def \invlink#1#2#3{ {{\bra{#1}#2\ket{#3}}\over{\braket{#1}{#3}}} }


\def \dyadic#1{\vbox{\ialign{##\crcr
     $\hfil
{\thinspace\scriptstyle\leftrightarrow}
\hfil$\crcr\noalign{\kern-.01pt\nointerlineskip}
     $\hfil\displaystyle{#1}\hfil$\crcr}}}

\def\centeronto#1#2{{\setbox0=\hbox{#1}\setbox1=\hbox{#2}\ifdim
\wd1>\wd0\kern.5\wd1\kern-.5\wd0\fi
\copy0\kern-.5\wd0\kern-.5\wd1\copy1\ifdim\wd0>\wd1
\kern.5\wd0\kern-.5\wd1\fi}}
\def\slash#1{\centeronto{$#1$}{$/$}}



\hyphenation{ap-pen-dix  spin-or  spin-ors}

\def \msreplacefermion{A.22}
\def \msreplacedotprod{A.21}


\def \slashsqr{A.6}

\def \fierznc{A.8}
\def \fierz{A.14}

\def \linkidnosum{A.15}
\def \linkidsummed{A.16}


\def \spinnorm{A.11}

\def \antisym{A.13a}



\def \anticommutator{A.3}
%
%
%
{\nopagenumbers
\hfill\hbox{
CLNS 91/1120}

\hfill\hbox{
September 1992}
\vfill
\baselineskip=\double
{\titlefont
        \centerline{GENERALIZED GLUON CURRENTS}
        \centerline{AND APPLICATIONS IN QCD}
}
\bigskip
\bigskip

\bigskip
\centerline{Gregory MAHLON$^{1}$ and Tung--Mow YAN$^{2}$}
\medskip

{\baselineskip .20in plus 2pt minus 2pt
\centerline{
{\it Newman Laboratory of Nuclear Studies,}
}
\centerline{
{\it Cornell University,
Ithaca, NY 14853, USA}
}
}
\bigskip
\centerline{Charles DUNN$^{3}$}
{\baselineskip .20 in plus 2pt minus 2pt
\centerline{{\it Jet Propulsion Laboratory, M/S 238-600,}}
\centerline{{\it 4800 Oak Grove Drive,}}
\centerline{{\it Pasadena, CA 91109, USA}}

}

\bigskip
\bigskip
\bigskip
\centerline{ABSTRACT}
{\narrower\medskip\baselineskip=.20in plus 2pt minus 2pt
{\abstractfont
We consider the process containing two quark lines and an arbitrary
number of gluons in a spinor helicity framework.
A current with two off-shell gluons appears in the amplitude.
We first study this modified gluon current using recursion
relations.  The recursion relation for the modified gluon
current is solved for the case of like-helicity gluons.
We apply the modified gluon current to compute the amplitude for
{\abstractitalfont q\=q}~$\rightarrow$
{\abstractitalfont q\=qgg}~$\cdots${\abstractitalfont g}\
in the like-helicity gluon case.}\smallskip}

\vfill

\noindent
\hrule
\smallskip

\noindent
$^1$ e-mail:  gdm@beauty.tn.cornell.edu

\noindent
$^2$ e-mail:  yan@lnssun5.tn.cornell.edu

\noindent
$^3$ e-mail:  ced@logos.jpl.nasa.gov

\eject}

\pageno=2 \bodypaging
\startrefs{refs1}

\chapter{INTRODUCTION}

In recent years, there has been a great interest in studying
the high energy scatterings in QCD.  An efficient approach in
the tree approximation is to formulate recursion relations for
currents with $n$ gluons in terms of those with fewer gluons
[\ref{F. A. Berends and W. T. Giele, Nucl. Phys. {\bf B306} (1988) 759;
C. Dunn and T.--M. Yan, Nucl. Phys. {\bf B352} (1991) 402;
C. Dunn, thesis, Cornell University (1990)}
\refname\QCDrecursions{\kern-.5em}].  The gluon currents which
have been studied so far contain only one off-shell gluon.  These are
appropriate for processes involving only gluons, or a single
quark line plus gluons.  However, as soon as we consider processes
with two or more quark lines, we encounter gluon currents with two
or more off-shell gluons.  These currents also satisfy recursion
relations similar to those for currents with one off-shell gluon.

In this paper we will study the process
$$
q \bar q \longrightarrow q \bar q gg\cdots g.
\eqlabel\monsteramplitude
$$
Because of the non-Abelian nature of QCD,
the number of Feynman diagrams for this process is
enormous, even at tree-level.  This is true even in cases that are
far simpler than the one being studied here.
Thus, a suitable means to organize and
simplify the calculation is required.
A fruitful concept in this respect has been the multispinor representation
of a vector field
[\ref{J. Schwinger, {\it Particles, Sources and Fields,}{ }Vol. I,
Addison Wesley, 1970;
Ann. Phys. {\bf 119}{ }(1979) 192}
\refname\SCHWINGER{\kern -.65em}].
Over the last 10 years or so,
a whole industry  devoted to the application of the spinor
technique to  multi-parton processes
has sprung up, employing the spinors in a number of different
contexts~[\ref{
The spinor technique was first introduced by the
CALCUL collaboration, in the context
of massless Abelian gauge theory:
P. De Causmaecker, R. Gastmans, W. Troost, and T.T. Wu,
Phys. Lett. {\bf 105B}{ }(1981) 215;
P. De Causmaecker, R. Gastmans, W. Troost and T.T. Wu,
Nucl. Phys. {\bf B206}{ }(1982) 53;
F. A. Berends, R. Kleiss, P. De Causmaecker, R. Gastmans,
W. Troost and T.T. Wu, Nucl. Phys. {\bf B206}{ }(1982) 61;
F.A. Berends, P. De Causmaecker, R. Gastmans, R. Kleiss,
W. Troost and T.T. Wu, Nucl. Phys. {\bf B239}{ }(1984) 382;
{\bf B239}{ }(1984) 395; {\bf B264}{ }(1986) 243; {\bf B264}{ }(1986) 265
\smallskip
\noexpand\item{ }
By now, many papers have been published on the subject.  A
partial list of references follows.
\smallskip
\noexpand\item{ }
P. De Causmaecker, thesis, Leuven University (1983);
R. Farrar and F. Neri, Phys. Lett. {\bf 130B}{ }(1983) 109;
R. Kleiss, Nucl. Phys. {\bf B241}{ }(1984) 61;
Z. Xu, D.H. Zhang and Z. Chang, Tsingua University
preprint TUTP-84/3, 84/4, and 84/5a (1984), and
Nucl. Phys. {\bf B291}{ }(1987) 392;
J.F. Gunion and Z. Kunszt, Phys. Lett. {\bf 161B}{ }(1985) 333;
F.A. Berends, P.H. Davereldt and R. Kleiss, Nucl. Phys. {\bf B253}{ }
(1985) 441;
R. Kleiss and W.J. Stirling, Nucl. Phys. {\bf B262}{ }(1985) 235;
J.F. Gunion and Z. Kunszt, Phys. Lett. {\bf 159B}{ }(1985) 167;
{\bf 161B}{ }(1985) 333;
S.J. Parke and T.R. Taylor, Phys. Rev. Lett. {\bf 56}{ }(1986) 2459;
Z. Kunszt, Nucl. Phys. {\bf B271}{ }(1986) 333;
J.F. Gunion and J. Kalinowski, Phys. Rev. {\bf D34}{ }(1986) 2119;
R. Kleiss and W.J. Stirling, Phys. Lett. {\bf 179B}{ }(1986) 159;
M. Mangano and S.J. Parke, Nucl. Phys. {\bf B299}{ }(1988) 673;
M. Mangano, S.J. Parke and Z. Xu, Nucl. Phys. {\bf B298}{ }
(1988) 653;
D.A. Kosower, B.--H. Lee and V.P. Nair, Phys. Lett. {\bf 201B}{ }
(1988) 85;
M. Mangano and S.J. Parke, Nucl. Phys. {\bf B299}{ }(1988) 673;
F.A. Berends and W.T. Giele, Nucl. Phys. {\bf B313}{ }(1989) 595;
M. Mangano, Nucl. Phys. {\bf B315}{ }(1989) 391;
D.A. Kosower, Nucl. Phys. {\bf B335}{ }(1990) 23;
D.A. Kosower, Phys. Lett. {\bf B254}{ }(1991) 439;
Z. Bern and D.A. Kosower, Nucl. Phys. {\bf B379}{ }(1992) 451;
C.S. Lam, McGill preprint McGill/92-32 (1992)},\kern-.01em
\ref{Many of the results for processes containing six or fewer
particles are collected in R. Gastmans and T.T. Wu,
The Ubiquitous Photon:  Helicity Method for QED and QCD
(Oxford University Press, New York, 1990)}\kern-.05em
].
An excellent guide to the many approaches
and methods developed for such calculations is the
review by Mangano and Parke~[\ref{M. Mangano and S. Parke,
Phys. Reports {\bf 200}{ }(1991) 301}~{\kern-.65em}~].
\refname\ManganoParke
We will organize our calculation
within the framework of  the recursion relations
presented in
reference~\cite\QCDrecursions.   The multispinor
representation allows us to treat quarks and gluons on an equal
footing, while the recursion relations supply connections
between currents with $n$ particles and those with $n-1$ and
$n-2$ particles.  Once we have the recursion relations, we can
use their solutions  as the starting point for our calculations.
Rather than a large number of Feynman diagrams showing all of
the possible gluon configurations, we consult a handful
of diagrams  built from the individual currents.
We will use Weyl-van der Waerden spinors
in this work (a summary of our conventions appears in the appendix).

The organization of this paper is as follows.  In section 2, we
briefly review the recursion relations for quarks and gluons
involving only one off-shell particle.  We also recount
the solutions to these relations for the case in which all of
the gluons have the same helicity.  In section 3, we study
the gluon recursion relation  for
a  gluon current with $n$ off-shell gluons plus
any number of on-shell gluons.
We then specialize to the case $n=2$ and
introduce a ``modified'' current.
This modified current will
have a ``special'' gluon, which is off mass-shell
and has a  modified ``polarization spinor.''
This recursion relation for the modified
current is  easily solved when
the modified ``polarization spinor'' for the special gluon
assumes a certain form, and the gluons have like helicity.
We present a closed form solution
for the modified gluon current for this special
case.  In section 4, we show how the modified gluon current fits
into a computation of the process (\monsteramplitude).  We then
do the actual calculation for  (\monsteramplitude) for the
case in which all of the gluons have the same helicity.
Mangano [\ref{M. Mangano, Nucl. Phys. {\bf B309}{ }(1988) 461}{]}
\refname\Mangano
has obtained an expression for this amplitude by
using the fact that each color configuration corresponds to
a gauge invariant amplitude.  Requiring
the correct collinear and soft limits  in the
amplitude leads to
the form published in reference~\cite\Mangano.
We find that we are able to reproduce Mangano's result
using an appropriate combination of quark and gluon currents
as the starting point.  Hopefully, the techniques and
details presented here will provide a stimulus for
the development of new methods to tackle even more
difficult problems.
The final section contains a few concluding remarks.

\chapter{The QCD recursion relations}
\andchapter{and their solutions}

In this section we will review the
recursion relations for quarks and gluons and their closed
form solutions for special helicity configurations presented in
reference \cite\QCDrecursions.
Each of the currents discussed below has only
one off-shell particle.


\section{The multi-gluon current}

We define the current $\Jhat^x_{\xi} (1,\ldots,n)$ for
one off-shell plus $n$ on-shell transverse gauge bosons in the
tree approximation.
By convention, all momenta will flow into the graph.
The $j$th gauge boson will have momentum $k_j$, with the off-shell
boson having momentum $k_{n+1}=-(k_1+k_2+\cdots+k_n)\equiv
-\kappa(1,n)$.  The color index of the off-shell boson is $x$,
while its Lorentz index is $\xi$.   We include the propagator
for the off-shell boson in the definition of
$\Jhat^x_{\xi} (1,\ldots,n)$.  This definition gives
$$
\eqalign{
\Jhat^x_{\xi}&(1,\ldots,n)=
\cr & =
{
{i}
\over
{\kappa^2(1,n)}
}
\permsum{1}{n}
\Biggl\{
g \sum_{j=1}^{n-1}
{
{1}
\over
{2! \thinspace j! \thinspace (n{-}j)!}
}
f^{x_1 x_2 x}
\cr & \qquad\qquad\qquad\qquad\times
{V^{\mu \nu}}_{\xi}\bigr(\kappa(1,j),\kappa(j{+}1,n),-\kappa(1,n)\bigr)
\cr & \qquad\qquad\qquad\qquad\times
\Jhat^{x_1}_{\mu}(1,\ldots,j)
\Jhat^{x_2}_{\nu}(j{+}1,\ldots,n)
\cr & \qquad\qquad +ig^2
\sum_{j=1}^{n-2} \sum_{\ell=j+1}^{n-1}
{
{1}
\over
{3! \thinspace j! \thinspace (j-\ell)! \thinspace (n-\ell)! }
}
{\sum_{{\cal{C}}(1\thinspace 2\thinspace 3)}}
f^{a_1 a_2 y} f^{y a_3 x}
\cr & \qquad\qquad\enspace\times
{K^{\alpha_1 \alpha_2 \alpha_3}}_{\xi}
\Jhat_{\alpha_1}^{a_1}(1,\ldots,j)
\Jhat_{\alpha_2}^{a_2}(j{+}1,\ldots,\ell)
\Jhat_{\alpha_3}^{a_3}(\ell{+}1,\ldots,n)
\Biggr\}.
}
\eqlabel\gluonstart
$$
In (\gluonstart) $f^{abc}$ is the structure constant of the group
$SU(N)$.  We denote  the fundamental matrix representation of
$SU(N)$ by $T^a$, hence,
$$
[T^a,T^b] = if^{abc} T^c.
\eqlabel\SUNcommutator
$$
The $T$'s are normalized such that
$$
tr\enspace T^a T^b = {1\over2}\delta^{ab},
\eqlabel\SUNnormalization
$$
and satisfy the completeness relation
$$
\sum_{x=1}^{N^2-1}
(T^x)_{ij}\thinspace (T^x)_{k\ell} =
{1\over2}
\biggl[
\delta_{jk}\delta_{i\ell}
-{1\over N}\delta_{ij}\delta_{k\ell}
\biggr].
\eqlabel\completeness
$$
The functions $V_{\mu \nu \xi}(k_1,k_2,k_3)$
and $K_{\alpha_1 \alpha_2 \alpha_3 \alpha_4}$ represent the
3-vertex and 4-vertex respectively:
$$
V_{\mu \nu \xi}(k_1,k_2,k_3) =
g_{\mu\nu}(k_1-k_2)_{\xi}
+g_{\nu\xi}(k_2-k_3)_{\mu}
+g_{\xi\mu}(k_3-k_1)_{\nu},
\eqlabel\threevertex
$$
$$
K_{\kappa \lambda \mu \nu} =
g_{\kappa \nu}g_{\lambda \mu} -
g_{\kappa \mu}g_{\lambda \nu}.
\eqlabel\fourvertex
$$
The symbol ${\cal{P}}(1\ldots n)$ denotes permutations among
$\{1,\ldots,n\}$, while ${\cal{C}}(1\ldots n)$ represents
cyclic permutations.

Berends and Giele \cite\QCDrecursions { }show that
$\Jhat^x_{\xi}(1,\ldots,n)$
satisfies the following factorization property and recursion relation:
$$
\Jhat^x_{\xi}(1,\ldots,n) =
2g^{n-1} \permsum{1}{n}
tr(\Omega[1,n]T^x) J_{\xi}(1,\ldots,n),
\newlettlabel\gluonrecursion
$$
$$
\eqalign{
J_{\xi}(1,\ldots,n)&=
{ 1\over{\kappa^2(1,n)} }
\Biggl(
\sum_{j=1}^{n-1}
\bigl[ J(1,\ldots, j),J(j{+}1,\ldots,n) \bigr]_{\xi}
\cr &
+
\sum_{j=1}^{n-2} \sum_{\ell=j+1}^{n-1}
\bigl\{ J(1,\ldots,j),J(j{+}1,\ldots,\ell),
        J(\ell{+}1,\ldots,n) \bigr\}_{\xi}
\Biggr),
}
\lett
$$
$$
J(1)=\eps(1).
\lett
$$
In (\gluonrecursion) we have the notations:
$$
\Omega[1,n] \equiv T^{a_1}\cdots T^{a_n},
\eqlabel\trace
$$
$$
\eqalign{
\bigl[J(1,\ldots&,j),J(j{+}1,\ldots,n)\bigr]_{\xi} =
\cr & =
2\kappa(j{+}1,n)\cdot J(1,\ldots,j) \thinspace
J_{\xi}(j{+}1,\ldots,n)
\cr & \qquad -
2\kappa(1,j)\cdot J(j{+}1,\ldots,n) \thinspace
J_{\xi}(1,\ldots,j)
\cr & \qquad +
J(1,\ldots,j)\cdot J(j{+}1,\ldots,n) \thinspace
[\kappa(1,j)-\kappa(j{+}1,n)]_{\xi},
}
\eqlabel\sqbrak
$$
and
$$
\eqalign{
\bigl\{J(1),J(2),J(3)\bigr\}_{\xi}&=
J(1)\cdot
[J(3)\thinspace J_{\xi}(2)
- J(2)\thinspace J_{\xi}(3)]
\cr & \qquad
-J(3)\cdot
[J(2)\thinspace J_{\xi}(1)
- J(1)\thinspace J_{\xi}(2)].
}
\eqlabel\curlbrak
$$

The current $J(1,\ldots,n)$ satisfies the following properties
\cite\QCDrecursions:
$$
J(1,2,\ldots,n)=(-1)^{n-1}J(n,n{-}1,\ldots,1),
\newlettlabel\usefulproperties
$$
$$
{\sum_{{\cal{C}}(1\ldots n)}}
J(1,\ldots,n)=0,
\lett
$$
$$
\sum_{j=1}^{n}
J(2,\ldots,j,1,j{+}1,\ldots,n) = 0,
\lett
$$
$$
\kappa(1,n)\thinspace\cdot\thinspace J(1,\ldots,n) = 0.
\lett
$$

The recursion relation (\gluonrecursion) may be solved easily
for any number of gauge bosons in two special helicity configurations.
The key ingredient \cite\QCDrecursions\
is the ability to choose gauge spinors such that
$$
\eps_{\alpha\dot\alpha}(i)
\bar\eps^{\dot\alpha\alpha}(j) = 0
\eqlabel\gaugecondition
$$
for any pair of gauge bosons $i$ and $j$.  Thus, to consider the situation
in which all of the gauge bosons have the same helicity
(positive, for concreteness), we choose
$$
\eps_{\alpha\dot\alpha}(j^{+}) =
{
{ u_{\alpha}(h) \bar u_{\dot\alpha}(k_j) }
\over
{ \braket{j}{h} }
},
\qquad
j=1,2,\ldots,n,
\eqlabel\gaugeallplus
$$
with the same arbitrary null-momentum $h$ for each particle.
We will not have occasion to utilize the solution for the situation
in which one of the gauge bosons has negative helicity:  hence, we
will not present it here.  The interested reader is referred to
reference \cite\QCDrecursions.

The gauge choice (\gaugeallplus) has
the following consequences for the currents:
$$
J(1,\ldots,j)\cdot J(\ell,\ldots,n) = 0,
\newlettlabel\niceprops
$$
$$
\bigl\{
J(1,\ldots,j),J(j{+}1,\ldots,\ell),J(\ell{+}1,\ldots,n)
\bigr\}
=0,
\lett
$$
for any values of $j$, $\ell$, and $n$.  Furthermore, the square
bracket function simplifies to
$$
\eqalign{
\bigl[J(1,\ldots,j),&J(j{+}1,\ldots,n)\bigr]_{\xi} =
\cr & =
2\kappa(j{+}1,n)\cdot J(1,\ldots,j) \thinspace
J_{\xi}(j{+}1,\ldots,n)
\cr & \quad -
2\kappa(1,j)\cdot J(j{+}1,\ldots,n) \thinspace
J_{\xi}(1,\ldots,j).
}
\eqlabel\sqbrak
$$

We may use these simplifications plus some help from the
Schouten identity (\fierznc) to write the recursion relation
(\gluonrecursion) in spinor form as
$$
\Jms_{\alpha\dot\alpha}(1,\ldots,n)=
{
{-\sqrt2}
\over
{\kappa^2(1,n)}
}
\sum_{j=1}^{n-1}
\Jms_{\alpha\dot\beta}(1,\ldots,j)
\Jbar^{\dot\beta\beta}(j{+}1,\ldots,n)
\kappa_{\beta\dot\alpha}(1,n),
\eqlabel\gluonrecursioneasy
$$
valid for
this helicity configuration.

The solution to (\gluonrecursioneasy) is \cite\QCDrecursions
$$
\Jms_{\alpha\dot\alpha}(1^{+},\ldots,n^{+})=
u_{\alpha}(h) u^{\beta}(h) \kappa_{\beta\dot\alpha}(1,n)
X(1^{+},\ldots,n^{+}),
\newlettlabel\gluesolnallplus
$$
$$
X(1^{+},\ldots,n^{+})=
{
{(-\sqrt2)^{n-1}}
\over
{ \bra{h}1,\ldots,n\ket{h} }
}
\lett
$$
for the case where all of the gluons have the same
helicity.


\section{The quark current}

We now review the fermion currents presented by Berends and Giele
\cite\QCDrecursions.   These currents consist of a quark line
plus $n$ gluons in the tree approximation.  One end of the quark
line will be off shell.  In the following, all momenta flow into
the diagram.  We will denote the momentum of the quark by $p$ and
its color index by $i$.  The antiquark will have momentum $q$ and
color index $j$.  In the case where the quark is off shell we have,
by definition:
$$
\eqalign{
\psibarhat_{ji}(q;1,\ldots,n)&=
\permsum{1}{n} \sum_{\ell=0}^{n-1}
{
{1}
\over
{\ell! \thinspace (n-\ell)!}
}
\psibarhat_{jm}(q;1,\ldots,\ell)
\cr & \qquad \times
(-ig)(T^x)_{mi}
\gamma^{\xi}
{\Jhat}^x_{\xi}(\ell{+}1,\ldots,n)
{
{-i}
\over
{\slash{q}+\slash{\kappa}(1,n)}
}.
}
\eqlabel\quarkstart
$$
This simplifies to
$$
\psibarhat_{ji}(q;1,\ldots,n)=
g^n \permsum{1}{n} (\Omega[1,n])_{ji}
\psibar(q;1,\ldots,n),
\newlettlabel\quarkrecursion
$$
where
$$
\eqalign{
\psibar&(q;1,\ldots,n)=
\cr & =
{
{-1}
\over
{ [q+\kappa(1,n)]^2 }
}
\sum_{\ell=0}^{n-1}
\psibar(q;1,\ldots,\ell)
\Jslash(\ell{+}1,\ldots,n)
[\slash{q}+ \slash\kappa(1,n) ].
}
\lett
$$
Since the helicity of the quark line is conserved in the massless
limit, (\quarkrecursion{b}) has two translations to Weyl spinors:
$$
\eqalign{
\psibar^{\alpha}&(q^{-};1,\ldots,n)=
\cr & =
{
{-\sqrt{2}}
\over
{ [q+\kappa(1,n)]^2 }
}
\sum_{\ell=0}^{n-1}
\psibar^{\beta}(q^{-};1,\ldots,\ell)
\Jms_{\beta\dot\alpha}(\ell{+}1,\ldots,n)
[\bar q+ \bar\kappa(1,n) ]^{\dot\alpha\alpha}
}
\newlettlabel\quarkrecursionms
$$
for a left-handed antiquark and
$$
\eqalign{
\psibar_{\dot\alpha}&(q^{+};1,\ldots,n)=
\cr & =
{
{-\sqrt{2}}
\over
{ [q+\kappa(1,n)]^2 }
}
\sum_{\ell=0}^{n-1}
\psibar_{\dot\beta}(q^{+};1,\ldots,\ell)
\Jbar^{\dot\beta\alpha}(\ell{+}1,\ldots,n)
[{q}+ \kappa(1,n) ]_{\alpha\dot\alpha}
}
\lett
$$
for a right-handed antiquark.

In analogous fashion, when the antiquark is off shell we have$$
\eqalign{
\psihat_{ji}(1,\ldots,n;p)&=
\permsum{1}{n} \sum_{\ell=1}^{n}
{
{1}
\over
{\ell! \thinspace (n-\ell)!}
}
{
{i}
\over
{\slash{\kappa}(1,n)+\slash{p}}
}
\cr & \qquad \times
(-ig)(T^x)_{jm}
\gamma^{\xi}
{\Jhat}^x_{\xi}(1,\ldots,\ell)
\psihat_{mi}(\ell{+}1,\ldots,n;p),
}
\eqlabel\antiquarkstart
$$
which simplifies to
$$
\psihat_{ji}(1,\ldots,n;p)=
g^n \permsum{1}{n} (\Omega[1,n])_{ji}
\psi(1,\ldots,n;p),
\newlettlabel\antiquarkrecursion
$$
where
$$
\eqalign{
\psi&(1,\ldots,n;p)=
\cr & =
{
{1}
\over
{ [\kappa(1,n)+p]^2 }
}
\sum_{\ell=1}^{n}
[\slash\kappa(1,n) +\slash{p}]
\Jslash(1,\ldots,\ell)
\psi(\ell{+}1,\ldots,n;p).
}
\lett
$$
In terms of Weyl spinors, (\antiquarkrecursion{b}) reads:
$$
\eqalign{
\psi_{\alpha}&(1,\ldots,n;p^{-})=
\cr & =
{
{\sqrt{2}}
\over
{ [\kappa(1,n)+p]^2 }
}
\sum_{\ell=1}^{n}
[\kappa(1,n) +p]_{\alpha\dot\alpha}
\Jbar^{\dot\alpha\beta}(1,\ldots,\ell)
\psi_{\beta}(\ell{+}1,\ldots,n;p^{-})
}
\newlettlabel\antiquarkrecursionms
$$
for a left-handed quark and
$$
\eqalign{
\psi^{\dot\alpha}&(1,\ldots,n;p^{+})=
\cr & =
{
{\sqrt{2}}
\over
{ [\kappa(1,n)+p]^2 }
}
\sum_{\ell=1}^{n}
[ \bar\kappa(1,n)+\bar p ]^{\dot\alpha\alpha}
\Jms_{\alpha\dot\beta}(1,\ldots,\ell)
\psi^{\dot\beta}(\ell{+}1,\ldots,n;p^{+})
}
\lett
$$
for a right-handed quark.  It is worth mentioning that
$$
\psi^{\dot\alpha}(1,\ldots,n;p^{+})= (-1)^n
{\vareps}^{\dot\alpha\dot\beta}
\psibar_{\dot\beta}(p^{+};n,n{-}1,\ldots,1)
\newlettlabel\chargeconjugate
$$
and
$$
\psi_{\alpha}(1,\ldots,n;p^{-})= (-1)^n
\psibar^{\beta}(p^{-};n,n{-}1,\ldots,1)
{\vareps}_{\beta\alpha}.
\lett
$$

The recursion relations (\quarkrecursionms) and (\antiquarkrecursionms)
are easily solved when all of the gluons have the same helicity.
The gauge choice for the gluons is the same as that used for the
pure gluonic current, namely (\gaugeallplus).  The solutions  for
$n\ge1$ are
$$
\psibar^{\alpha}(q^{-};1^{+},\ldots,n^{+}) =
-\sqrt2 \thinspace
u^{\alpha}(q) \braket{q}{h}
{
{\braket{h}{1}}
\over
{\braket{q}{1}}
}
X(1^{+},\ldots,n^{+})
\eqlabel\LHantiquarkallplus
$$
and
$$
\psibar_{\dot\alpha}(q^{+};1^{+},\ldots,n^{+}) =
\sqrt2 \thinspace
u^{\beta}(h)[q+\kappa(1,n)]_{\beta\dot\alpha}
{
{\braket{h}{1}}
\over
{\braket{q}{1}}
}
X(1^{+},\ldots,n^{+}),
\eqlabel\RHantiquarkallplus
$$
where the scalar function $X(1^{+},\ldots,n^{+})$ is given
by (\gluesolnallplus{b}).  For $n=0$ we have simply
$$
\psibar^{\alpha}(q^{-})=u^{\alpha}(q),
\eqlabel\LHantiquarkzero
$$
$$
\psibar_{\dot\alpha}(q^{+})=\bar u_{\dot\alpha}(q),
\eqlabel\RHantiquarkzero
$$
consistent with (\LHantiquarkallplus) and (\RHantiquarkallplus).
The off-shell antiquark currents are easily obtained from
(\chargeconjugate).

\chapter{Currents with Several Off-Shell Gluons}

In this section we will begin by presenting a generalization
of the Berends and Giele gluon recursion relation \cite\QCDrecursions\
which allows two or more of the gluons to be off shell.
We will then specialize to the situation in which one gluon
is off shell, one gluon has a ``generalized'' polarization spinor,
and the remaining gluons are on shell.  If all of the on-shell
gluons have the same helicity, the generalized recursion relation
simplifies, and we are able to solve it for arbitrary $n$.
As we shall see in Section 4, this quantity may be used to aid
in the computation of the process (\monsteramplitude).

\section{The recursion relation}

We define the current $\Ihat^x_{\xi}(1^*,2^*,\ldots,n^*)$ to
consist of the sum of all tree graphs with $n{+}1$ external
gluons.  The gluons labelled by $\{1,2,\ldots,n\}$ have momenta
$k_1,k_2,\ldots,k_n$, color indices $a_1,a_2,\ldots,a_n$, and
special polarization vectors $\eps(1^*),\eps(2^*),\ldots,\eps(n^*)$.
These special polarizations are included to avoid a proliferation
of indices:  in general,
$$
k_j \cdot \eps(j) \ne 0.
\eqlabel\nottransverse
$$
Furthermore, we allow for
$$
k_j^2 \ne 0.
\eqlabel\notonshell
$$
We have not included the propagator factors for theses gluons.
The last gluon has momentum $k_{n+1}=-\kappa(1,n)$,
color index $x$, and Lorentz index $\xi$.  We {\it do}\ include
the propagator for this gluon in the definition of $\Ihat$.

Because of (\nottransverse) and (\notonshell), it is easy to
obtain currents with any number of off-shell gluons from
$\Ihat^x_{\xi}(1^*,2^*,\ldots,n^*)$.  All that is required
is to remove the appropriate polarization vectors, and supply a
propagator for each off-shell gluon.  ``Normal'' polarization
vectors may be used to replace the polarizations of the remaining
gluons which are to be on shell.

An examination of the derivation given by Berends and Giele
\cite\QCDrecursions\ for the gluon recursion relation (\gluonrecursion)
reveals that the kinematic and color structures factorize
quite early in the derivation.  As a result, it is immediately
obvious that $\Ihat$ satisfies the same factorization
property as $\Jhat$ and  the recursion relation for $\Ihat$ has the same
form as (\gluonrecursion), namely,
$$
\Ihat^x_{\xi}(1^*,\ldots,n^*) =
2g^{n-1} \permsum{1}{n}
tr(\Omega[1,n]T^x) I_{\xi}(1^*,\ldots,n^*),
\newlettlabel\gengluonrecursion
$$
$$
\eqalign{
I_{\xi}(1^*,\ldots,n^*)&=
{ 1\over{\kappa^2(1,n)} }
\Biggl(
\sum_{j=1}^{n-1}
[\negthinspace[ I(1^*,\ldots, j^*),
I((j{+}1)^*,\ldots,n^*) ]\negthinspace]_{\xi}
\cr &
+
\sum_{j=1}^{n-2} \sum_{\ell=j+1}^{n-1}
\bigl\{ I(1^*,\ldots,j^*),I((j{+}1)^*,\ldots,\ell^*),
        I((\ell{+}1)^*,\ldots,n^*) \bigr\}_{\xi}
\Biggr),
}
\lett
$$
$$
I(1^*)=\eps(1^*).
\lett
$$
The only difference between (\gengluonrecursion b) and
(\gluonrecursion b)
is the appearance of a new square bracket function, defined
by
$$
\eqalign{
[\negthinspace[I(1^*,\ldots&,j^*),
I((j{+}1)^*,\ldots,n^*)]\negthinspace]_{\xi} =
\cr & =
[2\kappa(j{+}1,n)+\kappa(1,j)]\cdot I(1^*,\ldots,j^*) \thinspace
I_{\xi}((j{+}1)^*,\ldots,n^*)
\cr & \qquad -
[2\kappa(1,j)+\kappa(j{+}1,n)]\cdot I((j{+}1)^*,\ldots,n^*) \thinspace
I_{\xi}(1^*,\ldots,j^*)
\cr & \qquad +
I(1^*,\ldots,j^*)\cdot I((j{+}1)^*,\ldots,n^*) \thinspace
[\kappa(1,j)-\kappa(j{+}1,n)]_{\xi},
}
\eqlabel\gensqbrak
$$
The curly bracket function remains as in (\curlbrak).
A comparison of (\gensqbrak) with (\sqbrak) reveals that the
differences between the two are generated by the fact that
$I(1^*,\ldots,n^*)$ is not a conserved current, that is
$$
\kappa(1,n)\cdot I(1^*,\ldots,n^*) \ne 0.
\eqlabel\noCC
$$
In spite of the loss of current conservation,  $I$ still
satisfies
the following properties:
$$
I(1^*,2^*,\ldots,n^*) = (-1)^n I(n^*,\ldots,2^*,1^*),
\newlettlabel\genprops
$$
$$
{\sum_{{\cal{C}}(1\ldots n)}}
I(1^*,\ldots,n^*)=0,
\lett
$$
$$
\sum_{j=1}^{n}
I(2^*,\ldots,j^*,1^*,(j{+}1)^*,\ldots,n^*) = 0,
\lett
$$
corresponding to (\usefulproperties a)--(\usefulproperties c).

\section{The modified gluon current}

We will now consider
$\Ims_{\alpha\dot\alpha}(1^{*},2^{+},\ldots,n^{+})$,
a current with just one ``generalized'' polarization vector
and $n-1$ like-helicity on-shell gluons.
The polarization spinors for the gluons labelled $2,3,\ldots,n$
are given by (\gaugeallplus).  The ``generalized'' polarization
spinor for the first gluon is defined to be
$$
\Ims_{\alpha\dot\alpha}(1^{*}) \equiv
u_{\alpha}(h) u^{\beta}(h) k_{1\beta\dot\alpha}
\eqlabel\Isingle
$$
with
$$
k_1^2 \ne 0.
\eq
$$
As we shall see in Section 4, this form of $\Ims(1^*)$ appears
when we consider the computation of the process (\monsteramplitude).
Even though $k_1^2$ does not vanish, we still have the relation
$$
\bar k_1^{\dot\alpha\alpha} \Ims_{\alpha\dot\alpha}(1^{*})=0
\eqlabel\tranzverse
$$
as if $\Ims(1^{*})$ were a true polarization
spinor.  Because of (\tranzverse), and since all of the other
gluons have ``normal'' polarizations, $\Ims(1^*,2^{+},\ldots,n^{+})$
is a conserved current, in contrast to the more general
case leading to (\noCC).  Thus, the square bracket function in
(\gensqbrak) reduces to the form (\sqbrak).
Moreover, $\Ims(1^{*})$ is proportional to $u_{\alpha}(h)$.
This means that key properties
({\it cf.}{ }equations (\gaugecondition), (\niceprops),
and (\sqbrak))
leading to the simplified form of the recursion relation
(\gluonrecursioneasy)
for $\Jms$ still hold when one of the $\Jms$'s is replaced
by an $\Ims$.  Thus,
$$
\eqalign{
\Ims_{\alpha\dot\alpha}&(1^{*},2^{+},\ldots,n^{+})=
\cr & =
{
{-\sqrt2}
\over
{\kappa^2(1,n)}
}
\sum_{j=1}^{n-1}
\Ims_{\alpha\dot\beta}(1^{*},2^{+},\ldots,j^{+})
\Jbar^{\dot\beta\beta}\bigl((j{+}1)^{+},\ldots,n^{+}\bigr)
\kappa_{\beta\dot\alpha}(1,n),
}
\eqlabel\Irecursion
$$
with (\Isingle) giving the starting point
$ \Ims(1^{*}) $.

The next current, $\Ims(1^{*},2^{+})$, is found by direct calculation
from (\Irecursion) to be
$$
\Ims_{\alpha\dot\alpha}(1^{*},2^{+}) =
-\sqrt2 \thinspace
{
{u_{\alpha}(h) u^{\beta}(h) (k_1+k_2)_{\beta\dot\alpha} }
\over
{\braket{2}{h}}
}
\thinspace
{
{ u^{\gamma}(h)k_{1\gamma\dot\gamma}\bar u^{\dot\gamma}(k_2) }
\over
{ (k_1+k_2)^2 }
},
\eqlabel\Itwo
$$
which should be compared to
$$
\Jms_{\alpha\dot\alpha}(1^{+},2^{+}) =
-\sqrt2 \thinspace
{
{u_{\alpha}(h) u^{\beta}(h) (k_1+k_2)_{\beta\dot\alpha} }
\over
{\braket{2}{h}}
}
\thinspace
{
{1}
\over
{\braket{h}{1}\thinspace\braket{1}{2}}
},
\eqlabel\Jtwo
$$
as obtained from (\gluesolnallplus).  If it were true that
$k_1^2=0$, then we would have
$$
{
{ u^{\gamma}(h)k_{1\gamma\dot\gamma}\bar u^{\dot\gamma}(k_2) }
\over
{ (k_1+k_2)^2 }
} =
{
{\braket{h}{1} {\braket{2}{1}}^{*} }
\over
{\braket{2}{1} {\braket{2}{1}}^{*} }
}
={
{\braket{h}{1} }
\over
{\braket{2}{1} }
}.
\eq
$$
Thus, we see that
$$
\lim_{k_1^2\rightarrow0}
\Ims_{\alpha\dot\alpha}(1^{*},2^{+})
=
\bra{h}1\ket{h}
\Jms_{\alpha\dot\alpha}(1^{+},2^{+}).
\eqlabel\suggestive
$$
Equation (\suggestive) suggests  that we try the following ansatz for
$\Ims(1^{*},2^{+},\ldots,n^{+})$, valid for $n\ge~2$:
$$
\eqalign{
\Ims_{\alpha\dot\alpha}&(1^{*},2^{+},\ldots,n^{+}) =
\cr & =
(-\sqrt2)^{n-1}
{
{ u_{\alpha}(h) u^{\beta}(h) \kappa_{\beta\dot\alpha}(1,n) }
\over
{ \bra{2} 3,\ldots,n \ket{h} }
}
\Biggl[
{
{ u^{\gamma}(h)k_{1\gamma\dot\gamma}\bar u^{\dot\gamma}(k_2) }
\over
{ (k_1+k_2)^2 }
}
+ k_1^2 \Lambda_n
\Biggr].
}
\eqlabel\firstansatz
$$
The first term of (\firstansatz) reduces to
$\bra{h}1\ket{h}
\Jms_{\alpha\dot\alpha}(1^{+},2^{+},\ldots,n^{+})$
in the limit $k_1^2\rightarrow0$, while the second term,  with
its undetermined function $\Lambda_n$, vanishes in that limit.
It is obvious from (\Jtwo) that
$$
\Lambda_2=0.
\eqlabel\seedLambda
$$

We now prove (\firstansatz) by mathematical induction.  Assume
that (\firstansatz) is true for $\Ims(1^{*},2^{+},\ldots,\ell^{+})$
with $\ell<n$.  The recursion relation (\Irecursion) along with
the known solution (\gluesolnallplus) for $\Jms$ tell us
that
$$
\eqalign{
\Ims_{\alpha\dot\alpha}&(1^{*},2^{+},\ldots,n^{+})=
\cr & =
{-\sqrt2}
\biggl[
\Ims_{\alpha\dot\beta}(1^{*})
\Jbar^{\dot\beta\beta}\bigl(2^{+},\ldots,n^{+}\bigr)
\cr & \qquad\qquad +
\sum_{\ell=2}^{n-1}
\Ims_{\alpha\dot\beta}(1^{*},2^{+},\ldots,\ell^{+})
\Jbar^{\dot\beta\beta}\bigl((\ell{+}1)^{+},\ldots,n^{+}\bigr)
\biggr]
{
{\kappa_{\beta\dot\alpha}(1,n)}
\over
{\kappa^2(1,n)}
}
\cr & =
{
{ -(-\sqrt2)^{n-1} }
\over
{\kappa^2(1,n)}
}
{
{ u_{\alpha}(h) u^{\beta}(h)\kappa_{\beta\dot\alpha}(1,n) }
\over
{\bra{2}3,\ldots,n\ket{h} }
}
\Biggl\{
{
{ u^{\delta}(h)k_{1\delta\dot\delta}
\bar\kappa^{\dot\delta\gamma}(2,n)u_{\gamma}(h) }
\over
{ \braket{h}{2} }
}
\cr & \qquad
+
\sum_{\ell=2}^{n-1}
{
{\braket{\ell}{\ell{+}1}}
\over
{\bra{\ell}h\ket{\ell{+}1}}
}
u^{\delta}(h)\kappa_{\delta\dot\delta}(1,\ell)
\bar\kappa^{\dot\delta\eps}(\ell{+}1,n)u_{\eps}(h)
{
{ u^{\gamma}(h)k_{1\gamma\dot\gamma}\bar u^{\dot\gamma}(k_2) }
\over
{ (k_1+k_2)^2 }
}
\cr & \qquad+
\sum_{\ell=2}^{n-1}
{
{\braket{\ell}{\ell{+}1}}
\over
{\bra{\ell}h\ket{\ell{+}1}}
}
u^{\delta}(h)\kappa_{\delta\dot\delta}(1,\ell)
\bar\kappa^{\dot\delta\eps}(\ell{+}1,n)u_{\eps}(h)
k_1^2 \Lambda_\ell
\Biggr\}.
}
\eqlabel\messyone
$$
Let us denote the second term in the curly brackets of
(\messyone) by $S$.  Writing out the implicit $\kappa$-sums
we have
$$
S =
{
{ u^{\gamma}(h)k_{1\gamma\dot\gamma}\bar u^{\dot\gamma}(k_2) }
\over
{ (k_1+k_2)^2 }
}
\sum_{\ell=2}^{n-1} \sum_{i=1}^{\ell} \sum_{j=\ell+1}^{n}
{
{\braket{\ell}{\ell{+}1}}
\over
{\bra{\ell}h\ket{\ell{+}1}}
}
u^{\delta}(h)k_{i\delta\dot\delta}
\bar k_j^{\dot\delta\eps}u_{\eps}(h).
\eqlabel\Sstart
$$
Because $k_1^2\ne0$,
$$
k_{1\delta\dot\delta} \ne u_{\delta}(k_1) \bar u_{\dot\delta}(k_1),
\eqlabel\spinormess
$$
and we must treat the $i=1$ portion of (\Sstart) separately.  Doing
this  and changing the order of the summations produces
$$
\eqalign{
S   =
{
{ u^{\gamma}(h)k_{1\gamma\dot\gamma}\bar u^{\dot\gamma}(k_2) }
\over
{ (k_1+k_2)^2 }
}&
\Biggl[
\sum_{i=2}^{n-1} \sum_{j=i+1}^n \sum_{\ell=i}^{j-1}
{
{\braket{\ell}{\ell{+}1}}
\over
{\bra{\ell}h\ket{\ell{+}1}}
}
\braket{h}{i} {\braket{j}{i}}^{*} \braket{j}{h}
\cr &  +
\sum_{j=3}^n \sum_{\ell=2}^{j-1}
{
{\braket{\ell}{\ell{+}1}}
\over
{\bra{\ell}h\ket{\ell{+}1}}
}
u^{\delta}(h)k_{1\delta\dot\delta}\bar u^{\dot\delta}(k_j)
\braket{j}{h}
\Biggr].
}
\eqlabel\Stwo
$$
The sums on $\ell$ may be performed using (\linkidsummed), with
the result
$$
\eqalign{
S &  =
{
{ u^{\gamma}(h)k_{1\gamma\dot\gamma}\bar u^{\dot\gamma}(k_2) }
\over
{ (k_1+k_2)^2 }
}
\Biggl[
\sum_{i=2}^{n-1} \sum_{j=i+1}^n
{
{\braket{i}{j}}
\over
{\bra{i}h\ket{j}}
}
\braket{h}{i} {\braket{j}{i}}^{*} \braket{j}{h}
\cr & \qquad\qquad\qquad\qquad +
\sum_{j=3}^n
{
{\braket{2}{j}}
\over
{\bra{2}h\ket{j}}
}
u^{\delta}(h)k_{1\delta\dot\delta}\bar u^{\dot\delta}(k_j)
\braket{j}{h}
\Biggr]
\cr & =
{
{ u^{\gamma}(h)k_{1\gamma\dot\gamma}\bar u^{\dot\gamma}(k_2) }
\over
{ (k_1+k_2)^2 }
}
\Biggl[
-\kappa^2(2,n) -
{
{ u^{\delta}(h)k_{1\delta\dot\delta}
\bar\kappa(3,n)^{\dot\delta\eps}u_{\eps}(k_2) }
\over
{ \braket{h}{2} }
}
\Biggr].
}
\eqlabel\Sthree
$$
The Weyl equation allows us to extend the sum in the numerator
in the second term of (\Sthree) to $\kappa(2,n)$.  We
may then write $k_1=\kappa(1,n)-\kappa(2,n)$ to obtain
$$
\eqalign{
-u^{\delta}(h)k_{1\delta\dot\delta}
\bar\kappa(3,n)^{\dot\delta\eps}u_{\eps}(k_2)& =
-u^{\delta}(h)\kappa_{\delta\dot\delta}(1,n)
\bar\kappa(2,n)^{\dot\delta\eps}u_{\eps}(k_2)
\cr & \quad
+ \braket{h}{2} \kappa^2(2,n).
}
\eqlabel\numredo
$$
Substituting (\numredo) into (\Sthree) produces
$$
\eqalign{
S &  =-
{
{ u^{\gamma}(h)k_{1\gamma\dot\gamma}\bar u^{\dot\gamma}(k_2) }
\over
{ (k_1+k_2)^2 }
}
{
{u^{\delta}(h)\kappa_{\delta\dot\delta}(1,n)
\bar\kappa(2,n)^{\dot\delta\eps}u_{\eps}(k_2) }
\over
{ \braket{h}{2} }
}
\cr & = -
{
{u^{\delta}(h)\kappa_{\delta\dot\delta}(1,n)
\bar\kappa^{\dot\delta\eps}(2,n)k_{2\eps\dot\gamma}
\bar k_1^{\dot\gamma\gamma}u_{\gamma}(h) }
\over
{ \braket{h}{2} \thinspace
(k_1+k_2)^2 }
}.
}
\eqlabel\Sfour
$$
Repeated use of this trick yields
$$
\eqalign{
S & =
-{
{u^{\delta}(h)\kappa_{\delta\dot\delta}(1,n)
\bar\kappa^{\dot\delta\eps}(2,n)k_{2\eps\dot\gamma}
(\bar k_1 + \bar k_2)^{\dot\gamma\gamma}u_{\gamma}(h) }
\over
{ \braket{h}{2} \thinspace
(k_1+k_2)^2 }
}
\cr & =
-{
{u^{\delta}(h)\kappa_{\delta\dot\delta}(1,n)
\bar\kappa^{\dot\delta\gamma}(2,n)u_{\gamma}(h) }
\over
{ \braket{h}{2} }
}
\cr & \quad +
{
{u^{\delta}(h)\kappa_{\delta\dot\delta}(1,n)
\bar\kappa^{\dot\delta\eps}(2,n)k_{1\eps\dot\gamma}
(\bar k_1 + \bar k_2)^{\dot\gamma\gamma}u_{\gamma}(h) }
\over
{ \braket{h}{2} \thinspace
(k_1+k_2)^2 }
}
\cr & =
-{
{u^{\delta}(h)k_{1\delta\dot\delta}
\bar\kappa^{\dot\delta\gamma}(2,n)u_{\gamma}(h) }
\over
{ \braket{h}{2} }
}
\cr & \quad -
\kappa^2(1,n)
{
{u^{\delta}(h)k_{1\delta\dot\delta}
\bar u^{\delta}(k_2) }
\over
{ (k_1+k_2)^2 }
}
\cr & \quad -
k_1^2
{
{u^{\delta}(h)\kappa_{\delta\dot\gamma}(1,n)
(\bar k_1 + \bar k_2)^{\dot\gamma\gamma}u_{\gamma}(h) }
\over
{ \braket{h}{2} \thinspace
(k_1+k_2)^2 }
}.
}
\eqlabel\Sdone
$$
By inserting (\Sdone) back into (\messyone) we arrive at
$$
\eqalign{
\Ims_{\alpha\dot\alpha}&(1^{*},2^{+},\ldots,n^{+})=
\cr & =
{
{ (-\sqrt2)^{n-1}
u_{\alpha}(h) u^{\beta}(h)\kappa_{\beta\dot\alpha}(1,n) }
\over
{\bra{2}3,\ldots,n\ket{h} }
}
\Biggl\{
{
{u^{\delta}(h)k_{1\delta\dot\delta}
\bar u^{\delta}(k_2) }
\over
{ (k_1+k_2)^2 }
}
\cr &\qquad +
{
{ k_1^2}
\over
{\kappa^2(1,n)}
}
\Biggl[
{
{u^{\delta}(h)\kappa_{\delta\dot\gamma}(1,n)
(\bar k_1 + \bar k_2)^{\dot\gamma\gamma}u_{\gamma}(h) }
\over
{ \braket{h}{2} \thinspace
(k_1+k_2)^2 }
}
\cr &\qquad -
\sum_{\ell=2}^{n-1}
\Lambda_\ell \thinspace
{
{\braket{\ell}{\ell{+}1}}
\over
{\bra{\ell}h\ket{\ell{+}1}}
}
u^{\delta}(h)\kappa_{\delta\dot\delta}(1,\ell)
\bar\kappa^{\dot\delta\eps}(\ell{+}1,n)u_{\eps}(h)
\Biggr]
\Biggr\},
}
\eq
$$
which satisfies the ansatz (\firstansatz) if we insist
upon the following recursion relation for $\Lambda_n$:
$$
\eqalign{
\Lambda_n =
{
{1}
\over
{\kappa^2(1,n)}
}&
\Biggl[
{
{u^{\delta}(h)\kappa_{\delta\dot\gamma}(1,n)
(\bar k_1 + \bar k_2)^{\dot\gamma\gamma}u_{\gamma}(h) }
\over
{ \braket{h}{2} \thinspace
(k_1+k_2)^2 }
}
\cr & -
\sum_{\ell=2}^{n-1}
\Lambda_\ell \thinspace
{
{\braket{\ell}{\ell{+}1}}
\over
{\bra{\ell}h\ket{\ell{+}1}}
}
u^{\delta}(h)\kappa_{\delta\dot\delta}(1,\ell)
\bar\kappa^{\dot\delta\eps}(\ell{+}1,n)u_{\eps}(h)
\Biggr].
}
\eqlabel\LAMBDArecursion
$$

Since $\Lambda_2$ vanishes, (\LAMBDArecursion) tells us that
$$
\Lambda_3 =
{
{u^{\delta}(h) k_{3\delta\dot\gamma}
(\bar k_1 + \bar k_2)^{\dot\gamma\gamma}u_{\gamma}(h) }
\over
{ \braket{h}{2} \thinspace
(k_1+k_2)^2 \thinspace (k_1+k_2+k_3)^2}
}.
\eqlabel\LAMBDAthree
$$
Use of (\LAMBDAthree) in (\LAMBDArecursion) to obtain $\Lambda_4$
yields
$$
\eqalign{
\Lambda_4& =
{
{u^{\delta}(h) \kappa_{\delta\dot\gamma}(1,4)
\bar\kappa^{\dot\gamma\gamma}(1,2)u_{\gamma}(h) }
\over
{ \braket{h}{2} \thinspace
\kappa^2(1,2) \thinspace \kappa^2(1,4)}
}
\cr &\quad +
{
{u^{\delta}(h) \kappa_{\delta\dot\delta}(1,3)
\bar k_4^{\dot\delta\gamma}
k_{3\gamma\dot\gamma}
\bar\kappa^{\dot\gamma\eps}(1,2)
u_{\eps}(h) }
\over
{ \braket{h}{2} \thinspace
\kappa^2(1,2) \thinspace \kappa^2(1,3) \thinspace \kappa^2(1,4)}
}.
}
\eqlabel\LAMBDAfourstart
$$
The  numerator in the second term of (\LAMBDAfourstart) may be
rewritten as follows:
$$
\eqalign{
\num & \equiv
u^{\delta}(h) \kappa_{\delta\dot\delta}(1,3)
\bar k_4^{\dot\delta\gamma}
k_{3\gamma\dot\gamma}
\bar\kappa^{\dot\gamma\eps}(1,2)
u_{\eps}(h)
\cr & =
u^{\delta}(h) \kappa_{\delta\dot\delta}(1,4)
\bar k_4^{\dot\delta\gamma}
\kappa_{\gamma\dot\gamma}(1,3)
\bar\kappa^{\dot\gamma\eps}(1,2)
u_{\eps}(h)
\cr & \quad -
\kappa^2(1,2)
u^{\delta}(h) \kappa_{\delta\dot\delta}(1,3)
\bar k_4^{\dot\delta\gamma}
u_{\gamma}(h)
\cr & =
\kappa^2(1,4)
u^{\delta}(h)
k_{3\delta\dot\gamma}
\bar\kappa^{\dot\gamma\eps}(1,2)
u_{\eps}(h)
\cr & \quad -
\kappa^2(1,3)
u^{\delta}(h) \kappa_{\delta\dot\delta}(1,4)
\bar\kappa^{\dot\delta\eps}(1,2)
u_{\eps}(h)
\cr & \quad +
\kappa^2(1,2)
u^{\gamma}(h)
k_{4\gamma\dot\delta}
\bar\kappa^{\dot\delta\delta}(1,3)
u_{\delta}(h),
}
\eqlabel\numfix
$$
where we have made repeated use of the Weyl equation,
the antisymmetry of the spinor product, and added
and subtracted terms as required to complete the squares.
When (\numfix) is combined with (\LAMBDAfourstart)
we obtain
$$
\eqalign{
\Lambda_4 &=
{
{u^{\delta}(h) k_{3\delta\dot\gamma}
\bar\kappa^{\dot\gamma\gamma}(1,2)u_{\gamma}(h) }
\over
{ \braket{h}{2} \thinspace
\kappa^2(1,2) \thinspace \kappa^2(1,3)}
}
+
{
{u^{\delta}(h) k_{4\delta\dot\gamma}
\bar\kappa^{\dot\gamma\gamma}(1,3)u_{\gamma}(h) }
\over
{ \braket{h}{2} \thinspace
\kappa^2(1,3) \thinspace \kappa^2(1,4)}
}.
}
\eqlabel\LAMBDAfour
$$

Equations (\LAMBDAthree) and (\LAMBDAfour) suggest the ansatz
$$
\Lambda_n =
{ {1}\over{\braket{h}{2}} }
\sum_{j=3}^n
{
{u^{\delta}(h) k_{j\delta\dot\gamma}
\bar\kappa^{\dot\gamma\gamma}(1,j)u_{\gamma}(h) }
\over
{\kappa^2(1,j{-}1) \thinspace \kappa^2(1,j)}
}.
\eqlabel\LAMBDAsolution
$$
We prove this ansatz  by mathematical induction.
Assuming $\Lambda_\ell$ to be given by (\LAMBDAsolution) for
$\ell<n$, $\Lambda_n$ is determined by (\LAMBDArecursion) to be
$$
\eqalign{
\Lambda_n& =
{
{u^{\delta}(h)\kappa_{\delta\dot\gamma}(1,n)
\bar\kappa^{\dot\gamma\gamma}(1,2)u_{\gamma}(h) }
\over
{ \braket{h}{2} \thinspace
\kappa^2(1,2)\thinspace\kappa^2(1,n) }
}
\cr & \quad -
\sum_{\ell=3}^{n-1}
\sum_{j=3}^{\ell}
{
{\braket{\ell}{\ell{+}1}}
\over
{\bra{\ell}h\ket{\ell{+}1}}
}
{
{ u^{\delta}(h)\kappa_{\delta\dot\delta}(1,\ell)
\bar\kappa^{\dot\delta\eps}(\ell{+}1,n)u_{\eps}(h) }
\over
{ \braket{h}{2} \thinspace\kappa^2(1,n) }
}
\cr & \qquad\qquad\qquad\qquad\times
{
{u^{\beta}(h) k_{j\beta\dot\gamma}
\bar\kappa^{\dot\gamma\gamma}(1,j)
u_{\gamma}(h) }
\over
{\kappa^2(1,j{-}1) \thinspace\kappa^2(1,j)}
}.
}
\eqlabel\itsgettingugly
$$
Let us interchange the order of the summations in the second term
of  (\itsgettingugly) and break $\kappa(1,\ell)$ into
$\kappa(1,j)+\kappa(j{+}1,\ell)$:
$$
\eqalign{
\Lambda_n& =
{
{u^{\delta}(h)\kappa_{\delta\dot\gamma}(1,n)
\bar\kappa^{\dot\gamma\gamma}(1,2)u_{\gamma}(h) }
\over
{ \braket{h}{2} \thinspace
\kappa^2(1,2)\thinspace\kappa^2(1,n) }
}
\cr & \quad -
\sum_{j=3}^{n-1}
\sum_{\ell=j}^{n-1}
{
{\braket{\ell}{\ell{+}1}}
\over
{\bra{\ell}h\ket{\ell{+}1}}
}
{
{ u^{\delta}(h)\kappa_{\delta\dot\delta}(1,j)
\bar\kappa^{\dot\delta\eps}(\ell{+}1,n)u_{\eps}(h) }
\over
{ \braket{h}{2} \thinspace\kappa^2(1,n) }
}
\cr & \qquad\qquad\qquad\qquad\times
{
{u^{\beta}(h) k_{j\beta\dot\gamma}
\bar\kappa^{\dot\gamma\gamma}(1,j)
u_{\gamma}(h) }
\over
{\kappa^2(1,j{-}1) \thinspace\kappa^2(1,j)}
}
\cr & \quad -
\sum_{j=3}^{n-2}
\sum_{\ell=j+1}^{n-1}
{
{\braket{\ell}{\ell{+}1}}
\over
{\bra{\ell}h\ket{\ell{+}1}}
}
{
{ u^{\delta}(h)\kappa_{\delta\dot\delta}(j{+}1,\ell)
\bar\kappa^{\dot\delta\eps}(\ell{+}1,n)u_{\eps}(h) }
\over
{ \braket{h}{2} \thinspace\kappa^2(1,n) }
}
\cr & \qquad\qquad\qquad\qquad\times
{
{u^{\beta}(h) k_{j\beta\dot\gamma}
\bar\kappa^{\dot\gamma\gamma}(1,j)
u_{\gamma}(h) }
\over
{\kappa^2(1,j{-}1) \thinspace\kappa^2(1,j)}
}.
}
\eqlabel\itsgettingREALLYugly
$$
We have adjusted the limits in the third term of
(\itsgettingREALLYugly) to reflect the vanishing of
$\kappa(j{+}1,\ell)$ for $j=\ell$.  Denote the three
contributions in (\itsgettingREALLYugly) by
$\zee_1$, $\zee_2$, and $\zee_3$ respectively.

We examine $\zee_3$ first.  Writing out the $\ell$-dependent
$\kappa$-sums yields
$$
\eqalign{
\zee_3 & =
-
\sum_{j=3}^{n-2}
\sum_{\ell=j+1}^{n-1}
\sum_{i=j+1}^{\ell}
\sum_{m=\ell+1}^n
{
{\braket{\ell}{\ell{+}1}}
\over
{\bra{\ell}h\ket{\ell{+}1}}
}
{
{ u^{\delta}(h)k_{i\delta\dot\delta}
\bar k_m^{\dot\delta\eps}u_{\eps}(h) }
\over
{ \braket{h}{2} \thinspace\kappa^2(1,n) }
}
\cr & \qquad\qquad\qquad\qquad\qquad\qquad\times
{
{u^{\beta}(h) k_{j\beta\dot\gamma}
\bar\kappa^{\dot\gamma\gamma}(1,j)
u_{\gamma}(h) }
\over
{\kappa^2(1,j{-}1) \thinspace\kappa^2(1,j)}
}.
}
\eq
$$
Next, we interchange the sum over $\ell$ with both the sum
on $i$ and the sum on $m$ and use (\linkidsummed) to do the
$\ell$-summation:
$$
\eqalign{
\zee_3 & =
-
\sum_{j=3}^{n-2}
\sum_{m=j+2}^n
\sum_{i=j+1}^{m-1}
\sum_{\ell=i}^{m-1}
{
{\braket{\ell}{\ell{+}1}}
\over
{\bra{\ell}h\ket{\ell{+}1}}
}
{
{ u^{\delta}(h)k_{i\delta\dot\delta}
\bar k_m^{\dot\delta\eps}u_{\eps}(h) }
\over
{ \braket{h}{2} \thinspace\kappa^2(1,n) }
}
\cr & \qquad\qquad\qquad\qquad\qquad\quad\times
{
{u^{\beta}(h) k_{j\beta\dot\gamma}
\bar\kappa^{\dot\gamma\gamma}(1,j)
u_{\gamma}(h) }
\over
{\kappa^2(1,j{-}1) \thinspace\kappa^2(1,j)}
}
\cr & =
-
\sum_{j=3}^{n-2}
\sum_{m=j+2}^n
\sum_{i=j+1}^{m-1}
{
{\braket{i}{m}}
\over
{\bra{i}h\ket{m}}
}
{
{ \braket{h}{i}  {\braket{m}{i}}^{*} \braket{m}{h} }
\over
{ \braket{h}{2} \thinspace\kappa^2(1,n) }
}
\cr & \qquad\qquad\qquad\qquad\quad\times
{
{u^{\beta}(h) k_{j\beta\dot\gamma}
\bar\kappa^{\dot\gamma\gamma}(1,j)
u_{\gamma}(h) }
\over
{\kappa^2(1,j{-}1) \thinspace\kappa^2(1,j)}
}.
}
\eqlabel\thatwasntsobad
$$
Recognizing that $\braket{i}{m}{\braket{m}{i}}^{*}=-2k_i\cdot k_m$
allows us to write (\thatwasntsobad) as
$$
\eqalign{
\zee_3 &=
\sum_{j=3}^{n-1}
{
{ \kappa^2(j{+}1,n) \thinspace
u^{\beta}(h) k_{j\beta\dot\gamma}
\bar\kappa^{\dot\gamma\gamma}(1,j)
u_{\gamma}(h) }
\over
{ \braket{h}{2} \thinspace\kappa^2(1,n)
\thinspace
\kappa^2(1,j{-}1) \thinspace\kappa^2(1,j)}
},
}
\eqlabel\zeethree
$$
where we have extended the sum to $j=n-1$ because $k_n^2=0$.

Setting $\zee_3$ aside, we turn to $\zee_2$.  There is only
one $\ell$-dependent $\kappa$-sum here, so we write
$$
\eqalign{
\zee_2 &=
-
\sum_{j=3}^{n-1}
\sum_{\ell=j}^{n-1}
\sum_{m=\ell+1}^{n}
{
{\braket{\ell}{\ell{+}1}}
\over
{\bra{\ell}h\ket{\ell{+}1}}
}
{
{ u^{\delta}(h)\kappa_{\delta\dot\delta}(1,j)
\bar k_m^{\dot\delta\eps}u_{\eps}(h) }
\over
{ \braket{h}{2} \thinspace\kappa^2(1,n) }
}
\cr & \qquad\qquad\qquad\qquad\times
{
{u^{\beta}(h) k_{j\beta\dot\gamma}
\bar\kappa^{\dot\gamma\gamma}(1,j)
u_{\gamma}(h) }
\over
{\kappa^2(1,j{-}1) \thinspace\kappa^2(1,j)}
}.
}
\eq
$$
Before we can do the sum on $\ell$, we must interchange it
with the sum on $m$. Once  this is accomplished, we  may
use (\linkidsummed) to perform the sum:
$$
\eqalign{
\zee_2 &=
-
\sum_{j=3}^{n-1}
\sum_{m=j+1}^{n}
\sum_{\ell=j}^{m-1}
{
{\braket{\ell}{\ell{+}1}}
\over
{\bra{\ell}h\ket{\ell{+}1}}
}
{
{ u^{\delta}(h)\kappa_{\delta\dot\delta}(1,j)
\bar k_m^{\dot\delta\eps}u_{\eps}(h) }
\over
{ \braket{h}{2} \thinspace\kappa^2(1,n) }
}
{
{u^{\beta}(h) k_{j\beta\dot\gamma}
\bar\kappa^{\dot\gamma\gamma}(1,j)
u_{\gamma}(h) }
\over
{\kappa^2(1,j{-}1) \thinspace\kappa^2(1,j)}
}
\cr & =
-
\sum_{j=3}^{n-1}
\sum_{m=j+1}^{n}
{
{\braket{j}{m}}
\over
{\bra{j}h\ket{m}}
}
{
{ u^{\delta}(h)\kappa_{\delta\dot\delta}(1,j)
\bar u^{\dot\delta}(k_m) \braket{m}{h} }
\over
{ \braket{h}{2} \thinspace\kappa^2(1,n) }
}
{
{\braket{h}{j}  \bar u_{\dot\gamma}(k_j)
\bar\kappa^{\dot\gamma\gamma}(1,j)
u_{\gamma}(h) }
\over
{\kappa^2(1,j{-}1) \thinspace\kappa^2(1,j)}
}
\cr & =
\sum_{j=3}^{n-1}
{
{ u^{\delta}(h)\kappa_{\delta\dot\delta}(1,j)
\bar\kappa^{\dot\delta\beta}(j{+}1,n)
k_{j\beta\dot\gamma}
\bar\kappa^{\dot\gamma\gamma}(1,j)
u_{\gamma}(h) }
\over
{ \braket{h}{2} \thinspace\kappa^2(1,n)
\thinspace
\kappa^2(1,j{-}1) \thinspace\kappa^2(1,j) }
}.
}
\eqlabel\blah
$$
Writing $\kappa_{\delta\dot\delta}(1,j)=
\kappa_{\delta\dot\delta}(1,n)-
\kappa_{\delta\dot\delta}(j{+}1,n)$ yields
$$
\eqalign{
\zee_2 &=
\sum_{j=3}^{n-1}
{
{ u^{\delta}(h)\kappa_{\delta\dot\delta}(1,n)
\bar\kappa^{\dot\delta\beta}(j{+}1,n)
k_{j\beta\dot\gamma}
\bar\kappa^{\dot\gamma\gamma}(1,j)
u_{\gamma}(h) }
\over
{ \braket{h}{2} \thinspace\kappa^2(1,n)
\thinspace
\kappa^2(1,j{-}1) \thinspace\kappa^2(1,j) }
}
\cr & \quad -
\sum_{j=3}^{n-1}
{
{  \kappa^2(j{+}1,n)\thinspace u^{\beta}(h)
k_{j\beta\dot\gamma}
\bar\kappa^{\dot\gamma\gamma}(1,j)
u_{\gamma}(h) }
\over
{ \braket{h}{2} \thinspace\kappa^2(1,n)
\thinspace
\kappa^2(1,j{-}1) \thinspace\kappa^2(1,j) }
}.
}
\eqlabel\zeetwo
$$
We recognize the second term in (\zeetwo) as precisely $-\zee_3$:
hence, when we form $\zee_{23}\equiv\zee_2+\zee_3$ we are left
with only
$$
\zee_{23}=
\sum_{j=3}^{n-1}
{
{ u^{\delta}(h)\kappa_{\delta\dot\delta}(1,n)
\bar\kappa^{\dot\delta\beta}(j{+}1,n)
k_{j\beta\dot\gamma}
\bar\kappa^{\dot\gamma\gamma}(1,j)
u_{\gamma}(h) }
\over
{ \braket{h}{2} \thinspace\kappa^2(1,n)
\thinspace
\kappa^2(1,j{-}1) \thinspace\kappa^2(1,j) }
}.
\eqlabel\Ztwothree
$$
Let us call the numerator of (\Ztwothree) $\num_{23}$.

We may use the Weyl equation and  the clever addition and subtraction
of terms to write $\num_{23}$ as the combination of three
terms containing perfect squares of momentum sums:
$$
\eqalign{
\num_{23} &=
u^{\delta}(h)\kappa_{\delta\dot\delta}(1,n)
\bar\kappa^{\dot\delta\beta}(j{+}1,n)
k_{j\beta\dot\gamma}
\bar\kappa^{\dot\gamma\gamma}(1,j)
u_{\gamma}(h)
\cr & =
\kappa^2(1,n) \thinspace u^{\delta}(h)
k_{j\delta\dot\gamma}
\bar\kappa^{\dot\gamma\gamma}(1,j)
u_{\gamma}(h)
\cr & \quad -
u^{\delta}(h)\kappa_{\delta\dot\delta}(1,n)
\bar\kappa^{\dot\delta\beta}(1,j)
k_{j\beta\dot\gamma}
\bar\kappa^{\dot\gamma\gamma}(1,j{-}1)
u_{\gamma}(h)
\cr & =
\kappa^2(1,n) \thinspace u^{\delta}(h)
k_{j\delta\dot\gamma}
\bar\kappa^{\dot\gamma\gamma}(1,j)
u_{\gamma}(h)
\cr & \quad -
\kappa^2(1,j) \thinspace
u^{\delta}(h)\kappa_{\delta\dot\delta}(1,n)
\bar\kappa^{\dot\delta\gamma}(1,j{-}1)
u_{\gamma}(h)
\cr & \quad +
\kappa^2(1,j{-}1) \thinspace
u^{\delta}(h)\kappa_{\delta\dot\delta}(1,n)
\bar\kappa^{\dot\delta\beta}(1,j)
u_{\beta}(h).
}
\eqlabel\breaknum
$$
Insertion of (\breaknum) into (\Ztwothree) gives
$$
\eqalign{
\zee_{23}&=
\sum_{j=3}^{n-1}
{
{  u^{\delta}(h)
k_{j\delta\dot\gamma}
\bar\kappa^{\dot\gamma\gamma}(1,j)
u_{\gamma}(h) }
\over
{ \braket{h}{2}
\thinspace
\kappa^2(1,j{-}1) \thinspace\kappa^2(1,j) }
}
\cr & \quad -
\sum_{j=3}^{n-1}
{
{u^{\delta}(h)\kappa_{\delta\dot\delta}(1,n)
  }
\over
{ \braket{h}{2} \thinspace \kappa^2(1,n) }
}
\Biggl[
{
{ \bar\kappa^{\dot\delta\gamma}(1,j{-}1)u_{\gamma}(h) }
\over
{\kappa^2(1,j{-}1) }
}
-
{
{ \bar\kappa^{\dot\delta\gamma}(1,j)u_{\gamma}(h) }
\over
{\kappa^2(1,j) }
}
\Biggr].
}
\eqlabel\almostdone
$$
Most of the terms in the second sum in (\almostdone) cancel.
The remainder reads
$$
\eqalign{
\zee_{23}&=
\sum_{j=3}^{n-1}
{
{  u^{\delta}(h)
k_{j\delta\dot\gamma}
\bar\kappa^{\dot\gamma\gamma}(1,j)
u_{\gamma}(h) }
\over
{ \braket{h}{2}
\thinspace
\kappa^2(1,j{-}1) \thinspace\kappa^2(1,j) }
}
\cr & \quad -
{
{u^{\delta}(h)\kappa_{\delta\dot\delta}(1,n)
\bar\kappa^{\dot\delta\gamma}(1,2)u_{\gamma}(h) }
\over
{ \braket{h}{2}\thinspace \kappa^2(1,2)\thinspace \kappa^2(1,n) }
}
+
{
{u^{\delta}(h)\kappa_{\delta\dot\delta}(1,n)
\bar\kappa^{\dot\delta\gamma}(1,n{-}1)u_{\gamma}(h) }
\over
{ \braket{h}{2} \thinspace \kappa^2(1,n{-}1) \thinspace \kappa^2(1,n) }
}.
}
\eqlabel\onelaststep
$$
If we compare the second term of (\onelaststep) to $\zee_1$
({\it i.e.}
the first term of  (\itsgettingREALLYugly)),  we see that these two
terms cancel.  In addition, the last term of (\onelaststep) provides
a $j=n$ term for the sum appearing in the first term.  Hence,
$$
\eqalign{
\Lambda_n &= \zee_1 + \zee_{23}
\cr & =
\sum_{j=3}^{n}
{
{  u^{\delta}(h)
k_{j\delta\dot\gamma}
\bar\kappa^{\dot\gamma\gamma}(1,j)
u_{\gamma}(h) }
\over
{ \braket{h}{2}
\thinspace
\kappa^2(1,j{-}1) \thinspace\kappa^2(1,j) }
},
}
\eqlabel\whew
$$
proving the ansatz (\LAMBDAsolution).

If we re-examine (\firstansatz), we see that the first term in
the square brackets there corresponds to a $j=2$ term for the
sum in $\Lambda_n$.  We may thus combine (\firstansatz) and
(\whew) to produce the  compact form
$$
\eqalign{
\Ims_{\alpha\dot\alpha}&(1^{*},2^{+},\ldots,n^{+}) =
\cr & =
(-\sqrt2)^{n-1} k_1^2
{
{ u_{\alpha}(h) u^{\beta}(h) \kappa_{\beta\alpha}(1,n) }
\over
{ \bra{h}2,\ldots,n\ket{h} }
}
\sum_{j=2}^{n}
{
{  u^{\delta}(h)
k_{j\delta\dot\gamma}
\bar\kappa^{\dot\gamma\gamma}(1,j)
u_{\gamma}(h) }
\over
{\kappa^2(1,j{-}1) \thinspace\kappa^2(1,j) }
},
}
\eqlabel\compactI
$$
valid for $n\ge2$.
\section{General Modified Gluon Currents}

In order to calculate the amplitude for the process (\monsteramplitude),
we require knowledge of $\Ims(1^{+},2^{*},3^{+},\ldots,n^{+})$,
$\Ims(1^{+},2^{+},3^{*},4^{+},\ldots,n^{+})$, {\it etc.}
We begin by defining some useful notation.

We may separate the expression for $\Ims$ into a factor containing
the spinor dependence and a scalar function.  It is easily verified
from the appropriate (modified) form of the recursion relation
(\gluonrecursioneasy)
that we may write
$$
\eqalign{
\Ims_{\alpha\dot\alpha}&(1^{+},\ldots,m^{*},\ldots,n^{+}) =
\cr & =
(-\sqrt2)^{n-1}
u_{\alpha}(h) u^{\beta}(h) \kappa_{\beta\dot\alpha}(1,n)
\Xms(1,\ldots,m^{*},\ldots,n).
}
\eqlabel\factorizeI
$$
For the case already discussed, (\compactI) tells us that
$$
\Xms(1^{*},2,\ldots,n) =
{
{ -k_1^2 }
\over
{ \bra{h}2,\ldots,n\ket{h} }
}
\sum_{j=2}^{n}
\coef(1,2,\ldots,j),
\eqlabel\Xsimple
$$
where
$$
\coef(1,2,\ldots,j) =
{
{ u^{\gamma}(h)
\kappa_{\gamma\dot\gamma}(1,j)
\bar k_j^{\dot\gamma\delta}
u_{\delta}(h) }
\over
{ \kappa^2(1,j{-}1) \thinspace\kappa^2(1,j) }
}.
\eqlabel\cdef
$$
Note that $\coef(1,2,\ldots,j)$ is symmetric in its first $j-1$
arguments.

We may obtain $\Xms(1,2^{*},3,\ldots,n)$ from the relation
$$
\eqalign{
-\Xms(1,2^{*},3,\ldots,n)&=
\Xms(2^{*},1,3,\ldots,n)
\cr & \quad +
\sum_{i=3}^n
\Xms(2^{*},3,\ldots,i,1,i{+}1,\ldots,n),
}
\eqlabel\cyclicX
$$
a specialized form of  (\genprops c) with an overall factor removed.
Careful consideration of (\Xsimple) tells us that we have
$$
\Xms(2^{*},1,3,\ldots,n) =
{
{ -k_2^2 }
\over
{ \bra{h}3,\ldots,n\ket{h} }
}
{
{ \braket{h}{3} }
\over
{ \bra{h}1\ket{3} }
}
\Biggl[
\coef(2,1) + \sum_{\ell=3}^n \coef(1,2,\ldots,\ell)
\Biggr],
\newlettlabel\Xtwostuff
$$
$$
\eqalign{
\Xms&(2^{*},3,\ldots,i,1,i{+}1,\ldots,n) =
{
{ -k_2^2 }
\over
{ \bra{h}3,\ldots,n\ket{h} }
}
{
{ \braket{i}{i{+}1} }
\over
{ \bra{i}1\ket{i{+}1} }
}
\cr & \quad\times
\Biggl[
\sum_{\ell=3}^i \coef(2,3,\ldots,\ell)
+ \coef(2,3,\ldots,i,1)
+ \sum_{\ell=i+1}^n \coef(1,2,\ldots,\ell)
\Biggr].
}
\lett
$$
Equation (\Xtwostuff{b}) may even be used for $i=n$ if we
define $k_{n+1}\equiv h$ for that purpose.
Insertion of (\Xtwostuff) into (\cyclicX) yields
$$
\eqalign{
-\Xms&(1,2^{*},3,\ldots,n)=
\cr & =
{
{ -k_2^2 }
\over
{ \bra{h}3,\ldots,n\ket{h} }
}
\Biggl[
{
{ \braket{h}{3} }
\over
{ \bra{h}1\ket{3} }
}
\coef(2,1)
+
{
{ \braket{h}{3} }
\over
{ \bra{h}1\ket{3} }
}
\sum_{\ell=3}^{n}
\coef(1,2,\ldots,\ell)
\cr & \quad+
\sum_{i=3}^n \sum_{\ell=3}^i
{
{ \braket{i}{i{+}1} }
\over
{ \bra{i}1\ket{i{+}1} }
}
\coef(2,3,\ldots,\ell)
+
\sum_{i=3}^n
{
{ \braket{i}{i{+}1} }
\over
{ \bra{i}1\ket{i{+}1} }
}
\coef(2,3,\ldots,i,1)
\cr & \quad +
\sum_{i=3}^{n-1} \sum_{\ell=i+1}^n
{
{ \braket{i}{i{+}1} }
\over
{ \bra{i}1\ket{i{+}1} }
}
\coef(1,2,\ldots,\ell)
\Biggr].
}
\eq
$$
The sums on $i$ appearing in double sums may be performed
using (\linkidsummed). Reordering the double sums, doing the sum
on $i$ in those terms,
and changing the dummy summation
variable from $i$ to $\ell$ in the fourth term gives
$$
\eqalign{
-\Xms&(1,2^{*},3,\ldots,n)=
\cr & =
{
{ -k_2^2 }
\over
{ \bra{h}3,\ldots,n\ket{h} }
}
\Biggl[
{
{ \braket{h}{3} }
\over
{ \bra{h}1\ket{3} }
}
\coef(2,1)
+
{
{ \braket{h}{3} }
\over
{ \bra{h}1\ket{3} }
}
\sum_{\ell=3}^{n}
\coef(1,2,\ldots,\ell)
\cr & \quad+
\sum_{\ell=3}^n
{
{ \braket{\ell}{h} }
\over
{ \bra{\ell}1\ket{h} }
}
\coef(2,3,\ldots,\ell)
+
\sum_{\ell=3}^n
{
{ \braket{\ell}{\ell{+}1} }
\over
{ \bra{\ell}1\ket{\ell{+}1} }
}
\coef(2,3,\ldots,\ell,1)
\cr & \quad +
\sum_{\ell=4}^{n}
{
{ \braket{3}{\ell} }
\over
{ \bra{3}1\ket{\ell} }
}
\coef(1,2,\ldots,\ell)
\Biggr].
}
\eqlabel\runningoutofnames
$$
The last sum appearing in (\runningoutofnames) is trivially
extended to include $\ell=3$ since $\braket{3}{3}=0$.  We then
combine it with the second term in (\runningoutofnames)
by using (\linkidnosum):
$$
\link{h}{1}{3} + \link{3}{1}{\ell}
= \link{h}{1}{\ell}.
\eqlabel\onward
$$
We split up the coefficient of
$\coef(2,3,\ldots,\ell,1)$ by using (\linkidnosum) in reverse:
$$
\link{\ell}{1}{\ell{+}1}=
\link{\ell}{1}{h} + \link{h}{1}{\ell{+}1}.
\eqlabel\retreat
$$
Incorporation of the changes  to (\runningoutofnames)
implied by (\onward) and (\retreat) produces:
$$
\eqalign{
-\Xms&(1,2^{*},3,\ldots,n)=
\cr & =
{
{ -k_2^2 }
\over
{ \bra{h}3,\ldots,n\ket{h} }
}
\Biggl[
\link{h}{1}{3} \coef(2,1)
+
\sum_{\ell=3}^{n}
\link{h}{1}{\ell} \coef(1,2,\ldots,\ell)
\cr & \quad+
\sum_{\ell=3}^n
\link{\ell}{1}{h} \coef(2,3,\ldots,\ell)
+
\sum_{\ell=3}^n
\link{\ell}{1}{h} \coef(2,3,\ldots,\ell,1)
\cr & \quad +
\sum_{\ell=3}^{n}
\link{h}{1}{\ell{+}1} \coef(2,3,\ldots,\ell,1)
\Biggr].
}
\eqlabel\estodo
$$
We may shift the summation on the last term in (\estodo) by
1, so that $\ell$ runs from 4 to  $n+1$.  This causes the same factor
$\link{h}{1}{\ell}$ to appear here as in the other sums.
The $\ell=n{+}1$ term may be discarded since $\braket{h}{h}=0$.
The first term in (\estodo) is exactly what is required to restore
$\ell=3$ to the shifted sum.  Hence, the final result is
$$
\eqalign{
\Xms&(1,2^{*},3,\ldots,n)=
{
{ -k_2^2 }
\over
{ \bra{h}3,\ldots,n\ket{h} }
}
\sum_{\ell=3}^n
\link{\ell}{1}{h}
c_2(1,2,\ldots,\ell),
}
\eqlabel\sihaypreguntas
$$
where
$$
\eqalign{
c_2(1,2,\ldots,\ell) &\equiv
\coef(1,2,\ldots,\ell)
-\coef(2,3,\ldots,\ell)
\cr &  \quad
-\coef(\ell,\ell{-}1,\ldots,2,1)+\coef(\ell{-}1,\ell{-}2,\ldots,2,1).
}
\eqlabel\ctwodef
$$
The coefficient $c_2(1,2,\ldots,\ell)$ is fully symmetric under
permutation of all but its first and last arguments.  It is
immediately obvious from the definition (\ctwodef) plus the fact
that $\coef(1,2,\ldots,\ell)$ is symmetric in its first $\ell-1$
arguments that
$$
c_2(1,2,\ldots,\ell) = - c_2(\ell,\ell{-}1,\ldots,1).
\eqlabel\ctwoantisymmetry
$$

Next, we compute $\Xms(1,2,3^{*},4,\ldots,n)$ by using the sum rule
(\genprops c), written in the form
$$
\eqalign{
\Xms_{12} &\equiv -\Xms(1,2,3^{*},4,\ldots,n)-\Xms(2,1,3^{*},4,\ldots,n)
\cr & =
\Xms(2,3^{*},1,4,\ldots,n)
+\sum_{i=4}^n \Xms(2,3^{*},4,\ldots,i,1,i{+}1,\ldots,n).
}
\eqlabel\threestart
$$
After performing a series of steps similar in spirit to those
producing equations (\cyclicX)--(\sihaypreguntas), we obtain
$$
\eqalign{
\Xms_{12} =
{
{ k_3^2 }
\over
{ \bra{h}4,\ldots,n\ket{h} }
} &
\sum_{\ell=4}^n
\link{\ell}{1}{h}
\Biggl[
\link{\ell}{2}{h} c_2(2,1,3,4,\ldots,\ell)
\cr &
-\link{\ell}{2}{h} c_2(2,3,\ldots,\ell)
-\link{1}{2}{h} c_2(2,3,\ldots,\ell,1)
\cr &
+\link{1}{2}{h} c_2(2,3,\ldots,\ell{-}1,1)
\Biggr].
}
\eqlabel\threemid
$$
The terms in the square brackets may be simplified.  Denote these
terms by $\Xi$.  Using the definition (\ctwodef) we have
$$
\eqalign{
\Xi & =
\link{\ell}{2}{h}
\biggl\{
\coef(2,1,3,4,\ldots,\ell)-\coef(1,3,4,\ldots,\ell)
\cr & \qquad\qquad\quad
-\coef(1,3,4,\ldots,\ell,2)+\coef(1,3,4,\ldots,\ell{-}1,2)
\cr & \qquad\qquad\quad
-\coef(2,3,\ldots,\ell)+\coef(3,4,\ldots,\ell)
\cr & \qquad\qquad\quad
+\coef(3,4,\ldots,\ell,2)-\coef(3,4,\ldots,\ell{-}1,2)
\biggr\}
\cr &\quad
-\link{1}{2}{h}
\biggl\{
\coef(2,3,\ldots,\ell,1)-\coef(3,4,\ldots,\ell,1)
\cr & \qquad\qquad\quad
-\coef(3,4,\ldots,\ell,1,2)+\coef(3,4,\ldots,\ell,2)
\cr & \qquad\qquad\quad
-\coef(2,3,\ldots,\ell{-}1,1)+\coef(3,4,\ldots,\ell{-}1,1)
\cr & \qquad\qquad\quad
+\coef(3,4,\ldots,\ell{-}1,1,2)-\coef(3,4,\ldots,\ell{-}1,2)
\biggr\}.
}
\eqlabel\whatamess
$$
Judicious use of (\linkidnosum) and the symmetry of the $c$'s allows
us to rewrite this as
$$
\eqalign{
\Xi & =
\link{\ell}{2}{h}
\biggl\{
\coef(1,2,\ldots,\ell)-\coef(1,3,4,\ldots,\ell)
\cr & \qquad\qquad\quad
-\coef(2,3,\ldots,\ell)+\coef(3,4,\ldots,\ell)
\biggr\}
\cr &\quad
+\link{\ell}{2}{1}
\biggl\{
\coef(3,4,\ldots,\ell,2)
-\coef(3,4,\ldots,\ell{-}1,2)
\cr & \qquad\qquad\quad
-\coef(1,3,4,\ldots,\ell,2)
+\coef(1,3,4,\ldots,\ell{-}1,2)
\biggr\}
\cr &\quad
-\link{1}{2}{h}
\biggl\{
\coef(2,3,\ldots,\ell,1)-\coef(3,4,\ldots,\ell,1)
\cr & \qquad\qquad\quad
-\coef(2,3,\ldots,\ell{-}1,1)+\coef(3,4,\ldots,\ell{-}1,1)
\biggr\}.
}
\eqlabel\noimprovement
$$
Inserting (\noimprovement) back into (\threemid) produces
$$
\eqalign{
\Xms_{12} & =
{
{ k_3^2 }
\over
{ \bra{h}4,\ldots,n\ket{h} }
}
\sum_{\ell=4}^n
\Biggl[
{
{ {\braket{\ell}{h}}^2 }
\over
{ \bra{\ell}1\ket{h} \thinspace \bra{\ell}{2}\ket{h} }
}
\biggl\{
\coef(1,2,\ldots,\ell)-\coef(1,3,4,\ldots,\ell)
\cr & \qquad\qquad\qquad\qquad\qquad\qquad\qquad\quad
-\coef(2,3,\ldots,\ell)+\coef(3,4,\ldots,\ell)
\biggr\}
\cr & \quad\qquad\qquad
+
{
{ \braket{\ell}{h} }
\over
{ \bra{\ell}2,1\ket{h} }
}
\biggl\{
\coef(3,4,\ldots,\ell,2)-\coef(3,4,\ldots,\ell{-}1,2)
\cr & \qquad\qquad\qquad\qquad\qquad
-\coef(1,3,4,\ldots,\ell,2)+\coef(1,3,4,\ldots,\ell{-}1,2)
\biggr\}
\cr & \quad\qquad\qquad
-
{
{ \braket{\ell}{h} }
\over
{ \bra{\ell}1,2\ket{h} }
}
\biggl\{
\coef(2,3,\ldots,\ell,1)-\coef(3,4,\ldots,\ell,1)
\cr & \qquad\qquad\qquad\qquad\qquad
-\coef(2,3,\ldots,\ell{-}1,1)+\coef(3,4,\ldots,\ell{-}1,1)
\biggr\}.
}
\eqlabel\isitoveryet
$$
Finally, note that  we may use (\linkidnosum)
to show that
$$
{
{ {\braket{\ell}{h}}^2 }
\over
{ \bra{\ell}1\ket{h} \thinspace \bra{\ell}{2}\ket{h} }
} =
{
{ \braket{\ell}{h} }
\over
{ \bra{\ell}1,2\ket{h} }
} + {
{ \braket{\ell}{h} }
\over
{ \bra{\ell}2,1\ket{h} }
}.
\eq
$$

Recalling that
$$
\Xms_{12} =-\Xms(1,2,3^{*},4,\ldots,n)-\Xms(2,1,3^{*},4,\ldots,n),
\eqlabel\convoluted
$$
we must disentangle the two contributions present in (\isitoveryet).
This is accomplished by considering the $n=4$ case explicitly,
for which we are able to obtain an expression for $\Xms(1,2,3^{*},4)$
from the relation ({\it cf.}\ equation (\genprops a))
$$
\Xms(1,2,3^{*},4)=(-1)^3\Xms(4,3^{*},2,1).
\eq
$$
The result of this process is
$$
\eqalign{
\Xms(1,2,3^{*},4,\ldots,n) =
{
{ -k_3^2 }
\over
{ \bra{h}4,\ldots,n\ket{h} }
}
\sum_{\ell=4}^n &
\Biggl[
{
{ \braket{\ell}{h} }
\over
{ \bra{\ell}1,2\ket{h} }
}
c_2(1,2,\ldots,\ell)
\cr &
-
{
{ \braket{\ell}{h} }
\over
{ \bra{\ell}2,1\ket{h} }
}
c_2(2,3,\ldots,\ell)
\Biggr].
}
\eqlabel\ooo
$$
We may write (\ooo) in the suggestive form
$$
\eqalign{
\Xms&(1,2,3^{*},4,\ldots,n) =
\cr & =
{
{k_3^2}
\over
{ \bra{h}1,2\ket{h} \thinspace
\bra{h}4,5,\ldots,n\ket{h} }
}
\sum_{j=1}^2 \sum_{\ell=4}^n
\invlink{j}{h}{\ell} c_2(j,\ldots,\ell).
}
\eqlabel\XTHREE
$$
Likewise, (\sihaypreguntas) may be recast as
$$
\eqalign{
\Xms&(1,2^{*},3,\ldots,n) =
\cr & =
{
{k_2^2}
\over
{ \bra{h}1\ket{h} \thinspace
\bra{h}3,4,\ldots,n\ket{h} }
}
\sum_{j=1}^1 \sum_{\ell=4}^n
\invlink{j}{h}{\ell} c_2(j,\ldots,\ell).
}
\eqlabel\XTWO
$$
Equations (\XTHREE) and (\XTWO) lead us to the ansatz
$$
\eqalign{
\Xms&(1,\ldots,m^{*},\ldots,n) =
\cr & =
{
{k_m^2}
\over
{ \bra{h}1,\ldots,m{-}1\ket{h} \thinspace
\bra{h}m{+}1,\ldots,n\ket{h} }
}
\sum_{j=1}^{m-1} \sum_{\ell=m+1}^n
\invlink{j}{h}{\ell} c_2(j,\ldots,\ell),
}
\eqlabel\XGEN
$$
valid for $2\le m \le n-1$.


In order to prove that (\XGEN) is correct, we must return to the
recursion relation (\gluonrecursioneasy), which becomes
$$
\eqalign{
&\Ims_{\alpha\dot\alpha}(1^{+},\ldots,m^{*},\ldots,n^{+})=
\cr & =
{
{-\sqrt2}
\over
{\kappa^2(1,n)}
}
\sum_{i=1}^{m-1}
\Jms_{\alpha\dot\beta}(1^{+},\ldots,i^{+})
\Ibar^{\dot\beta\beta}\bigl((i{+}1)^{+},\ldots,m^{*},\ldots,n^{+}\bigr)
\kappa_{\beta\dot\alpha}(1,n)
\cr &
+
{
{-\sqrt2}
\over
{\kappa^2(1,n)}
}
\sum_{i=m}^{n-1}
\Ims_{\alpha\dot\beta}(1^{+},\ldots,m^{*},\ldots,i^{+})
\Jbar^{\dot\beta\beta}\bigl((i{+}1)^{+},\ldots,n^{+}\bigr)
\kappa_{\beta\dot\alpha}(1,n).
}
\eqlabel\genIrecursion
$$
Using (\factorizeI) to replace $\Ims$ by $\Xms$ and
(\gluesolnallplus) for the factors of $\Jms$ appearing
in (\genIrecursion) yields
$$
\eqalign{
&\Xms(1,\ldots,m^{*},\ldots,n)=
\cr & =
{
{-1}
\over
{\kappa^2(1,n)}
}
\Biggl\{
\sum_{i=1}^{m-1}
{
{ u^{\gamma}(h)\kappa_{\gamma\dot\gamma}(1,i)
\bar\kappa^{\dot\gamma\delta}(i{+}1,n)u_{\delta(h)} }
\over
{ \bra{h}1,\ldots,i\ket{h} }
}
\Xms(i{+}1,\ldots,m^{*},\ldots,n)
\cr & \qquad\qquad
+
\sum_{i=m}^{n-1}
{
{ u^{\gamma}(h)\kappa_{\gamma\dot\gamma}(1,i)
\bar\kappa^{\dot\gamma\delta}(i{+}1,n)u_{\delta(h)} }
\over
{ \bra{h}i{+}1,\ldots,n\ket{h} }
}
\Xms(1,\ldots,m^{*},\ldots,i)
\Biggr\}
}
\eqlabel\HELP
$$
after a bit of algebra.  We should like to prove that the ansatz
(\XGEN) is correct by induction.  Note that we already know
the correct form for $\Xms(1^{*},2,\ldots,n)$ (equation
(\Xsimple); we may use that in combination with
(\genprops a)
to obtain $\Xms(1,2,\ldots,n^{*})$:
$$
\eqalign{
\Xms&(1,2,\ldots,n^{*}) =
{
{k_n^2}
\over
{ \bra{h}1,\ldots,n{-}1\ket{h} }
}
\sum_{j=1}^{n-1}
\coef(n,n{-}1,\ldots,j).
}
\eqlabel\Xlaststarred
$$
Since the form of the solution is different when $m^{*}$ appears
first or last in the argument list, we must handle those terms
separately.  So, replacing the $\Xms$'s on the right hand side
of (\HELP)
with their values produces
$$
\eqalign{
&\Xms(1,\ldots,m^{*},\ldots,n) =
{
{ -k_m^2 }
\over
{ \kappa^2(1,n) }
}
{
{1}
\over
{ \bra{h}1,\ldots,m{-}1\ket{h} \thinspace
  \bra{h}m{+}1,\ldots,n\ket{h} }
}
\cr & \times\negthinspace\negthinspace
\Biggl\{
\sum_{i=m+1}^{n-1} \sum_{r=1}^{m-1} \sum_{s=m+1}^i
\negthinspace\negthinspace
u^{\gamma}(h)\kappa_{\gamma\dot\gamma}(1,n)
\bar\kappa^{\dot\gamma\delta}(i{+}1,n) u_{\delta}(h)
\link{i}{h}{i{+}1} \invlink{r}{h}{s} c_2(r,\ldots,s)
\cr & \quad+
u^{\gamma}(h)\kappa_{\gamma\dot\gamma}(1,m)
\bar\kappa^{\dot\gamma\delta}(m{+}1,n) u_{\delta}(h)
\sum_{r=1}^{m-1}
\coef(m,m{-}1,\ldots,r)
\cr & \quad -
u^{\gamma}(h)\kappa_{\gamma\dot\gamma}(1,m{-}1)
\bar\kappa^{\dot\gamma\delta}(m,n) u_{\delta}(h)
\negthinspace\negthinspace
\sum_{s=m+1}^{n}
\negthinspace\negthinspace
\coef(m,m{+}1,\ldots,s)
\cr & \quad +
\sum_{i=1}^{m-2} \sum_{r=i+1}^{m-1} \sum_{s=m+1}^n
\negthinspace\negthinspace
u^{\gamma}(h)\kappa_{\gamma\dot\gamma}(1,i)
\bar\kappa^{\dot\gamma\delta}(1,n) u_{\delta}(h)
\link{i}{h}{i{+}1} \invlink{r}{h}{s} c_2(r,\ldots,s)
\Biggr\}.
}
\eqlabel\eins
$$
We have used (\slashsqr) and (\antisym) to write
$$
\eqalign{
u^{\gamma}(h)\kappa_{\gamma\dot\gamma}(1,i)
\bar\kappa^{\dot\gamma\delta}(i{+}1,n) u_{\delta}(h)
&=u^{\gamma}(h)\kappa_{\gamma\dot\gamma}(1,n)
\bar\kappa^{\dot\gamma\delta}(i{+}1,n) u_{\delta}(h)
\cr &
=u^{\gamma}(h)\kappa_{\gamma\dot\gamma}(1,i)
\bar\kappa^{\dot\gamma\delta}(1,n) u_{\delta}(h),
}
\eq
$$
where the first line is used in the first term of (\eins) and the
second line
in the fourth term.

We begin to simplify (\eins) by evaluating the sums on $i$ appearing
in the first and fourth terms.  Consider the first term
(${\equiv}\Xms_1$) in the
curly brackets of (\eins).  Writing out the $i$-dependent
$\kappa$-sum as a sum on $j$ we have
$$
\eqalign{
\Xms_1&= \negthinspace\negthinspace
\sum_{i=m+1}^{n-1} \sum_{j=i+1}^n \sum_{r=1}^{m-1} \sum_{s=m+1}^i
\negthinspace\negthinspace
u^{\gamma}(h)\kappa_{\gamma\dot\gamma}(1,n)
k_j^{\dot\gamma\delta}u_{\delta}(h)
\link{i}{h}{i{+}1} \invlink{r}{h}{s} c_2(r,\ldots,s)
\cr & = \negthinspace
\sum_{r=1}^{m-1} \sum_{s=m+1}^{n-1} \sum_{j=s+1}^n \sum_{i=s}^{j-1}
u^{\gamma}(h)\kappa_{\gamma\dot\gamma}(1,n)
\bar u^{\dot\gamma}(k_j)\braket{j}{h}
\link{i}{h}{i{+}1} \invlink{r}{h}{s} c_2(r,\ldots,s).
}
\eq
$$
Using (\linkidsummed) to perform  the sum on $i$ yields
$$
\eqalign{
\Xms_1&= \negthinspace
\sum_{r=1}^{m-1} \sum_{s=m+1}^{n-1} \sum_{j=s+1}^n
\negthinspace\negthinspace
u^{\gamma}(h)\kappa_{\gamma\dot\gamma}(1,n)
\bar u^{\dot\gamma}(k_j)\braket{j}{h}
\link{s}{h}{j} \invlink{r}{h}{s} c_2(r,\ldots,s)
\cr & = \negthinspace
\sum_{r=1}^{m-1} \sum_{s=m+1}^{n-1}
\negthinspace\negthinspace
u^{\gamma}(h)\kappa_{\gamma\dot\gamma}(1,n)
\bar\kappa^{\dot\gamma\delta}(s{+}1,n) u_{\delta}(k_s)
{
{ \braket{h}{r} }
\over
{ \braket{r}{s} }
}
c_2(r,\ldots,s),
}
\eqlabel\zwei
$$
where in the second line we have cancelled common factors
and restored the implicit $\kappa$-sum from the explicit
sum on $j$.
Employing the same sequence of steps on the fourth term in the
curly brackets of (\eins) results in
$$
\Xms_4=\negthinspace
\sum_{r=2}^{m-1} \sum_{s=m+1}^{n}
\negthinspace\negthinspace
u^{\gamma}(k_r)\kappa_{\gamma\dot\gamma}(1,r{-}1)
\bar\kappa^{\dot\gamma\delta}(1,n) u_{\delta}(h)
{
{ \braket{s}{h} }
\over
{ \braket{r}{s} }
}
c_2(r,\ldots,s).
\eqlabel\drei
$$
Combining (\zwei) and (\drei) with  (\eins) produces
$$
\eqalign{
\Xms&(1,\ldots,m^{*},\ldots,n) =
{
{ -k_m^2 }
\over
{ \kappa^2(1,n) }
}
{
{1}
\over
{ \bra{h}1,\ldots,m{-}1\ket{h} \thinspace
  \bra{h}m{+}1,\ldots,n\ket{h} }
}
\cr & \times\negthinspace\negthinspace
\Biggl\{
\sum_{r=1}^{m-1} \sum_{s=m+1}^{n-1}
\negthinspace\negthinspace
u^{\gamma}(h)\kappa_{\gamma\dot\gamma}(1,n)
\bar\kappa^{\dot\gamma\delta}(s{+}1,n) u_{\delta}(k_s)
{
{ \braket{h}{r} }
\over
{ \braket{r}{s} }
}
c_2(r,\ldots,s)
\cr & \quad+
u^{\gamma}(h)\kappa_{\gamma\dot\gamma}(1,m)
\bar\kappa^{\dot\gamma\delta}(m{+}1,n) u_{\delta}(h)
\sum_{r=1}^{m-1}
\coef(m,m{-}1,\ldots,r)
\cr & \quad -
u^{\gamma}(h)\kappa_{\gamma\dot\gamma}(1,m{-}1)
\bar\kappa^{\dot\gamma\delta}(m,n) u_{\delta}(h)
\negthinspace\negthinspace
\sum_{s=m+1}^{n}
\negthinspace\negthinspace
\coef(m,m{+}1,\ldots,s)
\cr & \quad +
\sum_{r=2}^{m-1} \sum_{s=m+1}^{n}
\negthinspace\negthinspace
u^{\gamma}(k_r)\kappa_{\gamma\dot\gamma}(1,r{-}1)
\bar\kappa^{\dot\gamma\delta}(1,n) u_{\delta}(h)
{
{ \braket{s}{h} }
\over
{ \braket{r}{s} }
}
c_2(r,\ldots,s)
\Biggr\}.
}
\eqlabel\vier
$$
We now write $\kappa(s{+}1,n)$ as $\kappa(1,n)-\kappa(1,s)$ and
$\kappa(1,r{-}1)$ as $\kappa(1,n)-\kappa(s,n)$ in order to
extract explicit factors of $\kappa^2(1,n)$ where possible.
Rearranging slightly, we obtain
$$
\eqalign{
\Xms&(1,\ldots,m^{*},\ldots,n) =
{
{ -k_m^2 }
\over
{ \kappa^2(1,n) }
}
{
{1}
\over
{ \bra{h}1,\ldots,m{-}1\ket{h} \thinspace
  \bra{h}m{+}1,\ldots,n\ket{h} }
}
\cr & \times\negthinspace\negthinspace
\Biggl\{
\negthinspace
-\kappa(1,n)^2
\sum_{r=1}^{m-1} \sum_{s=m+1}^{n-1}
\negthinspace\negthinspace
\invlink{r}{h}{s}
c_2(r,\ldots,s)
\cr & \quad
-\kappa(1,n)^2
\sum_{r=2}^{m-1} \sum_{s=m+1}^{n}
\negthinspace\negthinspace
\invlink{r}{h}{s}
c_2(r,\ldots,s)
\cr & \quad -
\sum_{r=1}^{m-1} \sum_{s=m+1}^{n-1}
\negthinspace
u^{\gamma}(h)\kappa_{\gamma\dot\gamma}(1,n)
\bar\kappa^{\dot\gamma\delta}(1,s) u_{\delta}(k_s)
{
{ \braket{h}{r} }
\over
{ \braket{r}{s} }
}
c_2(r,\ldots,s)
\cr & \quad -
\sum_{r=2}^{m-1} \sum_{s=m+1}^{n}
\negthinspace\negthinspace
u^{\gamma}(k_r)\kappa_{\gamma\dot\gamma}(r,n)
\bar\kappa^{\dot\gamma\delta}(1,n) u_{\delta}(h)
{
{ \braket{s}{h} }
\over
{ \braket{r}{s} }
}
c_2(r,\ldots,s)
\cr & \quad +
u^{\gamma}(h)\kappa_{\gamma\dot\gamma}(1,m)
\bar\kappa^{\dot\gamma\delta}(m{+}1,n) u_{\delta}(h)
\sum_{r=1}^{m-1}
\coef(m,m{-}1,\ldots,r)
\cr & \quad -
u^{\gamma}(h)\kappa_{\gamma\dot\gamma}(1,m{-}1)
\bar\kappa^{\dot\gamma\delta}(m,n) u_{\delta}(h)
\negthinspace\negthinspace
\sum_{s=m+1}^{n}
\negthinspace\negthinspace
\coef(m,m{+}1,\ldots,s)
\Biggr\}.
}
\eqlabel\funf
$$
We may extend the limits of the first two sums from
$r=1$ to $m-1$ and  $s=m+1$ to $n$ by adding and
subtracting the appropriate terms.  The extra terms which
arise are precisely the terms needed to extend the second
pair of double sums in the same manner.  Thus, all four double sums
may be combined to give
$$
\eqalign{
\Xms&(1,\ldots,m^{*},\ldots,n) =
{
{ -k_m^2 }
\over
{ \kappa^2(1,n) }
}
{
{1}
\over
{ \bra{h}1,\ldots,m{-}1\ket{h} \thinspace
  \bra{h}m{+}1,\ldots,n\ket{h} }
}
\cr & \times\negthinspace\negthinspace
\Biggl\{
\negthinspace
-\sum_{r=1}^{m-1} \sum_{s=m+1}^{n}
\negthinspace
\Biggl[
2\kappa^2(1,n)
\invlink{r}{h}{s}
\cr & \qquad\qquad\qquad\quad
+ u^{\gamma}(h)\kappa_{\gamma\dot\gamma}(1,n)
\bar\kappa^{\dot\gamma\delta}(1,s) u_{\delta}(k_s)
{
{ \braket{h}{r} }
\over
{ \braket{r}{s} }
}
\cr & \qquad\qquad\qquad\quad +
u^{\gamma}(k_r)\kappa_{\gamma\dot\gamma}(r,n)
\bar\kappa^{\dot\gamma\delta}(1,n) u_{\delta}(h)
{
{ \braket{s}{h} }
\over
{ \braket{r}{s} }
}
\Biggr]
c_2(r,\ldots,s)
\cr & \quad +
u^{\gamma}(h)\kappa_{\gamma\dot\gamma}(1,m)
\bar\kappa^{\dot\gamma\delta}(m{+}1,n) u_{\delta}(h)
\sum_{r=1}^{m-1}
\coef(m,m{-}1,\ldots,r)
\cr & \quad -
u^{\gamma}(h)\kappa_{\gamma\dot\gamma}(1,m{-}1)
\bar\kappa^{\dot\gamma\delta}(m,n) u_{\delta}(h)
\negthinspace\negthinspace
\sum_{s=m+1}^{n}
\negthinspace\negthinspace
\coef(m,m{+}1,\ldots,s)
\Biggr\}.
}
\eqlabel\sechs
$$
Let us denote the argument of the double sum by $\Delta$:
$$
\eqalign{
\Delta \equiv&
\Biggl[
2\kappa^2(1,n)
\invlink{r}{h}{s}
+u^{\gamma}(h)\kappa_{\gamma\dot\gamma}(1,n)
\bar\kappa^{\dot\gamma\delta}(1,s) u_{\delta}(k_s)
{
{ \braket{h}{r} }
\over
{ \braket{r}{s} }
}
\cr & \enspace +
u^{\gamma}(k_r)\kappa_{\gamma\dot\gamma}(r,n)
\bar\kappa^{\dot\gamma\delta}(1,n) u_{\delta}(h)
{
{ \braket{s}{h} }
\over
{ \braket{r}{s} }
}
\Biggr]
c_2(r,\ldots,s)
}
\eqlabel\sieben
$$

In the second term of $\Delta$, we write $\kappa(1,s)$ as
$\kappa(1,n)-\kappa(s{+}1,n)$.  This gives
$$
\eqalign{
\Delta =&
\Biggl[
\kappa^2(1,n)
\invlink{r}{h}{s}
-u^{\gamma}(h)\kappa_{\gamma\dot\gamma}(1,n)
\bar\kappa^{\dot\gamma\delta}(s{+}1,n) u_{\delta}(k_s)
{
{ \braket{h}{r} }
\over
{ \braket{r}{s} }
}
\cr & \enspace +
u^{\gamma}(k_r)\kappa_{\gamma\dot\gamma}(r,n)
\bar\kappa^{\dot\gamma\delta}(1,n) u_{\delta}(h)
{
{ \braket{s}{h} }
\over
{ \braket{r}{s} }
}
\Biggr]
c_2(r,\ldots,s)
}
\eqlabel\siebenplus
$$
Using (\fierznc) to write
$$
u_{\delta}(k_s) \thinspace \braket{h}{r} =
-u_{\delta}(h)  \thinspace \braket{r}{s}
-u_{\delta}(k_r)\thinspace \braket{s}{h}
\eqlabel\acht
$$
in the second term of (\siebenplus) produces
$$\eqalign{
\Delta =&
\Biggl[
\kappa^2(1,n)
\invlink{r}{h}{s}
+u^{\gamma}(h)\kappa_{\gamma\dot\gamma}(1,n)
\bar\kappa^{\dot\gamma\delta}(s{+}1,n) u_{\delta}(h)
\cr & \enspace +
u^{\gamma}(h)\kappa_{\gamma\dot\gamma}(1,n)
\bar\kappa^{\dot\gamma\delta}(s{+}1,n) u_{\delta}(k_r)
{
{ \braket{s}{h} }
\over
{ \braket{r}{s} }
}
\cr &\enspace  +
u^{\gamma}(k_r)\kappa_{\gamma\dot\gamma}(r,n)
\bar\kappa^{\dot\gamma\delta}(1,n) u_{\delta}(h)
{
{ \braket{s}{h} }
\over
{ \braket{r}{s} }
}
\Biggr]
c_2(r,\ldots,s)
\cr & \negthinspace\negthinspace\negthinspace\negthinspace
\negthinspace\negthinspace = \negthinspace\negthinspace
\Biggl[
\kappa^2(1,n)
\invlink{r}{h}{s}
+u^{\gamma}(h)\kappa_{\gamma\dot\gamma}(1,n)
\bar\kappa^{\dot\gamma\delta}(s{+}1,n) u_{\delta}(h)
\cr &\enspace  +
u^{\gamma}(k_r)\kappa_{\gamma\dot\gamma}(r,s)
\bar\kappa^{\dot\gamma\delta}(1,n) u_{\delta}(h)
{
{ \braket{s}{h} }
\over
{ \braket{r}{s} }
}
\Biggr]
c_2(r,\ldots,s).
}
\eqlabel\neun
$$
We have used the antisymmetry of  the spinor product to accomplish
the matrix transposition required to produce the last line of
(\neun).

Let us examine the third term (${\equiv}\Delta_3$) in (\neun).
Using (\ctwodef) to write out the factor of $c_2(r,\ldots,s)$
we have
$$
\eqalign{
\Delta_3 &=
u^{\gamma}(k_r)\kappa_{\gamma\dot\gamma}(r,s)
\bar\kappa^{\dot\gamma\delta}(1,n) u_{\delta}(h)
{
{ \braket{s}{h} }
\over
{ \braket{r}{s} }
}
\cr & \quad \times
\biggl[
\coef(r,r{+}1,\ldots,s) - \coef(r{+}1,r{+}2,\ldots,s)
\cr & \qquad\enspace
-\coef(s,s{-}1,\ldots,r{+}1,r)+\coef(s{-}1,s{-}2,\ldots,r{+}1,r)
\biggr]
}
\eqlabel\zehn
$$
Employing (\fierznc) to write
$$
u^{\gamma}(k_r) \thinspace \braket{s}{h} =
-u^{\gamma}(h)  \thinspace \braket{r}{s}
+u^{\gamma}(k_s)\thinspace \braket{r}{h},
\eq
$$
produces
$$
\eqalign{
\Delta_3 &=
\biggl[
u^{\gamma}(k_s)\kappa_{\gamma\dot\gamma}(r,s)
\bar\kappa^{\dot\gamma\delta}(1,n) u_{\delta}(h)
{
{ \braket{r}{h} }
\over
{ \braket{r}{s} }
}
-u^{\gamma}(h)\kappa_{\gamma\dot\gamma}(r,s)
\bar\kappa^{\dot\gamma\delta}(1,n) u_{\delta}(h)
\biggr]
\cr & \qquad\enspace \times
\biggl[
\coef(r,r{+}1,\ldots,s) - \coef(r{+}1,r{+}2,\ldots,s)
\biggr]
\cr & \quad
+u^{\gamma}(k_r)\kappa_{\gamma\dot\gamma}(r,s)
\bar\kappa^{\dot\gamma\delta}(1,n) u_{\delta}(h)
{
{ \braket{s}{h} }
\over
{ \braket{r}{s} }
}
\cr & \qquad\enspace\times
\biggl[
\coef(s{-}1,s{-}2,\ldots,r)-\coef(s,s{-}1,\ldots,r)
\biggr].
}
\eq
$$
Inserting the definition (\cdef)  for  some of the $\coef$-functions
and rearranging a bit we arrive at
$$
\eqalign{
\Delta_3 &=
-u^{\gamma}(h)\kappa_{\gamma\dot\gamma}(r,s)
\bar\kappa^{\dot\gamma\delta}(1,n) u_{\delta}(h)
\biggl[
\coef(r,r{+}1,\ldots,s) - \coef(r{+}1,r{+}2,\ldots,s)
\biggr]
\cr & \quad
+\invlink{r}{h}{s}
\Biggl\{
{
{ u^{\eps}(h)\kappa_{\eps\dot\eps}(r{+}1,s)
\bar k_s^{\dot\eps\gamma} \kappa(r,s)_{\gamma\dot\gamma}
\bar\kappa^{\dot\gamma\delta}(1,n)u_{\delta}(h) }
\over
{ \kappa^2(r{+}1,s) \kappa^2(r{+}1,s{-}1) }
}
\cr & \qquad\qquad\quad\enspace
-
{
{ u^{\eps}(h)\kappa_{\eps\dot\eps}(r,s)
\bar k_s^{\dot\eps\gamma} \kappa(r,s)_{\gamma\dot\gamma}
\bar\kappa^{\dot\gamma\delta}(1,n)u_{\delta}(h) }
\over
{ \kappa^2(r,s) \kappa^2(r,s{-}1) }
}
\cr & \qquad\qquad\quad\enspace
+
{
{ u^{\eps}(h)\kappa_{\eps\dot\eps}(r,s)
\bar k_r^{\dot\eps\gamma} \kappa(r,s)_{\gamma\dot\gamma}
\bar\kappa^{\dot\gamma\delta}(1,n)u_{\delta}(h) }
\over
{ \kappa^2(r,s) \kappa^2(r{+}1,s) }
}
\cr & \qquad\qquad\quad\enspace
-
{
{ u^{\eps}(h)\kappa_{\eps\dot\eps}(r,s{-}1)
\bar k_r^{\dot\eps\gamma} \kappa(r,s)_{\gamma\dot\gamma}
\bar\kappa^{\dot\gamma\delta}(1,n)u_{\delta}(h) }
\over
{ \kappa^2(r,s{-}1) \kappa^2(r{+}1,s{-}1) }
}
\Biggr\}.
}
\eqlabel\elf
$$
The terms in the curly brackets may be simplified by noting that
$$
\eqalign{
\kappa_{\alpha\dot\alpha}(1,s) \bar k_s^{\dot\alpha\beta}
\kappa_{\beta\dot\beta}(1,s) &=
\kappa_{\alpha\dot\alpha}(1,s) \bar k_s^{\dot\alpha\beta}
\kappa_{\beta\dot\beta}(1,s{-}1)
\cr & =
\kappa_{\alpha\dot\alpha}(1,s)
[\bar\kappa(1,s)-\bar\kappa(1,s{-}1)]^{\dot\alpha\beta}
\kappa_{\beta\dot\beta}(1,s{-}1)
\cr & =
\kappa^2(1,s) \kappa_{\alpha\dot\beta}(1,s{-}1)
-\kappa^2(1,s{-}1) \kappa_{\alpha\dot\beta}(1,s).
}
\eqlabel\partialfractions
$$
Note that in order to use (\partialfractions) in the first and last
terms in the curly brackets of (\elf), we must extend one
of the $\kappa$-sums and compensate.  This relation is useful
in cancelling many of the denominators.  The result of applying
(\partialfractions) to all four terms in (\elf) is
$$
\eqalign{
\Delta_3 &=
-u^{\gamma}(h)\kappa_{\gamma\dot\gamma}(r,s)
\bar\kappa^{\dot\gamma\delta}(1,n) u_{\delta}(h)
\biggl[
\coef(r,r{+}1,\ldots,s) - \coef(r{+}1,r{+}2,\ldots,s)
\biggr]
\cr & \quad
+\invlink{r}{h}{s}
\Biggl[
{
{ u^{\eps}(h)\kappa_{\eps\dot\eps}(r{+}1,s)
\bar k_s^{\dot\eps\gamma} k_{r\gamma\dot\gamma}
\bar\kappa^{\dot\gamma\delta}(1,n)u_{\delta}(h) }
\over
{ \kappa^2(r{+}1,s) \kappa^2(r{+}1,s{-}1) }
}
\cr & \qquad\qquad\quad\enspace
-
{
{ u^{\eps}(h)\kappa_{\eps\dot\eps}(r,s{-}1)
\bar k_r^{\dot\eps\gamma} k_{s\gamma\dot\gamma}
\bar\kappa^{\dot\gamma\delta}(1,n)u_{\delta}(h) }
\over
{ \kappa^2(r,s{-}1) \kappa^2(r{+}1,s{-}1) }
}
\Biggr]
\cr & \quad
+\invlink{r}{h}{s}
\Biggl\{
{
{ u^{\eps}(h)\kappa_{\eps\dot\gamma}(r{+}1,s{-}1)
\bar\kappa^{\dot\gamma\delta}(1,n)u_{\delta}(h) }
\over
{ \kappa^2(r{+}1,s{-}1) }
}
\cr & \qquad\quad
-
{
{ u^{\eps}(h)\kappa_{\eps\dot\gamma}(r{+}1,s)
\bar\kappa^{\dot\gamma\delta}(1,n)u_{\delta}(h) }
\over
{ \kappa^2(r{+}1,s) }
}
-
{
{ u^{\eps}(h)\kappa_{\eps\dot\gamma}(r,s{-}1)
\bar\kappa^{\dot\gamma\delta}(1,n)u_{\delta}(h) }
\over
{ \kappa^2(r,s{-}1) }
}
\cr & \qquad\quad
+
{
{ u^{\eps}(h)\kappa_{\eps\dot\gamma}(r,s)
\bar\kappa^{\dot\gamma\delta}(1,n)u_{\delta}(h) }
\over
{ \kappa^2(r,s) }
}
+
{
{ u^{\eps}(h)\kappa_{\eps\dot\gamma}(r{+}1,s)
\bar\kappa^{\dot\gamma\delta}(1,n)u_{\delta}(h) }
\over
{ \kappa^2(r{+}1,s) }
}
\cr & \qquad\quad
-
{
{ u^{\eps}(h)\kappa_{\eps\dot\gamma}(r,s)
\bar\kappa^{\dot\gamma\delta}(1,n)u_{\delta}(h) }
\over
{ \kappa^2(r,s) }
}
-
{
{ u^{\eps}(h)\kappa_{\eps\dot\gamma}(r{+}1,s{-}1)
\bar\kappa^{\dot\gamma\delta}(1,n)u_{\delta}(h) }
\over
{ \kappa^2(r{+}1,s{-}1) }
}
\cr & \qquad\quad
+
{
{ u^{\eps}(h)\kappa_{\eps\dot\gamma}(r,s{-}1)
\bar\kappa^{\dot\gamma\delta}(1,n)u_{\delta}(h) }
\over
{ \kappa^2(r,s{-}1) }
}
\Biggr\}.
}
\eqlabel\zwolf
$$
The terms in curly brackets cancel among themselves, leaving just
$$
\eqalign{
\Delta_3 &=
-u^{\gamma}(h)\kappa_{\gamma\dot\gamma}(r,s)
\bar\kappa^{\dot\gamma\delta}(1,n) u_{\delta}(h)
\biggl[
\coef(r,r{+}1,\ldots,s) - \coef(r{+}1,r{+}2,\ldots,s)
\biggr]
\cr & \quad
-
{
{ u^{\eps}(h)\kappa_{\eps\dot\eps}(r{+}1,s)
\bar k_s^{\dot\eps\alpha}  u_{\alpha}(h) }
\over
{ \kappa^2(r{+}1,s) \kappa^2(r{+}1,s{-}1) }
}
u^{\gamma}(h) k_{r\gamma\dot\gamma}
\bar\kappa^{\dot\gamma\delta}(1,n)u_{\delta}(h)
\cr & \quad
-
{
{ u^{\eps}(h)\kappa_{\eps\dot\eps}(r,s{-}1)
\bar k_r^{\dot\eps\alpha} u_{\alpha}(h) }
\over
{ \kappa^2(r,s{-}1) \kappa^2(r{+}1,s{-}1) }
}
u^{\gamma}(h) k_{s\gamma\dot\gamma}
\bar\kappa^{\dot\gamma\delta}(1,n)u_{\delta}(h)
\cr & =
-u^{\gamma}(h)\kappa_{\gamma\dot\gamma}(r,s)
\bar\kappa^{\dot\gamma\delta}(1,n) u_{\delta}(h)
\biggl[
\coef(r,r{+}1,\ldots,s) - \coef(r{+}1,r{+}2,\ldots,s)
\biggr]
\cr & \quad
- u^{\gamma}(h) k_{r\gamma\dot\gamma}
\bar\kappa^{\dot\gamma\delta}(1,n)u_{\delta}(h)
\coef(r{+}1,r{+}2,\ldots,s)
\cr & \quad
- u^{\gamma}(h) k_{s\gamma\dot\gamma}
\bar\kappa^{\dot\gamma\delta}(1,n)u_{\delta}(h)
\coef(s{-}1,s{-}2,\ldots,r),
}
\eqlabel\dreizehn
$$
where we have simplified the remaining terms to extract factors of $\coef$.

Substitution of (\dreizehn) back into (\neun) to recover
$\Delta$, and then putting $\Delta$ back into (\sechs)
to obtain $\Xms$ once more gives us
$$
\eqalign{
\Xms&(1,\ldots,m^{*},\ldots,n) =
{
{ -k_m^2 }
\over
{ \kappa^2(1,n) }
}
{
{1}
\over
{ \bra{h}1,\ldots,m{-}1\ket{h} \thinspace
  \bra{h}m{+}1,\ldots,n\ket{h} }
}
\cr & \times\negthinspace\negthinspace
\Biggl\{
\negthinspace
-\kappa^2(1,n)
\sum_{r=1}^{m-1} \sum_{s=m+1}^{n}
\negthinspace
\invlink{r}{h}{s}
c_2(r,\ldots,s)
\cr & \quad +
u^{\gamma}(h)\kappa_{\gamma\dot\gamma}(1,m)
\bar\kappa^{\dot\gamma\delta}(m{+}1,n) u_{\delta}(h)
\sum_{r=1}^{m-1}
\coef(m,m{-}1,\ldots,r)
\cr & \quad -
u^{\gamma}(h)\kappa_{\gamma\dot\gamma}(1,m{-}1)
\bar\kappa^{\dot\gamma\delta}(m,n) u_{\delta}(h)
\negthinspace\negthinspace
\sum_{s=m+1}^{n}
\negthinspace\negthinspace
\coef(m,m{+}1,\ldots,s)
\cr & \quad +
\sum_{r=1}^{m-1} \sum_{s=m+1}^n
\biggl[
u^{\delta}(h)\kappa_{\delta\dot\gamma}(r,n)
\bar\kappa^{\dot\gamma\gamma}(1,n)u_{\gamma}(h)
\cr & \qquad\qquad\qquad\qquad\times
\bigl[\coef(r,r{+}1,\ldots,s)-\coef(r{+}1,r{+}2,\ldots,s)\bigr]
\cr & \enspace\qquad\qquad\qquad +
u^{\delta}(h)\kappa_{\delta\dot\gamma}(s{+}1,n)
\bar\kappa^{\dot\gamma\gamma}(1,n)u_{\gamma}(h)
\cr & \qquad\qquad\qquad\qquad\times
\bigl[\coef(s{-}1,s{-}2,\ldots,r)-\coef(s,s{-}1,\ldots,r)\bigr]
\cr & \enspace\qquad\qquad\qquad +
u^{\gamma}(h)k_{r\gamma\dot\gamma}
\bar\kappa^{\dot\gamma\delta}(1,n)u_{\delta}(h)
\coef(r{+}1,r{+}2,\ldots,s)
\cr & \enspace\qquad\qquad\qquad +
u^{\gamma}(h)k_{s\gamma\dot\gamma}
\bar\kappa^{\dot\gamma\gamma}(1,n)u_{\delta}(h)
\coef(s{-}1,s{-}2,\ldots,r)
\biggr]
\Biggr\}.
}
\eqlabel\vierzehn
$$
In addition to regrouping terms, we have made us of the symmetry
of the arguments of $\coef$ in obtaining (\vierzehn).  Consider the
second double sum in (\vierzehn), which we call $\delta$:
$$
\eqalign{
\delta & \equiv
\sum_{r=1}^{m-1} \sum_{s=m+1}^n \sum_{j=r}^n
u^{\delta}(h) k_{j\delta\dot\gamma}
\bar\kappa^{\dot\gamma\gamma}(1,n)u_{\gamma}(h)
\cr & \qquad\qquad\qquad\qquad\times
\bigl[\coef(r,r{+}1,\ldots,s)-\coef(r{+}1,r{+}2,\ldots,s)\bigr]
\cr & \quad +
\sum_{r=1}^{m-1} \sum_{s=m+1}^n \sum_{j=s+1}^n
u^{\delta}(h) k_{j\delta\dot\gamma}
\bar\kappa^{\dot\gamma\gamma}(1,n)u_{\gamma}(h)
\cr & \quad\qquad\qquad\qquad\qquad\times
\bigl[\coef(s{-}1,s{-}2,\ldots,r)-\coef(s,s{-}1,\ldots,r)\bigr]
\cr & \quad +
\sum_{r=1}^{m-1} \sum_{s=m+1}^n
\biggl[
u^{\gamma}(h)k_{r\gamma\dot\gamma}
\bar\kappa^{\dot\gamma\delta}(1,n)u_{\delta}(h)
\coef(r{+}1,r{+}2,\ldots,s)
\cr & \enspace\qquad\qquad\qquad +
u^{\gamma}(h)k_{s\gamma\dot\gamma}
\bar\kappa^{\dot\gamma\gamma}(1,n)u_{\delta}(h)
\coef(s{-}1,s{-}2,\ldots,r)
\biggr].
}
\eqlabel\funfzehn
$$
In (\funfzehn) we have written out the two implicit $\kappa$-sums
that do not cover the full range 1 to $n$.  We now
prepare to do the sum on
$r$ in the first term of (\funfzehn) and the sum on $s$ in the
second term:
$$
\eqalign{
\delta & =
\sum_{s=m+1}^n\sum_{j=1}^{m-1}  \sum_{r=1}^j
u^{\delta}(h) k_{j\delta\dot\gamma}
\bar\kappa^{\dot\gamma\gamma}(1,n)u_{\gamma}(h)
\cr & \qquad\qquad\qquad\qquad\times
\bigl[\coef(r,r{+}1,\ldots,s)-\coef(r{+}1,r{+}2,\ldots,s)\bigr]
\cr & \quad + \negthinspace\negthinspace
\sum_{s=m+1}^n \sum_{j=m}^{n}  \sum_{r=1}^{m-1}
u^{\delta}(h) k_{j\delta\dot\gamma}
\bar\kappa^{\dot\gamma\gamma}(1,n)u_{\gamma}(h)
\cr & \qquad\qquad\qquad\qquad\times
\bigl[\coef(r,r{+}1,\ldots,s)-\coef(r{+}1,r{+}2,\ldots,s)\bigr]
\cr & \quad +
\sum_{r=1}^{m-1} \sum_{j=m+2}^n \sum_{s=m+1}^{j-1}
u^{\delta}(h) k_{j\delta\dot\gamma}
\bar\kappa^{\dot\gamma\gamma}(1,n)u_{\gamma}(h)
\cr & \quad\qquad\qquad\qquad\qquad\times
\bigl[\coef(s{-}1,s{-}2,\ldots,r)-\coef(s,s{-}1,\ldots,r)\bigr]
\cr & \quad +
\sum_{r=1}^{m-1} \sum_{s=m+1}^n
\biggl[
u^{\gamma}(h)k_{r\gamma\dot\gamma}
\bar\kappa^{\dot\gamma\delta}(1,n)u_{\delta}(h)
\coef(r{+}1,r{+}2,\ldots,s)
\cr & \enspace\qquad\qquad\qquad +
u^{\gamma}(h)k_{s\gamma\dot\gamma}
\bar\kappa^{\dot\gamma\gamma}(1,n)u_{\delta}(h)
\coef(s{-}1,s{-}2,\ldots,r)
\biggr].
}
\eqlabel\sechszehn
$$
We are forced to break the first term into two pieces when
doing the summation interchange.  The indicated sums are
easy to do:  only the endpoints survive, leaving
$$
\eqalign{
\delta & =
\sum_{s=m+1}^n\sum_{j=1}^{m-1}
u^{\delta}(h) k_{j\delta\dot\gamma}
\bar\kappa^{\dot\gamma\gamma}(1,n)u_{\gamma}(h)
\bigl[\coef(1,2,\ldots,s)-\coef(j{+}1,j{+}2,\ldots,s)\bigr]
\cr & \quad + \negthinspace\negthinspace\negthinspace
\sum_{s=m+1}^n \sum_{j=m}^{n}
u^{\delta}(h) k_{j\delta\dot\gamma}
\bar\kappa^{\dot\gamma\gamma}(1,n)u_{\gamma}(h)
\bigl[\coef(1,2,\ldots,s)-\coef(m,m{+}1,\ldots,s)\bigr]
\cr & \quad +
\sum_{r=1}^{m-1} \sum_{j=m+2}^n
u^{\delta}(h) k_{j\delta\dot\gamma}
\bar\kappa^{\dot\gamma\gamma}(1,n)u_{\gamma}(h)
\cr & \quad\qquad\qquad\qquad\qquad\times
\bigl[\coef(m,m{-}1,\ldots,r)-\coef(j{-}1,j{-}2,\ldots,r)\bigr]
\cr & \quad +
\sum_{r=1}^{m-1} \sum_{s=m+1}^n
\biggl[
u^{\gamma}(h)k_{r\gamma\dot\gamma}
\bar\kappa^{\dot\gamma\delta}(1,n)u_{\delta}(h)
\coef(r{+}1,r{+}2,\ldots,s)
\cr & \enspace\qquad\qquad\qquad +
u^{\gamma}(h)k_{s\gamma\dot\gamma}
\bar\kappa^{\dot\gamma\gamma}(1,n)u_{\delta}(h)
\coef(s{-}1,s{-}2,\ldots,r)
\biggr].
}
\eqlabel\siebszehn
$$
The terms containing $\coef(1,2,\ldots,s)$ may be summed on
$j$ and combined to give
$$
\eqalign{
\sum_{s=m+1}^n
u^{\delta}(h) &
\bigl[\kappa(1,m{-}1)+\kappa(m,n)\bigr]_{\delta\dot\gamma}
\bar\kappa^{\dot\gamma\gamma}(1,n) u_{\gamma}(h) \coef(1,2,\ldots,s) =
\cr & =
\kappa^2(1,n) \thinspace \braket{h}{h}
\negthinspace\negthinspace\sum_{s=m+1}^n \coef(1,2,\ldots,s)
\cr & = 0.
}
\eq
$$
The remaining terms may be written as
$$
\eqalign{
\delta & =
-\sum_{j=1}^{m-1}\sum_{s=m+1}^n
u^{\delta}(h) k_{j\delta\dot\gamma}
\bar\kappa^{\dot\gamma\gamma}(1,n)u_{\gamma}(h)
\coef(j{+}1,j{+}2,\ldots,s)
\cr & \quad - \negthinspace\negthinspace
\sum_{s=m+1}^n
u^{\delta}(h) \kappa_{\delta\dot\gamma}(m,n)
\bar\kappa^{\dot\gamma\gamma}(1,n)u_{\gamma}(h)
\coef(m,m{+}1,\ldots,s)
\cr & \quad +
\sum_{r=1}^{m-1}
u^{\delta}(h) \kappa_{\delta\dot\gamma}(m{+}2,n)
\bar\kappa^{\dot\gamma\gamma}(1,n)u_{\gamma}(h)
\coef(m,m{-}1,\ldots,r)
\cr & \quad -
\sum_{r=1}^{m-1} \sum_{j=m+2}^n
u^{\delta}(h) k_{j\delta\dot\gamma}
\bar\kappa^{\dot\gamma\gamma}(1,n)u_{\gamma}(h)
\coef(j{-}1,j{-}2,\ldots,r)
\cr & \quad +
\sum_{r=1}^{m-1} \sum_{s=m+1}^n
\biggl[
u^{\gamma}(h)k_{r\gamma\dot\gamma}
\bar\kappa^{\dot\gamma\delta}(1,n)u_{\delta}(h)
\coef(r{+}1,r{+}2,\ldots,s)
\cr & \enspace\qquad\qquad\qquad +
u^{\gamma}(h)k_{s\gamma\dot\gamma}
\bar\kappa^{\dot\gamma\gamma}(1,n)u_{\delta}(h)
\coef(s{-}1,s{-}2,\ldots,r)
\biggr].
}
\eqlabel\achtzehn
$$
Everything except for the $s=m+1$ term of the second term in
the square brackets of the
final double sum  is cancelled by the first two double sums
in (\achtzehn).  This leaves only
$$
\eqalign{
\delta & =
- \negthinspace\negthinspace
\sum_{s=m+1}^n
u^{\delta}(h) \kappa_{\delta\dot\gamma}(m,n)
\bar\kappa^{\dot\gamma\gamma}(1,n)u_{\gamma}(h)
\coef(m,m{+}1,\ldots,s)
\cr & \quad +
\sum_{r=1}^{m-1}
u^{\delta}(h) \kappa_{\delta\dot\gamma}(m{+}2,n)
\bar\kappa^{\dot\gamma\gamma}(1,n)u_{\gamma}(h)
\coef(m,m{-}1,\ldots,r)
\cr & \quad +
\sum_{r=1}^{m-1}
u^{\gamma}(h)(k_{m+1})_{\gamma\dot\gamma}
\bar\kappa^{\dot\gamma\gamma}(1,n)u_{\delta}(h)
\coef(m,m{-}1,\ldots,r).
}
\eqlabel\neunzehn
$$
The third line of (\neunzehn) may be combined with  the second
line by extending $\kappa(m{+}2,n)$ to $\kappa(m{+}1,n)$.
We then transpose the order of  the matrix multiplication and use
(\slashsqr) to achieve
$$
\eqalign{
\delta & =
\sum_{s=m+1}^n
u^{\gamma}(h) \kappa_{\gamma\dot\gamma}(1,m{-}1)
\bar\kappa^{\dot\gamma\delta}(m,n)u_{\delta}(h)
\coef(m,m{+}1,\ldots,s)
\cr & \quad -
\sum_{r=1}^{m-1}
u^{\gamma}(h) \kappa_{\gamma\dot\gamma}(1,m)
\bar\kappa^{\dot\gamma\delta}(m{+}1,n)u_{\delta}(h)
\coef(m,m{-}1,\ldots,r).
}
\eqlabel\zwanzig
$$
A comparison of (\zwanzig) with (\vierzehn) shows that the contribution
from $\delta$ exactly cancels the contribution from the two
remaining
single sums appearing in  $\Xms$.  Thus, we have
$$
\eqalign{
\Xms&(1,\ldots,m^{*},\ldots,n) =
\cr & =
{
{k_m^2}
\over
{ \bra{h}1,\ldots,m{-}1,\ket{h} \thinspace
\bra{h}m{+}1,\ldots,n\ket{h} }
}
\sum_{r=1}^{m-1} \sum_{s=m+1}^n
\invlink{r}{h}{s} c_2(r,\ldots,s),
}
\eq
$$
proving the ansatz (\XGEN).

\chapter{LIKE-HELICITY GLUON PRODUCTION}

In this section we present the computation of the process
(\monsteramplitude) for the case where all of the gluons
have the same helicity.  We begin by examining the
currents which must be sewn together to form the amplitude
for (\monsteramplitude).  We will see that the modified
gluon current discussed in the previous section arises in
a natural manner within this context.  After evaluating the
color factor and the sub-leading contribution in $1\over N$,
we engage in a lengthy discussion of the main contribution.
In spite of a very complicated starting point, we will end
up with a relatively simple final result,
in agreement with Mangano \cite\Mangano.

\section{Preliminary considerations}

Figure\Flabel\monsterfig\ illustrates the form of the tree-level
Feynman graphs contributing
to (\monsteramplitude).  In terms of the currents introduced
in the second and third sections, we see that it consists of
two quark currents, two antiquark currents, and a gluon
current with {\it two} off-shell gluons.  It is convenient
to define a current
$({\GAMMA}_{\xi}^{x})_{ji}(p;1,\ldots,n;p')$
consisting of a quark line, $n$ on-shell gluons, and one off-shell gluon.
With all momenta directed inward, this current has an
antiquark of momentum $p$ and color index $j$, a quark
of momentum $p'$ and color index $i$, $n$ on-shell gluons
with momenta $k_\ell$ and color indices $a_\ell$, and
an off-shell gluon of momentum $-p-\kappa(1,n)-p'$,
Lorentz index $\xi$ and color index $x$.  We do not include
the propagator for the off-shell gluon in the definition
of $({\GAMMA}_{\xi}^{x})_{ji}$.  According to this definition
we have
$$
\eqalign{
&({\GAMMA}_{\xi}^{x})_{ji}(p;1,\ldots,n;p') =
\cr &\enspace =
\permsum{1}{n} \sum_{s=0}^n
{
{\psibarhat_{jk}(p;1,\ldots,s)}
\over
{s!}
}
\biggl[
-ig(T^x)_{km} \gamma_{\xi}
\biggr]
{
{{\psihat}_{mi}(s{+}1,\ldots,n;p')}
\over
{(n-s)!}
}.
}
\eqlabel\glueone
$$
Insertion of the color factorizations (\quarkrecursion a)
and (\antiquarkrecursion a) into (\glueone) produces the
form
$$
\eqalign{
({\GAMMA}_{\xi}^{x})_{ji}&(p;1,\ldots,n;p') =
\cr & = -ig^{n+1}
\permsum{1}{n} \sum_{s=0}^n
(\Omega[1,s] T^x \Omega[s{+}1,n] )_{ji}
\cr & \qquad\qquad\qquad\qquad\times
\psibar(p;1,\ldots,s)
\gamma_{\xi}
\psi(s{+}1,\ldots,n;p').
}
\eqlabel\gluednormal
$$

We now turn to the current with two off-shell gluons.  From
(\gluonrecursion a) we know that
$$
\eqalign{
\Jhat_{\xi}^x & (0,1,\ldots,n) =
\cr & =
2g^n \permsum{0}{n} tr(\Omega[0,n] T^x) J_{\xi}(0,1,\ldots,n)
\cr & =
2g^n \permsum{1}{n} \sum_{s=0}^n
tr( \Omega[1,s] T^{a_0} \Omega[s{+}1,n] T^x)
J_{\xi}(1,\ldots,s,0,s{+}1,\ldots,n).
}
\eqlabel\Jexpanded
$$
To obtain a formal expression for the current with two off-shell
gluons, we differentiate  (\Jexpanded) with respect to the
polarization vector of gluon number 0 and restore its propagator:
$$
\eqalign{
&\Ihat_{\xi\zeta}^{xz}(0^{*},1,\ldots,n) =
\cr &\enspace=
2g^n { {-i} \over {k_0^2} }
\permsum{1}{n} \sum_{s=0}^n
tr( \Omega[1,s] T^z \Omega[s{+}1,n] T^x)
{
{ \partial }
\over
{ \partial\eps^{\zeta}(0) }
}
J_{\xi}(1,\ldots,s,0,s{+}1,\ldots,n).
}
\eqlabel\trueIdef
$$

Let us denote by $\amp^{ij}_{k\ell}(p,q;1,\ldots,n;p',q')$
the amplitude for the process represented by Figure \monsterfig.
Putting the pieces together, we write
$$
\eqalign{
\amp^{ij}_{k\ell}&(p,q;1,\ldots,n;p',q') \equiv
\cr & =
\permsum{1}{n}
\sum_{v=0}^n \sum_{w=v}^n
{ {1}\over{v!} }
\Ihat_{\xi\zeta}^{xz}(0^{*},1,\ldots,v)
\cr & \qquad\qquad\times
{ {1}\over{(w{-}v)!} }
({\GAMMA}^{\xi x})^{ij}(p;v{+}1,\ldots,w;p')
\cr & \qquad\qquad\times
{ {1}\over{(n{-}w)!} }
({\GAMMA}^{\zeta z})_{k\ell}(q;w{+}1,\ldots,n;q'),
}
\eqlabel\monsterstart
$$
where momentum conservation tells us that
$$
\eqalign{
k_0 &= q+\kappa(w{+}1,n)+q'
\cr & =
-p -\kappa(1,w)-p'.
}
\eqlabel\momentumkey
$$

We are able to evaluate (\monsterstart) in the case where
all of the radiated gluons have the same helicity.
Since the methods illustrated below handle all three of
the possible quark-antiquark helicity combinations with no
extra complications, we
select the amplitude with  two right-handed quark
lines for further study.
Clearly, this helicity combination will involve
$\psibar^{\alpha}\bigl(q^{-};(w{+}1)^{+},\ldots,s^{+}\bigr)$
for various values
of $s$, as implied by the form of (\gluednormal).  However,
we know from (\LHantiquarkallplus) that this current may be
made to vanish for $n\ge1$ by the gauge choice $h=q$.
Thus, the sum on $s$ in (\gluednormal) may be replaced by a single
term:
$$
\eqalign{
&({\GAMMA}^{\zeta z})_{k\ell}
\bigl(q^{-};(w{+}1)^{+},\ldots,n^{+};q'^{+}\bigr) =
\cr &\quad = -ig^{n-w+1}
\permsum{w+1}{n}
( T^z \Omega[w{+}1,n] )_{k\ell}
\psibar(q^{-})
\gamma^{\zeta}
\psi\bigl((w{+}1)^{+},\ldots,n^{+};q'^{+}\bigr).
}
\eqlabel\gluedshort
$$
Unfortunately, the same reduction is not permissible on the
other quark line.  Once $h=q$  is chosen for one of the currents,
consistency forces us to choose $h=q$ in all of the other currents,
since individual terms in the sum (\monsterstart) are not
gauge-invariant even though the sum as a whole is.

Combination of  (\gluednormal), (\trueIdef) and (\gluedshort)
with (\monsterstart)
produces
$$
\eqalign{
\amp^{ij}_{k\ell}&(p^{-},q^{-};1^{+},\ldots,n^{+};p'^{+},q'^{+}) \equiv
\cr &= 2ig^{n+2}
\permsum{1}{n}
\sum_{v=0}^n \sum_{w=v}^n
\sum_{u=0}^v \sum_{t=v}^w
tr( \Omega[1,u] T^z \Omega[u{+}1,v] T^x )
\cr & \qquad\qquad\qquad\times
( \Omega[v{+}1,t] T^x \Omega[t{+}1,w] )^{ij}
( T^z \Omega[w{+}1,n] )_{k\ell}
\cr & \qquad\qquad\qquad\times
{ {1}\over{k_0^2} }
{
{\partial}
\over
{ \partial\eps^{\zeta}(0^{+}) }
}
J_{\xi}\bigl(1^{+},\ldots,u^{+},0^{+},(u{+}1)^{+},\ldots,v^{+}\bigr)
\cr & \qquad\qquad\qquad\times
\psibar\bigl(p^{-};(v{+}1)^{+},\ldots,t^{+}\bigr)
\gamma^{\xi} \psi\bigl((t{+}1)^{+},\ldots,w^{+};p'^{+}\bigr)
\cr & \qquad\qquad\qquad\times
\psibar(q^{-}) \gamma^{\zeta}
\psi\bigl((w{+}1)^{+},\ldots,n^{+};q'^{+}\bigr).
}
\eqlabel\monsterlorentz
$$
The multispinor replacement rules (\msreplacefermion) tell
us that this is equivalent to
$$
\eqalign{
\amp^{ij}_{k\ell}&(p^{-},q^{-};1^{+},\ldots,n^{+};p'^{+},q'^{+}) =
\cr &= 4ig^{n+2}
\permsum{1}{n}
\sum_{v=0}^n \sum_{w=v}^n
\sum_{u=0}^v \sum_{t=v}^w
tr( \Omega[1,u] T^z \Omega[u{+}1,v] T^x )
\cr & \qquad\qquad\qquad\times
( \Omega[v{+}1,t] T^x \Omega[t{+}1,w] )^{ij}
( T^z \Omega[w{+}1,n] )_{k\ell}
\cr & \qquad\qquad\qquad\times
{ {1}\over{k_0^2} }
{
{\partial}
\over
{ \partial\bar\eps^{\dot\beta\beta}(0^{+}) }
}
\Jms_{\alpha\dot\alpha}
\bigl(1^{+},\ldots,u^{+},0^{+},(u{+}1)^{+},\ldots,v^{+}\bigr)
\cr & \qquad\qquad\qquad\times
\psibar^{\alpha}\bigl(p^{-};(v{+}1)^{+},\ldots,t^{+}\bigr)
\psi^{\dot\alpha}\bigl((t{+}1)^{+},\ldots,w^{+};p'^{+}\bigr)
\cr & \qquad\qquad\qquad\times
\psibar^{\beta}(q^{-})
\psi^{\dot\beta}\bigl((w{+}1)^{+},\ldots,n^{+};q'^{+}\bigr).
}
\eqlabel\monsterMS
$$
{}From (\gluesolnallplus b), (\chargeconjugate a),
(\RHantiquarkallplus), and (\LHantiquarkzero)
we see that
$$
\eqalign{
\psibar^{\beta}(q^{-})&
\psi^{\dot\beta}\bigl((w{+}1)^{+},\ldots,n^{+};q'^{+}\bigr) =
\cr & =
{
{(-\sqrt2)^{n-w}
u^{\beta}(q)
[\bar\kappa(w{+}1,n)+\bar q']^{\dot\beta\gamma}
u_{\gamma}(q)}
\over
{ \bra{q} w{+}1,\ldots,n \ket{q'} }
}.
}
\eqlabel\connect
$$
We may use the Weyl equation to add $\bar q^{\dot\beta\gamma}$ to
the expression in square brackets in (\connect).  But then we have
$$
\eqalign{
\psibar^{\beta}(q^{-})
\psi^{\dot\beta}\bigl((w{+}1)^{+},\ldots,n^{+};q'^{+}\bigr)& =
(-\sqrt2)^{n-w}
{
{u^{\beta}(q)
\bar k_0^{\dot\beta\gamma}
u_{\gamma}(q)}
\over
{ \bra{q} w{+}1,\ldots,n \ket{q'} }
}
\cr & =
(-\sqrt2)^{n-w}
{
{-\Ibar^{\dot\beta\beta}(0^{*})}
\over
{ \bra{q} w{+}1,\ldots,n \ket{q'} }
},
}
\eq
$$
where $\Ibar^{\dot\beta\beta}(0^{*})$ is precisely the type of
``generalized'' polarization spinor introduced in the last
section ({\it cf.}{ }equation (\Isingle)).  With this identification
made, we see that equation (\monsterMS) tells us to take the
usual gluon current, remove the polarization spinor for the
gluon labelled by ``0'', and replace it with a ``generalized''
polarization spinor, with $k_0^2 \ne 0$.  The result is
exactly the modified gluon current discussed in the previous section.
Hence,
$$
\eqalign{
\amp^{ij}_{k\ell}&(p^{-},q^{-};1^{+},\ldots,n^{+};p'^{+},q'^{+}) =
\cr &=- 4ig^{n+2}
\permsum{1}{n}
\sum_{v=0}^n \sum_{w=v}^n
\sum_{u=0}^v \sum_{t=v}^w
tr( \Omega[1,u] T^z \Omega[u{+}1,v] T^x )
\cr & \qquad\qquad\qquad\times
( \Omega[v{+}1,t] T^x \Omega[t{+}1,w] )^{ij}
( T^z \Omega[w{+}1,n] )_{k\ell}
\cr & \qquad\qquad\qquad\times
{ {1}\over{k_0^2} }
\Ims_{\alpha\dot\alpha}
\bigl(1^{+},\ldots,u^{+},0^{*},(u{+}1)^{+},\ldots,v^{+}\bigr)
\cr & \qquad\qquad\qquad\times
\psibar^{\alpha}\bigl(p^{-};(v{+}1)^{+},\ldots,t^{+}\bigr)
\psi^{\dot\alpha}\bigl((t{+}1)^{+},\ldots,w^{+};p'^{+}\bigr)
\cr & \qquad\qquad\qquad\times
(-\sqrt2)^{n-w}
{
{1}
\over
{ \bra{q} w{+}1,\ldots,n \ket{q'} }
}.
}
\eq
$$
Inserting the appropriate expressions from
section 2 for the remaining quark currents gives us
$$
\eqalign{
&\amp^{ij}_{k\ell}(p^{-},q^{-};1^{+},\ldots,n^{+};p'^{+},q'^{+}) =
\cr &\quad=
-4ig^{n+2} \negthinspace\negthinspace
\permsum{1}{n}
\sum_{v=0}^n \sum_{w=v}^n
\sum_{u=0}^v \sum_{t=v}^w
tr( \Omega[1,u] T^z \Omega[u{+}1,v] T^x )
\cr & \qquad\qquad\qquad\times
( \Omega[v{+}1,t] T^x \Omega[t{+}1,w] )^{ij}
( T^z \Omega[w{+}1,n] )_{k\ell}
\cr & \qquad\qquad\qquad\times
{ {1}\over{k_0^2} }
\Ims_{\alpha\dot\alpha}
\bigl(1^{+},\ldots,u^{+},0^{*},(u{+}1)^{+},\ldots,v^{+}\bigr)
\cr & \qquad\qquad\qquad\times
{
{(-\sqrt2)^{n-v} \braket{p}{q} u^{\alpha}(p)
[\bar\kappa(t{+}1,w) +\bar p']^{\dot\alpha\beta} u_{\beta}(q)}
\over
{ \bra{p}v{+}1,\ldots,t\ket{q}
\thinspace\bra{q} t{+}1,\ldots,w \ket{p'}
\thinspace\bra{q} w{+}1,\ldots,n \ket{q'}  }
}.
}
\eqlabel\launchpoint
$$
\section{Evaluation of the color factor}

We turn our attention to the color factor in (\launchpoint),
which reads
$$
\eqalign{
C^{ij}_{k\ell}&\equiv tr( \Omega[1,u] T^z \Omega[u{+}1,v] T^x )
( \Omega[v{+}1,t] T^x \Omega[t{+}1,w] )^{ij}
( T^z \Omega[w{+}1,n] )_{k\ell}
\cr & =
( \Omega[1,u] T^z \Omega[u{+}1,v] )^{pq}
( T^z \Omega[w{+}1,n] )_{k\ell}
\cr & \qquad\times
( \Omega[v{+}1,t] )^{ir}
(\Omega[t{+}1,w])^{sj}
(T^x)^{qp}(T^x)^{rs}.
}
\eqlabel\colorstart
$$
In order to evaluate $C^{ij}_{k\ell}$,
we only need to know the completeness relation for $SU(N)$,
as given by (\completeness).  Evaluating the (implicit)
sum on $x$ gives
$$
\eqalign{
C^{ij}_{k\ell}&=
{1\over2}
(\Omega[v{+}1,t]\Omega[1,u]T^z\Omega[u{+}1,v]\Omega[t{+}1,w])^{ij}
(T^z\Omega[w{+}1,n])_{k\ell}
\cr & \quad
-{1\over{2N}} tr(\Omega[1,u]T^z\Omega[u{+}1,v])
(\Omega[v{+}1,w])^{ij}
(T^z\Omega[w{+}1,n])_{k\ell}.
}
\eqlabel\midstage
$$
The second term in (\midstage) contains a trace.  Since the trace
is invariant under cyclic permutations, we may extract a factor
$$
{\sum_{{\cal{C}}(0\ldots v)}}
\Ims_{\alpha\dot\alpha}
\bigl(1^{+},\ldots,u^{+},0^{*},(u{+}1)^{+},\ldots,v^{+}\bigr)
= 0, \qquad v\ne0,
\eqlabel\ridit
$$
from the permutation sum in (\launchpoint).  Equation (\ridit),
the form of (\usefulproperties b) satisfied by the modified
current, tells us that there is no  contribution
to the amplitude from this term, except perhaps at $v=0$.
When $v=0$, however, the trace contains  just a single color matrix,
$$
tr\enspace T^z =0.
\eq
$$
Hence, the only non-vanishing contribution
to (\midstage) is generated by the first term and we may write
$$
\eqalign{
C^{ij}_{k\ell}&=
{1\over2}
(\Omega[v{+}1,t]\Omega[1,u])^{ip}
(\Omega[u{+}1,v]\Omega[t{+}1,w])^{qj}
\cr & \qquad\qquad\times
(\Omega[w{+}1,n])_{r\ell}
(T^z)_{pq}(T^z)_{kr}
\cr & =
{1\over4}
{(\Omega[v{+}1,t]\Omega[1,u]\Omega[w{+}1,n])^{i}}_{\ell}
{(\Omega[u{+}1,v]\Omega[t{+}1,w])_{k}}^{j}
\cr & \quad -
{1\over{4N}}
(\Omega[v{+}1,t]\Omega[1,v]\Omega[t{+}1,w])^{ij}
(\Omega[w{+}1,n])_{k\ell},
}
\eqlabel\killagain
$$
where we have performed the sum on $z$ to obtain the second line.
Notice that the ${1\over N}$ term in (\killagain) is independent
of $u$.  Thus, the sum on $u$ appearing in (\launchpoint) reads
$$
\sum_{u=0}^v
\Ims_{\alpha\dot\alpha}
\bigl(1^{+},\ldots,u^{+},0^{*},(u{+}1)^{+},\ldots,v^{+}\bigr)
= 0,
\qquad v\ne 0,
\eq
$$
according to (\usefulproperties c).  Thus, we see that the only term
which gives a contribution has $v=0$, producing
$$
\eqalign{
C^{ij}_{k\ell}&=
{1\over4}
{(\Omega[v{+}1,t]\Omega[1,u]\Omega[w{+}1,n])^{i}}_{\ell}
{(\Omega[u{+}1,v]\Omega[t{+}1,w])_{k}}^{j}
\cr & \quad -
{{\delta_{v0}}\over{4N}}
(\Omega[1,w])^{ij}
(\Omega[w{+}1,n])_{k\ell}.
}
\eqlabel\colorend
$$
We now insert (\colorend) into (\launchpoint), yielding
$$
\eqalign{
&\amp^{ij}_{k\ell}(p^{-},q^{-};1^{+},\ldots,n^{+};p'^{+},q'^{+}) =
\cr &\quad=
-ig^{n+2} \negthinspace\negthinspace
\permsum{1}{n}
\sum_{v=0}^n \sum_{w=v}^n
\sum_{u=0}^v \sum_{t=v}^w
{(\Omega[v{+}1,t]\Omega[1,u]\Omega[w{+}1,n])^{i}}_{\ell}
\cr & \qquad\qquad\qquad\times
{(\Omega[u{+}1,v]\Omega[t{+}1,w])_{k}}^{j}
\cr & \qquad\qquad\qquad\times
{ {1}\over{k_0^2} }
\Ims_{\alpha\dot\alpha}
\bigl(1^{+},\ldots,u^{+},0^{*},(u{+}1)^{+},\ldots,v^{+}\bigr)
\cr & \qquad\qquad\qquad\times
{
{(-\sqrt2)^{n-v} \braket{p}{q} u^{\alpha}(p)
[\bar\kappa(t{+}1,w)+\bar p'  ]^{\dot\alpha\beta} u_{\beta}(q)}
\over
{ \bra{p}v{+}1,\ldots,t\ket{q}
\thinspace\bra{q} t{+}1,\ldots,w \ket{p'}
\thinspace\bra{q} w{+}1,\ldots,n \ket{q'}  }
}
\cr &\quad\quad
+{{1}\over{N}}ig^{n+2} \negthinspace\negthinspace
\permsum{1}{n}
\sum_{w=0}^n \sum_{t=0}^w
(\Omega[1,w])^{ij}
(\Omega[w{+}1,n])_{k\ell}
{ {1}\over{k_0^2} }
\Ims_{\alpha\dot\alpha}(0^{*})
\cr & \qquad\qquad\qquad\times
{
{(-\sqrt2)^{n} \braket{p}{q} u^{\alpha}(p)
[\bar\kappa(t{+}1,w)+\bar p'  ]^{\dot\alpha\beta} u_{\beta}(q)}
\over
{ \bra{p}1,\ldots,t\ket{q}
\thinspace\bra{q} t{+}1,\ldots,w \ket{p'}
\thinspace\bra{q} w{+}1,\ldots,n \ket{q'}  }
}.
}
\eqlabel\breakuppoint
$$


\section{Evaluation of the $1/N$ contribution}

The simplest piece of (\breakuppoint) to evaluate is the $1\over N$
contribution, which we will denote by $\amp_5$.  Inserting the
explicit form (\Isingle) for $\Ims(0^{*})$ produces
$$
\eqalign{
\amp_5&=
{{i}\over{2N}}(-g\sqrt2)^{n+2} \negthinspace\negthinspace
\permsum{1}{n}
\sum_{w=0}^n \sum_{t=0}^w
(\Omega[1,w])^{ij}
(\Omega[w{+}1,n])_{k\ell}
\cr & \qquad\qquad\qquad\times
{ {1}\over{k_0^2} }
{
{ {\braket{p}{q}}^2  u^{\gamma}(q)k_{0\gamma\dot\alpha}
[\bar\kappa(t{+}1,w)+\bar p']^{\dot\alpha\beta} u_{\beta}(q)}
\over
{ \bra{p}1,\ldots,t\ket{q}
\thinspace\bra{q} t{+}1,\ldots,w \ket{p'}
\thinspace\bra{q} w{+}1,\ldots,n \ket{q'}  }
}.
}
\eqlabel\Mfivestart
$$

We note that the color factors are independent of $t$.  In order
to do the sum on $t$ in as compact a manner as possible we
employ the following mnemonic device.  We write the implicit
$\kappa$ sum appearing in (\Mfivestart) as an explicit sum:
$$
[\bar\kappa(t{+}1,w)+ \bar p']^{\dot\alpha\beta} \equiv
\sum_{s=t{+}1}^{w{+}1} \bar k_s^{\dot\alpha\beta},
\eqlabel\redefkey
$$
with the understanding that when $s=w{+}1$ we write $p'$ instead
of $k_{w{+}1}$.  The same kinds of considerations allow us to write
$$
{
{1}
\over
{ \bra{p}1,\ldots,t\ket{q}
\thinspace\bra{q} t{+}1,\ldots,w \ket{p'} }
}
=
\link{t}{q}{t{+}1}
{
{1}
\over
{ \bra{p}1,\ldots,w\ket{p'} }
}
\eqlabel\keyredef
$$
for all values of $t$.  In (\keyredef) we have the additional
stipulation that  when $t=0$ we should write $p$ rather
than $k_0$.   As a result, we do not have to spend the
extra time to consider the endpoints of the summation separately,
provided that we take care to follow the pattern set by
(\redefkey) and (\keyredef).   At first glance, this procedure
may sound {\it ad hoc;}{ }however, it is well-defined and always
follows the pattern just outlined:  when joining two denominator
``strings'', the correct value for ``out of range'' summation variables
({\it i.e.}{ }those which are not included in the range of gluon
momenta represented in the ``string'') is determined by the
quark momentum appearing at the corresponding end of the
``string''.
Since this type of structure
occurs many times in this calculation, the savings generated
by this trick is considerable.

Using (\redefkey) and (\keyredef), we see that (\Mfivestart)
contains
$$
\eqalign{
\sigma_5&\equiv
\sum_{t=0}^w \sum_{s=t{+}1}^{w+1}
\link{t}{q}{t{+}1} u^{\gamma}(q)k_{0\gamma\dot\alpha}
\bar k_s^{\dot\alpha\beta} u_{\beta}(q)
\cr & =
\sum_{s=1}^{w+1} \sum_{t=0}^{s-1}
\link{t}{q}{t{+}1} u^{\gamma}(q)k_{0\gamma\dot\alpha}
\bar u^{\dot\alpha}(k_s) \braket{s}{q},
}
\eq
$$
where we have interchanged the order of the summations and used
(\spinnorm) to obtain the second line.  We may perform the sum
on $t$ using (\linkidsummed) to obtain
$$
\eqalign{
\sigma_5&=
\sum_{s=1}^{w+1}
\link{p}{q}{s} u^{\gamma}(q)k_{0\gamma\dot\alpha}
\bar u^{\dot\alpha}(k_s) \braket{s}{q},
}
\eqlabel\samplesum
$$
replacing $k_{t=0}$ with $p$ as described in the previous paragraph.
Cancelling the common factor of $\braket{s}{q}$ from (\samplesum)
allows us to do the remaining sum:
$$
\eqalign{
\sigma_5&=
{ {1}\over{\braket{p}{q}} }
u^{\gamma}(q)k_{0\gamma\dot\alpha}
[\bar \kappa(1,w) + \bar p']^{\dot\alpha\beta}
u_{\beta}(p).
}
\eqlabel\going
$$
The  Weyl equation permits us to add $\bar p^{\dot\alpha\beta}$
to the sum in (\going) for free.  But, according to
(\momentumkey), the resulting sum is just $-\bar k_0^{\dot\alpha\beta}$.
Thus
$$
\eqalign{
\sigma_5&=-
{ {1}\over{\braket{p}{q}} }
u^{\gamma}(q)k_{0\gamma\dot\alpha}
k_0^{\dot\alpha\beta}
u_{\beta}(p)
\cr & =
k_0^2,
}
\eqlabel\gone
$$
where we have used (\slashsqr).  Putting (\gone) back into
(\Mfivestart) gives
$$
\eqalign{
\amp_5&=
{{i}\over{2N}}(-g\sqrt2)^{n+2} \negthinspace\negthinspace
\permsum{1}{n}
\sum_{w=0}^n
(\Omega[1,w])^{ij}
(\Omega[w{+}1,n])_{k\ell}
\cr & \qquad\qquad\qquad\qquad\qquad\times
{
{ {\braket{p}{q}}^2  }
\over
{ \bra{p}1,\ldots,w\ket{p'}
\thinspace\bra{q} w{+}1,\ldots,n \ket{q'}  }
}.
}
\eqlabel\Mfive
$$
\section{Evaluation of the $N$-independent contribution}

The $N$-independent contribution to the amplitude comes from
the first term in (\breakuppoint).  The seemingly
complicated nature of the color
factor appearing in this term suggests that the labelling
scheme for the gluons appearing in Figure~\monsterfig\ is not
optimal  for this contribution.  Instead, to obtain
a simpler expression, we use the labels suggested by
Figure\Flabel\tamedfig.  This makes the $N$-independent
contribution read
$$
\eqalign{
&\amp^{ij}_{k\ell}(p^{-},q^{-};1^{+},\ldots,n^{+};p'^{+},q'^{+}) =
\cr &\quad=
-ig^{n+2} \negthinspace\negthinspace
\permsum{1}{n}
\sum_{b=0}^n \sum_{c=b}^n
\sum_{d=c}^n \sum_{e=d}^n
{(\Omega[1,d])^{i}}_{\ell}
{(\Omega[d{+}1,n])_{k}}^{j}
\cr & \qquad\qquad\qquad\times
{ {1}\over{k_0^2} }
\Ims_{\alpha\dot\alpha}
\bigl((b{+}1)^{+},\ldots,c^{+},0^{*},(d{+}1)^{+},\ldots,e^{+}\bigr)
\cr & \qquad\qquad\qquad\times
{
{(-\sqrt2)^{n+b-c+d-e} \braket{p}{q} u^{\alpha}(p)
[\bar\kappa(e{+}1,n)+\bar p'  ]^{\dot\alpha\beta} u_{\beta}(q)}
\over
{ \bra{p}1,\ldots,b\ket{q}
\thinspace\bra{q} e{+}1,\ldots,n \ket{p'}
\thinspace\bra{q} c{+}1,\ldots,d \ket{q'}  }
},
}
\eqlabel\NEWrelabel
$$
where the new limits are most easily obtained by examining
Figure \tamedfig.  The momentum of  the zeroth gluon  now reads
$$
\eqalign{
k_0 &= q+q'+\kappa(c{+}1,d)
\cr &
= -p-\kappa(1,c)-\kappa(d{+}1,n)-p'
}
\eqlabel\KZEROnew
$$

Because the explicit form
of the modified gluon current depends on the location of
the gluon labelled by $0^{*}$, we write this term as the
sum of four contributions.  If $c=b$ and $e=d$, then
$\Ims(0^{*})$ appears:
$$
\eqalign{
\amp_1&\equiv
-ig^{n+2} \negthinspace\negthinspace
\permsum{1}{n}
\sum_{b=0}^n \sum_{d=b}^n
{(\Omega[1,d])^{i}}_{\ell}
{(\Omega[d{+}1,n])_{k}}^{j}
\cr & \qquad\qquad\times
{ {1}\over{k_0^2} }
{
{(-\sqrt2)^{n} \braket{p}{q} u^{\alpha}(p)
\Ims_{\alpha\dot\alpha}(0^{*})
[\bar\kappa(d{+}1,n)+\bar p']^{\dot\alpha\beta} u_{\beta}(q)}
\over
{ \bra{p}1,\ldots,b\ket{q}
\thinspace\bra{q} d{+}1,\ldots,n \ket{p'}
\thinspace\bra{q} b{+}1,\ldots,d \ket{q'}  }
}.
}
\eqlabel\Monedef
$$
This term corresponds to single gluon exchange.
Next, we write the terms that have $0^{*}$ as the first of
multiple arguments of $\Ims$.  These are the remaining $c=b$
contributions:
$$
\eqalign{
\amp_2 & \equiv
-ig^{n+2} \negthinspace\negthinspace
\permsum{1}{n}
\sum_{b=0}^{n-1} \sum_{d=b}^{n-1} \sum_{e=d{+}1}^n
{(\Omega[1,d])^{i}}_{\ell}
{(\Omega[d{+}1,n])_{k}}^{j}
\cr & \qquad\qquad\qquad\times
{ {1}\over{k_0^2} }
\Ims_{\alpha\dot\alpha}(0^{*},(d{+}1)^{+},\ldots,e^{+})
\cr & \qquad\qquad\qquad\times
{
{(-\sqrt2)^{n+d-e} \braket{p}{q} u^{\alpha}(p)
[\bar\kappa(e{+}1,n)+\bar p']^{\dot\alpha\beta} u_{\beta}(q)}
\over
{ \bra{p}1,\ldots,b\ket{q}
\thinspace\bra{q} e{+}1,\ldots,n \ket{p'}
\thinspace\bra{q} b{+}1,\ldots,d \ket{q'}  }
}.
}
\eqlabel\Mtwodef
$$
Note that we start at $e=d+1$  and adjust the $b$ and
$d$ summation ranges accordingly, since $e=d$  (with $c=b$)
is already given in (\Monedef).   The third piece contains
multiple-argument $\Ims$'s that have $0^{*}$ appearing
last.  These are the $e=d$ contributions not accounted for
in the previous two terms:
$$
\eqalign{
\amp_3 & \equiv
-ig^{n+2} \negthinspace\negthinspace
\permsum{1}{n}
\sum_{b=0}^{n-1} \sum_{c=b+1}^n \sum_{d=c}^n
{(\Omega[1,d])^{i}}_{\ell}
{(\Omega[d{+}1,n])_{k}}^{j}
\cr & \qquad\qquad\qquad\times
{ {1}\over{k_0^2} }
\Ims_{\alpha\dot\alpha}((b{+}1)^{+},\ldots,c^{+},0^{*})
\cr & \qquad\qquad\qquad\times
{
{(-\sqrt2)^{n+b-c} \braket{p}{q} u^{\alpha}(p)
[\bar\kappa(d{+}1,n)+\bar p']^{\dot\alpha\beta} u_{\beta}(q)}
\over
{ \bra{p}1,\ldots,b\ket{q}
\thinspace\bra{q} d{+}1,\ldots,n \ket{p'}
\thinspace\bra{q} c{+}1,\ldots,d \ket{q'}  }
}.
}
\eqlabel\Mthreedef
$$
The sum on $c$ begins at $b+1$ since $c=b$ (with $e=d$)
appears in (\Monedef).  As a consequence, the range of
$b$ must be adjusted.
In all of the remaining terms,  $0^{*}$
is neither the first nor the last argument of $\Ims$:
$$
\eqalign{
\amp_4 & \equiv
-ig^{n+2} \negthinspace\negthinspace
\permsum{1}{n}
\sum_{b=0}^{n-2} \sum_{c=b+1}^{n-1}
\sum_{d=c}^{n-1} \sum_{e=d+1}^n
{(\Omega[1,d])^{i}}_{\ell}
{(\Omega[d{+}1,n])_{k}}^{j}
\cr & \qquad\qquad\qquad\times
{ {1}\over{k_0^2} }
\Ims_{\alpha\dot\alpha}
\bigl((b{+}1)^{+},\ldots,c^{+},0^{*},(d{+}1)^{+},\ldots,e^{+}\bigr)
\cr & \qquad\qquad\qquad\times
{
{(-\sqrt2)^{n+b-c+d-e} \braket{p}{q} u^{\alpha}(p)
[\bar\kappa(e{+}1,n)+\bar p']^{\dot\alpha\beta} u_{\beta}(q)}
\over
{ \bra{p}1,\ldots,b\ket{q}
\thinspace\bra{q} e{+}1,\ldots,n \ket{p'}
\thinspace\bra{q} c{+}1,\ldots,d \ket{q'}  }
}.
}
\eqlabel\Mfourdef
$$
Once more we have adjusted the summation ranges to avoid
contributions already accounted for.  We will evaluate
each of $\amp_1$ through $\amp_4$ in order.


\subsection{Evaluation of $\amp_1$}

We insert the expression (\Isingle) for $\Ims(0^{*})$
into equation (\Monedef) to obtain the contribution from $\amp_1$:
$$
\eqalign{
\amp_1&=
{{-i}\over{2}}
(-g\sqrt2)^{n+2} \negthinspace\negthinspace
\permsum{1}{n}
\sum_{b=0}^n \sum_{d=b}^n
{(\Omega[1,d])^{i}}_{\ell}
{(\Omega[d{+}1,n])_{k}}^{j}
\cr & \qquad\qquad\times
{ {1}\over{k_0^2} }
{
{{\braket{p}{q}}^2 u^{\alpha}(q)
k_{0\alpha\dot\alpha}
[\bar\kappa(d{+}1,n)+\bar p']^{\dot\alpha\beta} u_{\beta}(q)}
\over
{ \bra{p}1,\ldots,b\ket{q}
\thinspace\bra{q} b{+}1,\ldots,d \ket{q'}
\thinspace\bra{q} d{+}1,\ldots,n \ket{p'}  }
}.
}
\eqlabel\Monestart
$$
Using (\KZEROnew) with $c=b$ we see that the numerator in (\Monestart)
reads
$$
\eqalign{
\num_1&\equiv u^{\alpha}(q)
k_{0\alpha\dot\alpha}
[\bar\kappa(d{+}1,n)+\bar p']^{\dot\alpha\beta} u_{\beta}(q)
\cr & =
-u^{\alpha}(q)
[p+\kappa(1,b)+\kappa(d{+}1,n)+p']_{\alpha\dot\alpha}
[\bar\kappa(d{+}1,n)+\bar p']^{\dot\alpha\beta} u_{\beta}(q)
\cr & =
-u^{\alpha}(q)
[p+\kappa(1,b)]_{\alpha\dot\alpha}
[\bar\kappa(d{+}1,n)+\bar p']^{\dot\alpha\beta} u_{\beta}(q),
}
\eqlabel\Monenum
$$
where we have used (\slashsqr) to extract a term
$[\kappa(d{+}1,n)+p']^2\braket{q}{q}=0$.  Inserting
(\Monenum) into (\Monestart) and interchanging the
order of the summations produces
$$
\eqalign{
\amp_1&=
{{i}\over{2}}
(-g\sqrt2)^{n+2} \negthinspace\negthinspace
\permsum{1}{n}
\sum_{d=0}^n \sum_{b=0}^d
{(\Omega[1,d])^{i}}_{\ell}
{(\Omega[d{+}1,n])_{k}}^{j}
\cr & \qquad\qquad\quad\times
{
{{\braket{p}{q}}^2 u^{\alpha}(q)
[p+\kappa(1,b)]_{\alpha\dot\alpha}
[\bar\kappa(d{+}1,n)+\bar p']^{\dot\alpha\beta} u_{\beta}(q)}
\over
{ \bra{p}1,\ldots,b\ket{q}
\thinspace\bra{q} b{+}1,\ldots,d \ket{q'}
\thinspace\bra{q} d{+}1,\ldots,n \ket{p'}  }
}
\cr & \qquad\qquad\quad\times
{ {1}\over{[q+q'+\kappa(b{+}1,d)]^2} }.
}
\eqlabel\Monedone
$$
No further simplification of $\amp_1$ is possible because
the propagator depends on both $b$ and $d$.


\subsection{Evaluation of $\amp_2$}

Equation (\Mtwodef) for $\amp_2$ contains
$\Ims(0^{*},(d{+}1)^{+},\ldots,e^{+})$ for $e\ge d+1$.  Application of
(\factorizeI) and (\Xsimple) produces
$$
\eqalign{
\amp_2 & =
{{i}\over{2}}
(-g\sqrt2)^{n+2} \negthinspace\negthinspace
\permsum{1}{n}
\sum_{b=0}^{n-1} \sum_{d=b}^{n-1} \sum_{e=d+1}^n \sum_{y=d+1}^e
{(\Omega[1,d])^{i}}_{\ell}
{(\Omega[d{+}1,n])_{k}}^{j}
\cr & \qquad\times
{
{ {\braket{p}{q}}^2 u^{\alpha}(q)
[ k_0 + \kappa(d{+}1,e) ]_{\alpha\dot\alpha}
[\bar\kappa(e{+}1,n)+\bar p']^{\dot\alpha\beta} u_{\beta}(q)}
\over
{ \bra{p}1,\ldots,b\ket{q}
\thinspace\bra{q} b{+}1,\ldots,d \ket{q'}
\thinspace\bra{q} d{+}1,\ldots,e \ket{q}
\thinspace\bra{q} e{+}1,\ldots,n \ket{p'}  }
}
\cr & \qquad\times
{ \coef(0,d{+}1,\ldots,y) }.
}
\eqlabel\Mtwostart
$$

We begin by observing that
$$
\eqalign{
\coef(0,d{+}1,\ldots,y) &=
{
{ u^{\gamma}(q)
[k_0 + \kappa(d{+}1,y)]_{\gamma\dot\gamma}
\bar k_y^{\dot\gamma\delta}
u_{\delta}(q) }
\over
{ [k_0+\kappa(d{+}1,y{-}1)]^2 \thinspace [k_0 +\kappa(d{+}1,y)]^2 }
}
\cr & =
{
{ u^{\gamma}(q)
[q{+}q'+\kappa(b{+}1,y)]_{\gamma\dot\gamma}
\bar k_y^{\dot\gamma\delta}
u_{\delta}(q) }
\over
{ [q{+}q'+\kappa(b{+}1,y{-}1)]^2
\thinspace [q{+}q'+\kappa(b{+}1,y)]^2 }
}
\cr & =
\coef(q{+}q',b{+}1,\ldots,y),
}
\eqlabel\Mtwocrewrite
$$
as defined by (\cdef) and in accordance with (\KZEROnew).
The numerator of (\Mtwostart) may be written as
$$
\eqalign{
\num_2 & \equiv
u^{\alpha}(q)
[k_0+\kappa(d{+}1,e)]_{\alpha\dot\alpha}
[\bar\kappa(e{+}1,n)+\bar p']^{\dot\alpha\beta} u_{\beta}(q)
\cr & =
u^{\alpha}(q)
[q+ q'+\kappa(b{+}1,e)]_{\alpha\dot\alpha}
[\bar\kappa(e{+}1,n)+\bar p']^{\dot\alpha\beta} u_{\beta}(q).
}
\eqlabel\oooooo
$$
We use (\slashsqr) and momentum conservation to rewrite (\oooooo) as
$$
\eqalign{
\num_2 & =
u^{\alpha}(q)
[q+ q'+\kappa(b{+}1,n)+p']_{\alpha\dot\alpha}
[\bar\kappa(e{+}1,n)+\bar p']^{\dot\alpha\beta} u_{\beta}(q)
\cr & =
-u^{\alpha}(q)
[p+\kappa(1,b)]_{\alpha\dot\alpha}
[\bar\kappa(e{+}1,n)+\bar p']^{\dot\alpha\beta} u_{\beta}(q).
}
\eqlabel\numadjusttwo
$$
Inserting (\Mtwocrewrite) and (\numadjusttwo) back into
(\Mtwostart) yields
$$
\eqalign{
\amp_2 & =
{{-i}\over{2}}
(-g\sqrt2)^{n+2} \negthinspace\negthinspace
\permsum{1}{n}
\sum_{b=0}^{n-1} \sum_{d=b}^{n-1}
\sum_{e=d+1}^{n} \sum_{y=d+1}^e
{(\Omega[1,d])^{i}}_{\ell}
{(\Omega[d{+}1,n])_{k}}^{j}
\cr & \qquad\times
{
{ {\braket{p}{q}}^2 u^{\alpha}(q)
[p+\kappa(1,b)]_{\alpha\dot\alpha}
[\bar\kappa(e{+}1,n)+\bar p']^{\dot\alpha\beta} u_{\beta}(q)}
\over
{ \bra{p}1,\ldots,b\ket{q}
\thinspace\bra{q} b{+}1,\ldots,d \ket{q'}
\thinspace\bra{q}  d{+}1,\ldots,e\ket{q}
\thinspace\bra{q} e{+}1,\ldots,n \ket{p'} }
}
\cr & \qquad\times
\coef(q{+}q',b{+}1,\ldots,y).
}
\eqlabel\Mtwopresum
$$

At this stage we may perform the sum on $e$.  Isolating
the $e$-dependent portion of  (\Mtwopresum) we have
$$
\sigma_2 \equiv \sum_{e=d+1}^n  \sum_{y=d+1}^e \sum_{f=e+1}^{n+1}
\link{e}{q}{e{+}1}
\bar k_f^{\dot\alpha\beta}u_{\beta}(q),
\eqlabel\DSUM
$$
where $k_d$ is understood to mean $q$ and $k_{n{+}1}$
really means $p'$.  The factor $\link{e}{q}{e{+}1}$
allows us to write $\bra{q} d{+}1,\ldots,n\ket{p'}$ in the
denominator of $\amp_2$.  Interchanging the order of
the sums and using (\linkidsummed) produces
$$
\eqalign{
\sigma_2 & =
\sum_{y=d+1}^n \sum_{f=y+1}^{n+1}\sum_{e=y}^{f-1}
\link{e}{q}{e{+}1}
\bar u^{\dot\alpha}(k_f) \braket{f}{q}
\cr & =
\sum_{y=d+1}^n \sum_{f=y+1}^{n+1}
\link{y}{q}{f}
\bar u^{\dot\alpha}(k_f) \braket{f}{q}
\cr & =
\sum_{y=d+1}^n
{ {1} \over { \braket{y}{q} } }
[\bar \kappa(y{+}1,n) + \bar p']^{\dot\alpha\beta}
u_{\alpha}(k_{y}).
}
\eqlabel\sigmatwo
$$
When we insert (\sigmatwo) back into (\Mtwopresum) and re-order the
sums, we obtain
$$
\eqalign{
\amp_2 & =
{{-i}\over{2}}
(-g\sqrt2)^{n+2} \negthinspace\negthinspace
\permsum{1}{n}
\sum_{d=0}^{n-1}
\sum_{b=0}^d \sum_{y=d+1}^n
{(\Omega[1,d])^{i}}_{\ell}
{(\Omega[d{+}1,n])_{k}}^{j}
\cr & \qquad\times
{
{ {\braket{p}{q}}^2 u^{\alpha}(q)
[p+\kappa(1,b)]_{\alpha\dot\alpha}
[\bar\kappa(y{+}1,n)+\bar p']^{\dot\alpha\beta} u_{\beta}(k_{y})}
\over
{ \bra{p}1,\ldots,b\ket{q}
\thinspace\bra{q} b{+}1,\ldots,d \ket{q'}
\thinspace\bra{q}  d{+}1,\ldots,n\ket{p'} }
}
\cr & \qquad\times
{ {1} \over { \braket{y}{q} } }
\coef(q{+}q',b{+}1,\ldots,y).
}
\eqlabel\Mtwoysum
$$

Let us examine some of the factors in (\Mtwoysum), namely the
combination
$$
\Xi_2 \equiv
{ {1}\over {\braket{y}{q}} }
u^{\alpha}(q)
[p+\kappa(1,b)]_{\alpha\dot\alpha}
[\bar\kappa(y{+}1,n)+\bar p']^{\dot\alpha\beta} u_{\beta}(k_{y})
\coef(q{+}q',b{+}1,\ldots,y).
\eq
$$
Supplying the definition of $\coef$ from (\cdef) yields
$$
\Xi_2 =
{
{ u^{\alpha}(q)
[p+\kappa(1,b)]_{\alpha\dot\alpha}
[\bar\kappa(y{+}1,n)+\bar p']^{\dot\alpha\beta}
k_{y\beta\dot\beta}
[\bar q + \bar q' + \bar\kappa(b{+}1,y) ]^{\dot\beta\gamma}
u_{\gamma}(q) }
\over
{ [q+q'+\kappa(b{+}1,y{-}1)]^2 \thinspace
[q+q'+\kappa(b{+}1,y)]^2 }
}.
\eqlabel\XItwodef
$$
We may rewrite the numerator appearing in (\XItwodef) as a
combination of terms containing perfect squares.
The key ingredients are using the Weyl equation to
extend or shorten sums as appropriate, applying
momentum conservation to factors, and adding
zero (cleverly written) to various factors.
For example,
using the Weyl equation to write $\bar\kappa(y,n)$
for $\bar\kappa(y{+}1,n)$ and
writing
$$
k_y=[q+q'+\kappa(b{+}1,y)]-[q+q'+\kappa(b{+}1,y{-}1)]
\eq
$$
produces
$$
\eqalign{
\num_2 &\equiv
u^{\alpha}(q)
[p+\kappa(1,b)]_{\alpha\dot\alpha}
[\bar\kappa(y,n)+\bar p']^{\dot\alpha\beta}
k_{y\beta\dot\beta}
[\bar q + \bar q' + \bar\kappa(b{+}1,y) ]^{\dot\beta\gamma}
u_{\gamma}(q)
\cr & =
[q+q'+\kappa(b{+}1,y)]^2 \thinspace
u^{\alpha}(q)
[p+\kappa(1,b)]_{\alpha\dot\alpha}
[\bar\kappa(y,n)+\bar p']^{\dot\alpha\gamma}
u_{\gamma}(q)
\cr & \quad
-u^{\alpha}(q)
[p+\kappa(1,b)]_{\alpha\dot\alpha}
[\bar\kappa(y,n)+\bar p']^{\dot\alpha\beta}
\cr & \qquad\quad\times
[q+q'+\kappa(b{+}1,y{-}1)]_{\beta\dot\beta}
[\bar q + \bar q' + \bar\kappa(b{+}1,y) ]^{\dot\beta\gamma}
u_{\gamma}(q).
}
\eq
$$
Momentum conservation lets us write
$$
-[\kappa(y,n)+p']=[p+\kappa(1,b)]+[q+q'+\kappa(b{+}1,y{-}1)],
\eq
$$
yielding
$$
\eqalign{
\num_2 & =
[q+q'+\kappa(b{+}1,y)]^2 \thinspace
u^{\alpha}(q)
[p+\kappa(1,b)]_{\alpha\dot\alpha}
[\bar\kappa(y,n)+\bar p']^{\dot\alpha\gamma}
u_{\gamma}(q)
\cr & \quad
+ [p+\kappa(1,b)]^2 \thinspace
u^{\alpha}(q)
[q+q'+\kappa(b{+}1,y{-}1)]_{\alpha\dot\beta}
[\bar q + \bar q' + \bar\kappa(b{+}1,y) ]^{\dot\beta\gamma}
u_{\gamma}(q)
\cr & \quad
+[q+q'+\kappa(b{+}1,y{-}1)]^2 \thinspace
u^{\alpha}(q)
[p+\kappa(1,b)]_{\alpha\dot\beta}
[\bar q + \bar q' + \bar\kappa(b{+}1,y) ]^{\dot\beta\gamma}
u_{\gamma}(q).
}
\eq
$$
Finally, we use momentum conservation on the factor
$[\bar q + \bar q' + \bar\kappa(b{+}1,y) ]^{\dot\beta\gamma}$
appearing in the third term and shorten some of the momentum
sums by extracting pieces proportional to $\braket{q}{q}$.
The result of this manipulation is
$$
\eqalign{
\num_2 & =
[q+q'+\kappa(b{+}1,y)]^2  \thinspace
u^{\alpha}(q)
[p+\kappa(1,b)]_{\alpha\dot\alpha}
[\bar\kappa(y,n)+\bar p']^{\dot\alpha\gamma}
u_{\gamma}(q)
\cr & \quad
-[q+q'+\kappa(b{+}1,y{-}1)]^2 \thinspace
u^{\alpha}(q)
[p+\kappa(1,b)]_{\alpha\dot\beta}
[\bar\kappa(y{+}1,n)+\bar p' ]^{\dot\beta\gamma}
u_{\gamma}(q)
\cr & \quad
+ [p+\kappa(1,b)]^2 \thinspace
u^{\alpha}(q)
[q+q'+\kappa(b{+}1,y{-}1)]_{\alpha\dot\beta}
\bar k_y^{\dot\beta\gamma}
u_{\gamma}(q),
}
\eqlabel\XItwodone
$$
where we have grouped the terms in a suggestive manner.

When (\XItwodone) is inserted into (\XItwodef), we observe
that there are denominator cancellations in the first two
terms, and that the third term contains a factor of
$\coef(q{+}q',b{+}1,\ldots,y)$:
$$
\eqalign{
\Xi_2 &=
\Biggl[
{
{u^{\alpha}(q)
[p+\kappa(1,b)]_{\alpha\dot\alpha}
[\bar\kappa(y,n)+\bar p']^{\dot\alpha\gamma}
u_{\gamma}(q)  }
\over
{ [q+q'+\kappa(b{+}1,y{-}1)]^2 }
}
\cr & \quad
-{
{u^{\alpha}(q)
[p+\kappa(1,b)]_{\alpha\dot\beta}
[\bar\kappa(y{+}1,n)+\bar p' ]^{\dot\beta\gamma}
u_{\gamma}(q)  }
\over
{[q+q'+\kappa(b{+}1,y)]^2 }
}
\Biggr]
\cr & \quad
+[p+\kappa(1,b)]^2 \thinspace \coef(q{+}q',b{+}1,\ldots,y).
}
\eqlabel\ughhhhhh
$$
Combining (\ughhhhhh) with (\Mtwoysum), we notice that the
terms grouped in brackets may be immediately summed over $y$,
giving
$$
\eqalign{
\amp_2 & = -
{{i}\over{2}}
(-g\sqrt2)^{n+2} \negthinspace\negthinspace
\permsum{1}{n}
\sum_{d=0}^{n-1}
\sum_{b=0}^d
{(\Omega[1,d])^{i}}_{\ell}
{(\Omega[d{+}1,n])_{k}}^{j}
\cr & \qquad\quad\times
{
{ {\braket{p}{q}}^2}
\over
{ \bra{p}1,\ldots,b\ket{q}
\thinspace\bra{q} b{+}1,\ldots,d \ket{q'}
\thinspace\bra{q}  d{+}1,\ldots,n\ket{p'} }
}
\cr & \qquad\quad\times
\Biggl[
{
{u^{\alpha}(q)
[p+\kappa(1,b)]_{\alpha\dot\alpha}
[\bar\kappa(d{+}1,n)+\bar p']^{\dot\alpha\gamma}
u_{\gamma}(q)  }
\over
{ [q+q'+\kappa(b{+}1,d)]^2 }
}
\cr & \qquad\qquad\quad
-{
{u^{\alpha}(q)
[p+\kappa(1,b)]_{\alpha\dot\beta}
\bar p'^{\dot\beta\gamma}
u_{\gamma}(q)  }
\over
{[q+q'+\kappa(b{+}1,n)]^2 }
}
\Biggr]
\cr & \quad-
{{i}\over{2}}
(-g\sqrt2)^{n+2} \negthinspace\negthinspace
\permsum{1}{n}
\sum_{d=0}^{n-1}
\sum_{b=0}^d \sum_{y=d+1}^n
{(\Omega[1,d])^{i}}_{\ell}
{(\Omega[d{+}1,n])_{k}}^{j}
\cr & \qquad\quad\times
{
{ {\braket{p}{q}}^2
[p+\kappa(1,b)]^2 \thinspace \coef(q{+}q',b{+}1,\ldots,y)}
\over
{ \bra{p}1,\ldots,b\ket{q}
\thinspace\bra{q} b{+}1,\ldots,d \ket{q'}
\thinspace\bra{q}  d{+}1,\ldots,n\ket{p'} }
}.
}
\eqlabel\Mtwonearlydone
$$
The quantity in square brackets in the first contribution
to (\Mtwonearlydone) vanishes when $d=n$; hence, we may extend
the sum to include that point.  So, the result for $\amp_2$ reads
$$
\eqalign{
\amp_2 & = -
{{i}\over{2}}
(-g\sqrt2)^{n+2} \negthinspace\negthinspace
\permsum{1}{n}
\sum_{d=0}^{n}
\sum_{b=0}^d
{(\Omega[1,d])^{i}}_{\ell}
{(\Omega[d{+}1,n])_{k}}^{j}
\cr & \qquad\qquad\times
{
{ {\braket{p}{q}}^2
u^{\alpha}(q)
[p+\kappa(1,b)]_{\alpha\dot\alpha}
[\bar\kappa(d{+}1,n)+\bar p']^{\dot\alpha\gamma}
u_{\gamma}(q)}
\over
{ \bra{p}1,\ldots,b\ket{q}
\thinspace\bra{q} b{+}1,\ldots,d \ket{q'}
\thinspace\bra{q}  d{+}1,\ldots,n\ket{p'} }
}
\cr & \qquad\qquad\times
{
{1}
\over
{ [q+q'+\kappa(b{+}1,d)]^2 }
}
\cr & \quad +
{{i}\over{2}}
(-g\sqrt2)^{n+2} \negthinspace\negthinspace
\permsum{1}{n}
\sum_{d=0}^{n}
\sum_{b=0}^d
{(\Omega[1,d])^{i}}_{\ell}
{(\Omega[d{+}1,n])_{k}}^{j}
\cr & \qquad\qquad\times
{
{ {\braket{p}{q}}^2
u^{\alpha}(q)
[p+\kappa(1,b)]_{\alpha\dot\beta}
\bar p'^{\dot\beta\gamma}
u_{\gamma}(q)}
\over
{ \bra{p}1,\ldots,b\ket{q}
\thinspace\bra{q} b{+}1,\ldots,d \ket{q'}
\thinspace\bra{q}  d{+}1,\ldots,n\ket{p'} }
}
\cr & \qquad\qquad\times
{
{ 1 }
\over
{[p+p'+\kappa(1,b)]^2 }
}
\cr & \quad-
{{i}\over{2}}
(-g\sqrt2)^{n+2} \negthinspace\negthinspace
\permsum{1}{n}
\sum_{d=0}^{n-1}
\sum_{b=0}^d \sum_{y=d+1}^n
{(\Omega[1,d])^{i}}_{\ell}
{(\Omega[d{+}1,n])_{k}}^{j}
\cr & \qquad\qquad\times
{
{ {\braket{p}{q}}^2
[p+\kappa(1,b)]^2 \thinspace \coef(q{+}q',b{+}1,\ldots,y)}
\over
{ \bra{p}1,\ldots,b\ket{q}
\thinspace\bra{q} b{+}1,\ldots,d \ket{q'}
\thinspace\bra{q}  d{+}1,\ldots,n\ket{p'} }
}.
}
\eqlabel\Mtwodone
$$
Note that the first term in (\Mtwodone) exactly cancels the
contribution from $\amp_1$ ({\it cf.}{ }equation (\Monedone)).

\subsection{Evaluation of $\amp_3$}

Equation (\Mthreedef) for $\amp_3$ contains
$\Ims((b{+}1)^{+},\ldots,c^{+},0^{*})$ for $c\ge b+1$.  Application of
(\factorizeI) and (\Xlaststarred) produces
$$
\eqalign{
\amp_3 & =
-{{i}\over{2}}
(-g\sqrt2)^{n+2} \negthinspace\negthinspace
\permsum{1}{n}
\sum_{b=0}^{n-1} \sum_{c=b+1}^n \sum_{d=c}^n \sum_{y=b+1}^c
{(\Omega[1,d])^{i}}_{\ell}
{(\Omega[d{+}1,n])_{k}}^{j}
\cr & \qquad\quad\times
{
{ {\braket{p}{q}}^2 u^{\alpha}(q)
[k_0+\kappa(b{+}1,c)]_{\alpha\dot\alpha}
[\bar\kappa(d{+}1,n)+\bar p']^{\dot\alpha\beta} u_{\beta}(q)}
\over
{ \bra{p}1,\ldots,b\ket{q}
\thinspace \bra{q}b{+}1,\ldots,c\ket{q}
\thinspace\bra{q} c{+}1,\ldots,d \ket{q'}
\thinspace\bra{q} d{+}1,\ldots,n \ket{p'}  }
}
\cr & \qquad\quad\times
\coef(0,c,c{-}1,\ldots,y).
}
\eqlabel\Mthreestart
$$
The numerator of (\Mthreestart) contains
$$
\eqalign{
\num_3  & \equiv
u^{\alpha}(q)
[k_0+\kappa(b{+}1,c)]_{\alpha\dot\alpha}
[\bar\kappa(d{+}1,n)+\bar p']^{\dot\alpha\beta} u_{\beta}(q)
\cr & =
u^{\alpha}(q)
[-p-\kappa(1,c)-\kappa(d{+}1,n)-p'+\kappa(b{+}1,c)]_{\alpha\dot\alpha}
[\bar\kappa(d{+}1,n)+\bar p']^{\dot\alpha\beta} u_{\beta}(q)
\cr & =
-u^{\alpha}(q)[p+\kappa(1,b)]_{\alpha\dot\alpha}
[\bar\kappa(d{+}1,n)+\bar p']^{\dot\alpha\beta} u_{\beta}(q),
}
\eqlabel\uuuuuu
$$
where we have used  (\KZEROnew) for $k_0$ and
shortened the sum in the first factor by using (\slashsqr)
to remove $[\kappa(d{+}1,n)+p']^2 \thinspace \braket{q}{q}=0$.
We note in passing that
$$
\coef(0,c,c{-}1,\ldots,y) =
\coef(q{+}q',d,d{-}1,\ldots,y).
\eqlabel\pppppp
$$
Inserting (\uuuuuu) and (\pppppp) back into (\Mthreestart)
and re-ordering some of the sums produces
$$
\eqalign{
\amp_3 &=
{{i}\over{2}}
(-g\sqrt2)^{n+2} \negthinspace\negthinspace
\permsum{1}{n}
\sum_{d=1}^n \sum_{y=1}^d \sum_{b=0}^{y-1} \sum_{c=y}^d
{(\Omega[1,d])^{i}}_{\ell}
{(\Omega[d{+}1,n])_{k}}^{j}
\cr & \qquad\quad\times
{
{ {\braket{p}{q}}^2 u^{\alpha}(q)
[p+\kappa(1,b)]_{\alpha\dot\alpha}
[\bar\kappa(d{+}1,n)+\bar p']^{\dot\alpha\beta} u_{\beta}(q)}
\over
{ \bra{p}1,\ldots,b\ket{q}
\thinspace \bra{q}b{+}1,\ldots,c\ket{q}
\thinspace\bra{q} c{+}1,\ldots,d \ket{q'}
\thinspace\bra{q} d{+}1,\ldots,n \ket{p'}  }
}
\cr & \qquad\quad\times
\coef(q{+}q',d,d{-}1,\ldots,y).
}
\eqlabel\Mthreenewvars
$$
The sum on $c$ appearing in (\Mthreenewvars) is almost trivial.
Applying (\linkidsummed) produces
$$
\eqalign{
\amp_3 &=
{{i}\over{2}}
(-g\sqrt2)^{n+2} \negthinspace\negthinspace
\permsum{1}{n}
\sum_{d=1}^n \sum_{y=1}^d \sum_{b=0}^{y{-}1}
{(\Omega[1,d])^{i}}_{\ell}
{(\Omega[d{+}1,n])_{k}}^{j}
\cr & \qquad\quad\times
{
{ {\braket{p}{q}}^2 u^{\alpha}(q)
[p+\kappa(1,b)]_{\alpha\dot\alpha}
[\bar\kappa(d{+}1,n)+\bar p']^{\dot\alpha\beta} u_{\beta}(q)}
\over
{ \bra{p}1,\ldots,b\ket{q}
\thinspace\bra{q} b{+}1,\ldots,d \ket{q'}
\thinspace\bra{q} d{+}1,\ldots,n \ket{p'}  }
}
\cr & \qquad\quad\times
\link{y}{q}{q'} \coef(q{+}q',d,d{-}1,\ldots,y).
}
\eqlabel\Mthreeprebsum
$$
The key factors in (\Mthreeprebsum),
$$
\sigma_3\equiv
\sum_{b=0}^{y{-}1} \sum_{a=0}^b
u^{\alpha}(q)k_{a\alpha\dot\alpha}
\link{b}{q}{b{+}1},
\eqlabel\sigmathreedef
$$
are summed following the procedure used on $\sigma_2$
({\it cf.}{ }equations (\DSUM)--(\sigmatwo)).
The result is
$$
\sigma_3 =
{ {1}\over{\braket{q}{y}} }
u^{\alpha}(k_y)[p+\kappa(1,y{-}1)]_{\alpha\dot\alpha}.
\eqlabel\sigmathree
$$
Inserting (\sigmathree) into (\Mthreeprebsum) results in
$$
\eqalign{
\amp_3 &=
{{i}\over{2}}
(-g\sqrt2)^{n+2} \negthinspace\negthinspace
\permsum{1}{n}
\sum_{d=1}^n \sum_{y=1}^d
{(\Omega[1,d])^{i}}_{\ell}
{(\Omega[d{+}1,n])_{k}}^{j}
\cr & \qquad\quad\times
{
{ {\braket{p}{q}}^2 u^{\alpha}(k_y)
[p+\kappa(1,y{-}1)]_{\alpha\dot\alpha}
[\bar\kappa(d{+}1,n)+\bar p']^{\dot\alpha\beta} u_{\beta}(q)}
\over
{ \bra{p}1,\ldots,d \ket{q'}
\thinspace\bra{q} d{+}1,\ldots,n \ket{p'}  }
}
\cr & \qquad\quad\times
{
{ \braket{y}{q'} }
\over
{ \braket{y}{q}\thinspace\braket{q}{q'}\thinspace\braket{q}{y} }
}
\coef(q{+}q',d,d{-}1,\ldots,y).
}
\eqlabel\Mthreeprebsum
$$
Momentum conservation allows us to split (\Mthreeprebsum) into
two contributions by writing
$$
[p+\kappa(1,y{-}1)]_{\alpha\dot\alpha}
= -[\kappa(d{+}1,n)+p']_{\alpha\dot\alpha}
-[q+q'+\kappa(y,d)]_{\alpha\dot\alpha}.
\eqlabel\splitMthree
$$
The first term of (\splitMthree) produces
$$
\eqalign{
\amp_{3{\rm{A}}} &\equiv
-{{i}\over{2}}
(-g\sqrt2)^{n+2} \negthinspace\negthinspace
\permsum{1}{n}
\sum_{d=1}^n \sum_{y=1}^d
{(\Omega[1,d])^{i}}_{\ell}
{(\Omega[d{+}1,n])_{k}}^{j}
\cr & \qquad\quad\times
{
{ {\braket{p}{q}}^2  \thinspace
[\kappa(d{+}1,n)+p']^2}
\over
{ \bra{p}1,\ldots,d \ket{q'}
\thinspace\bra{q} d{+}1,\ldots,n \ket{p'}  }
}
\cr & \qquad\quad\times
{
{ \braket{y}{q'} }
\over
{ \braket{q}{q'}\thinspace\braket{q}{y} }
}
\coef(q{+}q',d,d{-}1,\ldots,y),
}
\eqlabel\MthreeA
$$
which we set aside for the moment.  Later, we will find that
this contribution is cancelled by a portion of $\amp_4$.
The  other term  generated by (\splitMthree) produces
$$
\eqalign{
\amp_{3{\rm{B}}} &\equiv
-{{i}\over{2}}
(-g\sqrt2)^{n+2} \negthinspace\negthinspace
\permsum{1}{n}
\sum_{d=1}^n \sum_{y=1}^d
{(\Omega[1,d])^{i}}_{\ell}
{(\Omega[d{+}1,n])_{k}}^{j}
\cr & \qquad\quad\times
{
{ {\braket{p}{q}}^2 u^{\alpha}(k_y)
[q+q'+\kappa(y,d)]_{\alpha\dot\alpha}
[\bar\kappa(d{+}1,n)+\bar p']^{\dot\alpha\beta} u_{\beta}(q)}
\over
{ \bra{p}1,\ldots,d \ket{q'}
\thinspace\bra{q} d{+}1,\ldots,n \ket{p'}  }
}
\cr & \qquad\quad\times
{
{ \braket{y}{q'} }
\over
{ \braket{y}{q}\thinspace\braket{q}{q'}\thinspace\braket{q}{y} }
}
\coef(q{+}q',d,d{-}1,\ldots,y).
}
\eqlabel\MthreeBdef
$$

At this stage we would like to treat
$$
\eqalign{
\Xi_{3{\rm{B}}} &\equiv
u^{\alpha}(k_y)
[q+q'+\kappa(y,d)]_{\alpha\dot\alpha}
[\bar\kappa(d{+}1,n)+\bar p']^{\dot\alpha\beta} u_{\beta}(q)
\cr & \quad\times
{
{ \braket{y}{q'} }
\over
{ \braket{y}{q}\thinspace\braket{q}{q'}\thinspace\braket{q}{y} }
}
\coef(q{+}q',d,d{-}1,\ldots,y)
}
\eqlabel\XIthreeBdef
$$
in the same manner as  $\Xi_2$.  Thus, we insert the
definition (\cdef) of $c$ into (\XIthreeBdef) to obtain
$$
\eqalign{
\Xi_{3{\rm{B}}} &=
{
{ u^{\delta}(q)[q+q'+\kappa(y,d)]_{\delta\dot\delta}
\bar k_y^{\dot\delta\alpha}
[q+q'+\kappa(y{+}1,d)]_{\alpha\dot\alpha}
[\bar\kappa(d{+}1,n)+\bar p']^{\dot\alpha\beta} u_{\beta}(q)}
\over
{ [q+q'+\kappa(y{+}1,d)]^2 \thinspace [q+q'+\kappa(y,d)]^2 }
}
\cr & \quad\times
{
{ \braket{y}{q'} }
\over
{ \thinspace\braket{q}{q'}\thinspace\braket{q}{y} }
}
}
\eqlabel\XIthreego
$$
We use
$$
k_y=[q+q'+\kappa(y,d)]-[q+q'+\kappa(y{+}1,d)]
\eq
$$
to rewrite the numerator of (\XIthreego):
$$
\eqalign{
\num_{3{\rm{B}}} &\equiv
u^{\delta}(q)[q+q'+\kappa(y,d)]_{\delta\dot\delta}
\bar k_y^{\dot\delta\alpha}
[q+q'+\kappa(y{+}1,d)]_{\alpha\dot\alpha}
[\bar\kappa(d{+}1,n)+\bar p']^{\dot\alpha\beta} u_{\beta}(q)
\cr & =
[q+q'+\kappa(y,d)]^2
u^{\delta}(q)
[q+q'+\kappa(y{+}1,d)]_{\delta\dot\alpha}
[\bar\kappa(d{+}1,n)+\bar p']^{\dot\alpha\beta} u_{\beta}(q)
\cr & \quad
-[q+q'+\kappa(y{+}1,d)]^2
u^{\delta}(q)[q+q'+\kappa(y,d)]_{\delta\dot\delta}
[\bar\kappa(d{+}1,n)+\bar p']^{\dot\delta\beta} u_{\beta}(q).
}
\eq
$$
Thus,
$$
\eqalign{
\Xi_{3{\rm{B}}} &=
{
{ \braket{y}{q'} }
\over
{ \thinspace\braket{q}{q'}\thinspace\braket{q}{y} }
}
\Biggl[
{
{u^{\delta}(q)
[q+q'+\kappa(y{+}1,d)]_{\delta\dot\delta}
}
\over
{ [q+q'+\kappa(y{+}1,d)]^2  }
}
\cr & \qquad\qquad\qquad\enspace
-
{
{u^{\delta}(q)[q+q'+\kappa(y,d)]_{\delta\dot\delta}
 }
\over
{ [q+q'+\kappa(y,d)]^2 }
}
\Biggr]
[\bar\kappa(d{+}1,n)+\bar p']^{\dot\delta\beta} u_{\beta}(q).
}
\eqlabel\specialsplitstart
$$

It is useful to consider the sum on $y$
pending in the expression for $\amp_{3{\rm{B}}}$.
All of the $y$-dependence is contained in $\Xi_{3{\rm{B}}}$;
the sum may be written as
$$
\eqalign{
\sum_{y=1}^d \Xi_{3{\rm{B}}} &=
\sum_{y=1}^d
{
{ \braket{y}{q'} }
\over
{ \thinspace\braket{q}{q'}\thinspace\braket{q}{y} }
}
\biggl[ f(y{+}1) - f(y) \biggr],
}
\eq
$$
with $f(y)$ given by
$$
f(y) \equiv
{
{u^{\delta}(q)[q+q'+\kappa(y,d)]_{\delta\dot\delta}
[\bar\kappa(d{+}1,n)+\bar p']^{\dot\delta\beta} u_{\beta}(q) }
\over
{ [q+q'+\kappa(y,d)]^2 }
}.
\eqlabel\fdef
$$
Shifting the second sum by one produces
$$
\sum_{y=1}^d \Xi_{3{\rm{B}}} =
-\sum_{y=1}^d \link{y}{q}{q'} f(y{+}1)
-\sum_{y=0}^{d-1} \link{q'}{q}{y{+}1} f(y{+}1).
\eqlabel\shifted
$$
The two terms in (\shifted) may be combined using (\linkidnosum),
with the exception of the endpoints of the summation:
$$
\eqalign{
\sum_{y=1}^d \Xi_{3{\rm{B}}} &=
-\sum_{y=1}^{d-1}\link{y}{q}{y{+}1} f(y{+}1)
-\link{d}{q}{q'} f(d{+}1)
- \link{q'}{q}{1} f(1).
}
\eqlabel\hsjkafd
$$
At this stage, we recall that the ``natural'' assignment for
$k_{d+1}$ is $q'$, according to the rules  outlined in section 4.3.
Thus, the second term in (\hsjkafd)  may be absorbed into the first,
giving
$$
\eqalign{
\sum_{y=1}^d \Xi_{3{\rm{B}}} &=
-\sum_{y=1}^{d}\link{y}{q}{y{+}1} f(y{+}1)
+ \link{1}{q}{q'} f(1).
}
\eqlabel\XIthreeBdone
$$
Before combining (\MthreeBdef), (\fdef), and (\XIthreeBdone),
we note that the numerator of $f(y{+}1)$ may be rewritten
as
$$
\eqalign{
u^{\delta}(q)[&q+q'+\kappa(y{+}1,d)]_{\delta\dot\delta}
[\bar\kappa(d{+}1,n)+\bar p']^{\dot\delta\beta} u_{\beta}(q)
= \cr & =
-u^{\delta}(q)[p+\kappa(1,y)]_{\delta\dot\delta}
[\bar\kappa(d{+}1,n)+\bar p']^{\dot\delta\beta} u_{\beta}(q)
}
\eq
$$
by applying momentum conservation and (\slashsqr).
Hence, our final result for $\amp_{3{\rm{B}}}$ reads
$$
\eqalign{
\amp_{3{\rm{B}}}& =
-{{i}\over{2}}
(-g\sqrt2)^{n+2} \negthinspace\negthinspace
\permsum{1}{n}
\sum_{d=1}^n \sum_{y=1}^d
{(\Omega[1,d])^{i}}_{\ell}
{(\Omega[d{+}1,n])_{k}}^{j}
\cr & \qquad\quad\times
{
{ {\braket{p}{q}}^2 u^{\alpha}(q)
[p+\kappa(1,y)]_{\alpha\dot\alpha}
[\bar\kappa(d{+}1,n)+\bar p']^{\dot\alpha\beta} u_{\beta}(q)}
\over
{ \bra{p}1,\ldots,y\ket{q}
\bra{q} y{+}1,\ldots,d \ket{q'}
\thinspace\bra{q} d{+}1,\ldots,n \ket{p'}  }
}
\cr & \qquad\quad\times
{
{ 1 }
\over
{ [q+q'+\kappa(y{+}1,d)]^2 }
}
\cr & \quad +
{{i}\over{2}}
(-g\sqrt2)^{n+2} \negthinspace\negthinspace
\permsum{1}{n}
\sum_{d=1}^n
{(\Omega[1,d])^{i}}_{\ell}
{(\Omega[d{+}1,n])_{k}}^{j}
\cr & \qquad\quad\times
{
{ {\braket{p}{q}}^2 u^{\alpha}(q)
p_{\alpha\dot\alpha}
[\bar\kappa(d{+}1,n)+\bar p']^{\dot\alpha\beta} u_{\beta}(q)}
\over
{ \bra{p}1,\ldots,d \ket{q'}
\thinspace\bra{q} d{+}1,\ldots,n \ket{p'}  }
}
\cr & \qquad\quad\times
\link{1}{q}{q'}
{
{ 1 }
\over
{[q+q'+\kappa(1,d)]^2 }
}.
}
\eqlabel\MthreeB
$$

\vfill \eject


\subsection{Evaluation of $\amp_4$}

We come at last to the final contribution, $\amp_4$.  As indicated
by its definition (\Mfourdef), $\amp_4$ contains
$\Ims\bigl((b{+}1)^{+},\ldots,c^{+},0^{*},(d{+}1)^{+},\ldots,e^{+}\bigr)$.
As a consequence, it is the most complicated piece which
must be dealt with.  However, as we shall see, the procedure
is straightforward and employs the same techniques already
discussed.

We start by inserting (\factorizeI) and (\XGEN) into
(\Mfourdef).  After some regrouping, we obtain
$$
\eqalign{
\amp_4 &=
\negthinspace
{{-i}\over{2}}
(-g\sqrt2)^{n+2} \negthinspace\negthinspace
\negthinspace\negthinspace
\permsum{1}{n}
\sum_{b=0}^{n-2} \sum_{c=b+1}^{n-1}
\sum_{d=c}^{n-1} \sum_{e=d+1}^n
\sum_{y=b+1}^c \sum_{z=d+1}^e
\negthinspace\negthinspace
{(\Omega[1,d])^{i}}_{\ell}
{(\Omega[d{+}1,n])_{k}}^{j}
\cr & \qquad\enspace\times
{
{{\braket{p}{q}}^2 u^{\alpha}(q)
[\kappa(b{+}1,c)+k_0+\kappa(d{+}1,e)]_{\alpha\dot\alpha}
[\bar\kappa(e{+}1,n)+\bar p']^{\dot\alpha\beta} u_{\beta}(q)}
\over
{ \bra{p}1,\ldots,b\ket{q}
\thinspace\bra{q}b{+}1,\ldots,c\ket{q}
\thinspace\bra{q} c{+}1,\ldots,d \ket{q'}}
}
\cr & \qquad\enspace\times
{
{c_2(y,\ldots,c,0,d{+}1,\ldots,z)}
\over
{\bra{q}d{+}1,\ldots,e\ket{q}
\thinspace\bra{q} e{+}1,\ldots,n \ket{p'} }
}
\thinspace
\invlink{y}{q}{z}.
}
\eqlabel\Mfourstart
$$

Equation (\KZEROnew) tells us that
$$
\kappa(b{+}1,c)+k_0+\kappa(d{+}1,e) =
q+q'+\kappa(b{+}1,e).
\eqlabel\NEWNUM
$$
There are also
the consequences of the
``0'' appearing in $c_2(y,\ldots,c,0,d{+}1,\ldots,z)$ to consider.
First, note that the ranges $\{y,\ldots,c\}$ and
$\{d{+}1,\ldots,e\}$ always contain at least one
element each.    Hence, ``0'' is never
the first or last argument of $c_2$.  Now recall that
$c_2$ is symmetric in all but its
first and last arguments.  As a consequence, we may replace
the single argument ``0'' by the multiple arguments
``$q{+}q',c{+}1,c{+}2,\ldots,d$'' implied by (\KZEROnew).
Thus,
$$
\eqalign{
c_2(y,\ldots,c,0,d{+}1,\ldots,z)&=
c_2(y,\ldots,c,q{+}q',c{+}1,\ldots,d,d{+}1,\ldots,z)
\cr &
=c_2(y,q{+}q',y{+}1,\ldots,z),
}
\eqlabel\modifiedC
$$
where we have taken advantage of the symmetry in the arguments to
obtain the last line.  Combination of (\NEWNUM) and (\modifiedC)
with (\Mfourstart) produces the following expression after
interchanging some of the summations:
$$
\eqalign{
\amp_4 &=
\negthinspace
{{-i}\over{2}}
(-g\sqrt2)^{n+2} \negthinspace\negthinspace
\negthinspace\negthinspace
\permsum{1}{n}
\sum_{b=0}^{n-2} \sum_{c=b+1}^{n-1}
\sum_{d=c}^{n-1}
\sum_{y=b+1}^c \sum_{z=d+1}^n \sum_{e=z}^n
\negthinspace
{(\Omega[1,d])^{i}}_{\ell}
{(\Omega[d{+}1,n])_{k}}^{j}
\cr & \qquad\enspace\times
{
{{\braket{p}{q}}^2 u^{\alpha}(q)
[q+q'+\kappa(b{+}1,e)]_{\alpha\dot\alpha}
[\bar\kappa(e{+}1,n)+\bar p']^{\dot\alpha\beta} u_{\beta}(q)}
\over
{ \bra{p}1,\ldots,b\ket{q}
\thinspace\bra{q}b{+}1,\ldots,c\ket{q}
\thinspace\bra{q} c{+}1,\ldots,d \ket{q'}}
}
\cr & \qquad\enspace\times
{
{c_2(y,q{+}q',y{+}1,\ldots,z)}
\over
{\bra{q}d{+}1,\ldots,e\ket{q}
\thinspace\bra{q} e{+}1,\ldots,n \ket{p'} }
}
\thinspace
\invlink{y}{q}{z}.
}
\eqlabel\Mfournewvars
$$

Consider the sum on $e$ that appears in (\Mfournewvars).  We examine
the following factors that appear in $\amp_4$:
$$
\sigma_{4e} \equiv
\sum_{e=z}^n \negthinspace
u^{\alpha}(q)
[q+q'+\kappa(b{+}1,e)]_{\alpha\dot\alpha}
[\bar\kappa(e{+}1,n)+\bar p']^{\dot\alpha\beta} u_{\beta}(q)
\link{e}{q}{e{+}1},
\eqlabel\sigmafouredef
$$
which allows us to join $\bra{q}d{+}1,\ldots,e\ket{q}$
to $\bra{q}e{+}1,\ldots,n\ket{p'}$ with the understanding
that $k_{n+1}$ is read as $p'$.  Because of (\slashsqr) and
the antisymmetry of the spinor inner product, we may extend
the  first sum in (\sigmafouredef) to
$$
[q+q'+\kappa(b{+}1,n)+p']_{\alpha\dot\alpha}
 = -[p+\kappa(1,b)]_{\alpha\dot\alpha}.
\eq
$$
Applying this to (\sigmafouredef) and writing the remaining
$e$-dependent $\kappa$-sum explicitly produces
$$
\sigma_{4e} =
-
\sum_{e=z}^n \sum_{f=e+1}^{n+1}
u^{\alpha}(q)
[p+\kappa(1,b)]_{\alpha\dot\alpha}
\bar k_f^{\dot\alpha\beta} u_{\beta}(q)
\link{e}{q}{e{+}1}.
\eq
$$
Interchanging the order of summation and applying (\linkidsummed)
as usual produces
$$
\eqalign{
\sigma_{4e} &=
- \sum_{f=z+1}^{n{+}1} \sum_{e=z}^{f-1}
u^{\alpha}(q)
[p+\kappa(1,b)]_{\alpha\dot\alpha}
\bar k_f^{\dot\alpha\beta} u_{\beta}(q)
\link{e}{q}{e{+}1}
\cr & =
- \sum_{f=z+1}^{n{+}1}
u^{\alpha}(q)
[p+\kappa(1,b)]_{\alpha\dot\alpha}
\bar u^{\dot\alpha}(k_f) \braket{f}{q}
\link{z}{q}{f}
\cr & =
{ {-1}\over{\braket{z}{q}} }
u^{\alpha}(q)
[p+\kappa(1,b)]_{\alpha\dot\alpha}
[\bar\kappa(z,n)+\bar p']^{\dot\alpha\beta}
u_{\beta}(k_{z}).
}
\eqlabel\sigmafouredone
$$
Inserting (\sigmafouredone)  back into
(\Mfournewvars) and rearranging the summations  produces
$$
\eqalign{
\amp_4 &=
{{-i}\over{2}}
(-g\sqrt2)^{n+2} \negthinspace\negthinspace
\permsum{1}{n}
\sum_{d=1}^{n-1} \sum_{b=0}^{d-1}
\sum_{c=b+1}^{d} \sum_{y=b+1}^{c}
\sum_{z=d+1}^{n}
\negthinspace
{(\Omega[1,d])^i}_{\ell}
{(\Omega[d{+}1,n])_{k}}^{j}
\cr & \qquad\times
{
{{\braket{p}{q}}^2 u^{\alpha}(q)
[p+\kappa(1,b)]_{\alpha\dot\alpha}
[\bar\kappa(z{+}1,n)+\bar p']^{\dot\alpha\beta}
u_{\beta}(k_{z})}
\over
{ \bra{p}1,\ldots,b\ket{q}
\thinspace\bra{q}b{+}1,\ldots,c\ket{q}
\thinspace\bra{q} c{+}1,\ldots,d\ket{q'}
\thinspace\bra{q}d{+}1,\ldots,n \ket{p'}}
}
\cr & \qquad\times
{
{ \braket{y}{q} }
\over
{ \braket{y}{z} }
}
c_2(y,q{+}q',y{+}1,\ldots,z).
}
\eqlabel\MfourYZ
$$
Equation (\MfourYZ) has only ``trivial'' $c$-dependence;
we supply a factor $\link{c}{q}{c{+}1}$ to join the
corresponding pair
of denominator ``strings'' and write
$$
\eqalign{
\sum_{c=b+1}^{d} \sum_{y=b+1}^{c} \sum_{z=d+1}^{n}
\link{c}{q}{c{+}1} & =
\sum_{y=b+1}^d \sum_{z=d+1}^n \sum_{c=y}^d
\link{c}{q}{c{+}1} \cr & =
\sum_{y=b+1}^d \sum_{z=d+1}^n
\link{y}{q}{q'}.
}
\eq
$$
Hence, $\amp_4$ becomes
$$
\eqalign{
\amp_4 &=
{{-i}\over{2}}
(-g\sqrt2)^{n+2} \negthinspace\negthinspace
\permsum{1}{n}
\sum_{d=1}^{n-1} \sum_{b=0}^{d-1}  \sum_{y=b+1}^d
\sum_{z=d+1}^n
\negthinspace
{(\Omega[1,d])^i}_{\ell}
{(\Omega[d{+}1,n])_{k}}^{j}
\cr & \qquad\times
{
{{\braket{p}{q}}^2 u^{\alpha}(q)
[p+\kappa(1,b)]_{\alpha\dot\alpha}
[\bar\kappa(z{+}1,n)+\bar p']^{\dot\alpha\beta}
u_{\beta}(k_{z})}
\over
{ \bra{p}1,\ldots,b\ket{q}
\thinspace\bra{q}b{+}1,\ldots,d\ket{q'}
\thinspace\bra{q}d{+}1,\ldots,n \ket{p'}}
}
\cr & \qquad\times
{
{ \braket{y}{q'} }
\over
{ \braket{y}{z}  \braket{q}{q'}}
}
c_2(y,q{+}q',y{+}1,\ldots,z).
}
\eqlabel\onemoretogo
$$
The sum on $b$ in (\onemoretogo) may be performed as well.  The
relevant factors are
$$
\sigma_{4b} \equiv
\sum_{b=0}^{d-1}  \sum_{y=b+1}^d
\sum_{z=d+1}^n \sum_{a=0}^b
\link{b}{q}{b{+}1}
u^{\alpha}(q)
k_{a\alpha\dot\alpha},
\eq
$$
where we have written the $b$-dependent $\kappa$-sum as a
sum on $a$, $k_0\equiv p$.  Interchanging the order of summation
and applying (\linkidsummed), we obtain
$$
\eqalign{
\sigma_{4b} &=
\sum_{y=1}^{d} \sum_{z=d+1}^n
\sum_{a=0}^{y-1} \sum_{b=a}^{y-1}
\link{b}{q}{b{+}1}
\braket{q}{a} \bar u_{\dot\alpha}(k_a)
\cr & =
\sum_{y=1}^{d} \sum_{z=d+1}^n
{ {1}\over{ \braket{q}{y} } }
u^{\alpha}(k_y) [p+\kappa(1,y{-}1)]_{\alpha\dot\alpha}.
}
\eq
$$
At this stage, we have
$$
\eqalign{
\amp_4 &=
{{-i}\over{2}}
(-g\sqrt2)^{n+2} \negthinspace\negthinspace
\permsum{1}{n}
\sum_{d=1}^{n-1} \sum_{y=1}^d \sum_{z=d+1}^n
\negthinspace
{(\Omega[1,d])^i}_{\ell}
{(\Omega[d{+}1,n])_{k}}^{j}
\cr & \qquad\times
{
{{\braket{p}{q}}^2 u^{\alpha}(k_y)
[p+\kappa(1,y{-}1)]_{\alpha\dot\alpha}
[\bar\kappa(z{+}1,n)+\bar p']^{\dot\alpha\beta}
u_{\beta}(k_{z})}
\over
{ \bra{p}1,\ldots,d\ket{q'}
\thinspace\bra{q}d{+}1,\ldots,n \ket{p'}}
}
\cr & \qquad\times
{
{ \braket{y}{q'} }
\over
{ \braket{q}{y} \braket{y}{z}  \braket{q}{q'}}
}
\biggl[
\coef(q{+}q',y,y{+}1,\ldots,z) - \coef(q{+}q',y{+}1,y{+}2,\ldots,z)
\cr & \qquad\qquad\qquad\qquad\qquad\negthinspace
-\coef(q{+}q',z,z{-}1,\ldots,y) + \coef(q{+}q',z{-}1,z{-}2,\ldots,y)
\biggr],
}
\eqlabel\timetobreakup
$$
where we have made use of  (\ctwodef) and
the symmetry properties of  the $c$'s  to write out
$c_2(y,q{+}q',y{+}1,\ldots,z)$.

In order to make further progress with equation (\timetobreakup),
we must insert explicit expressions for the $c$ functions.
Let us begin by defining
$$
\eqalign{
\Xi_4 &\equiv
\sum_{d=1}^{n-1} \sum_{y=1}^d \sum_{z=d+1}^n
\negthinspace
{ u^{\alpha}(k_y)
[p+\kappa(1,y{-}1)]_{\alpha\dot\alpha}
[\bar\kappa(z{+}1,n)+\bar p']^{\dot\alpha\beta}
u_{\beta}(k_{z})}
\cr & \qquad\times
{
{ \braket{y}{q'} }
\over
{ \braket{q}{y} \braket{y}{z}  \braket{q}{q'}}
}
\biggl[
\coef(q{+}q',y,y{+}1,\ldots,z) - \coef(q{+}q',y{+}1,y{+}2,\ldots,z)
\cr & \qquad\qquad\qquad\qquad\qquad\negthinspace
-\coef(q{+}q',z,z{-}1,\ldots,y) + \coef(q{+}q',z{-}1,z{-}2,\ldots,y)
\biggr].
}
\eqlabel\XIfourdef
$$
We have retained the sum on $d$ since
we will be adjusting its range at a later stage.  It is understood
that there are $d$-dependent factors appearing
{\it inside}{ }this sum, although they are not written
out explicitly.  In addition, we will designate
the contributions from each of the four $\coef$-functions
appearing in (\XIfourdef)
by $\Xi_{4{\rm{A}}}$, $\Xi_{4{\rm{B}}}$, $\Xi_{4{\rm{C}}}$,
and $\Xi_{4{\rm{D}}}$ respectively.

Consider $\Xi_{4{\rm{A}}}$.  If we insert the definition
(\cdef),  we get
$$
\eqalign{
\Xi_{4{\rm{A}}} =
\sum_{d=1}^{n-1} \sum_{y=1}^d \sum_{z=d+1}^n
\negthinspace &
{
{ \braket{y}{q'}\braket{z}{q} }
\over
{ \braket{q}{y} \braket{y}{z}  \braket{q}{q'}}
}
\thinspace
u^{\alpha}(k_y)
[p+\kappa(1,y{-}1)]_{\alpha\dot\alpha}
\cr & \times
{
{[\bar\kappa(z{+}1,n)+\bar p']^{\dot\alpha\beta}
k_{z\beta\dot\gamma}
[\bar q+\bar q'+\bar\kappa(y,z)]^{\dot\gamma\gamma}
u_{\gamma}(q)}
\over
{ [q+q'+\kappa(y,z{-}1)]^2 [q+q'+\kappa(y,z)]^2 }
}.
}
\eqlabel\XIfourAdef
$$
We now proceed to transform
$$
\num_{4{\rm{A}}} \equiv
u^{\alpha}(k_y)
[p+\kappa(1,y{-}1)]_{\alpha\dot\alpha}
[\bar\kappa(z,n)+\bar p']^{\dot\alpha\beta}
k_{z\beta\dot\gamma}
[\bar q+\bar q'+\bar\kappa(y,z)]^{\dot\gamma\gamma}
u_{\gamma}(q),
\eq
$$
the numerator of (\XIfourAdef),
using the same
tricks outlined in the discussion of
$\Xi_2$ ({\it cf.}{ } equations (\XItwodef)--(\XItwodone)).
As in that case, our goal is to produce a series of terms
containing inverse propagator factors in order to cancel as many
denominators as possible.  The result of this series of manipulations
is
$$
\eqalign{
\num_{4{\rm{A}}}& =
[q+q'+\kappa(y,z)]^2  \thinspace
u^{\alpha}(k_y)
[p+\kappa(1,y{-}1)]_{\alpha\dot\alpha}
[\bar\kappa(z,n)+\bar p']^{\dot\alpha\beta}
u_{\beta}(q)
\cr & \quad
-[q+q'+\kappa(y,z{-}1)]^2  \thinspace
u^{\alpha}(k_y)
[p+\kappa(1,y{-}1)]_{\alpha\dot\alpha}
[\bar\kappa(z{+}1,n)+\bar p']^{\dot\alpha\gamma}
u_{\gamma}(q)
\cr & \quad
+[p+\kappa(1,y{-}1)]^2  \thinspace
u^{\alpha}(k_y)
[q+q'+\kappa(y,z{-}1)]_{\alpha\dot\gamma}
\bar k_z^{\dot\gamma\gamma}
u_{\gamma}(q).
}
\eqlabel\YGHET
$$

Application of (\YGHET) to (\XIfourAdef) yields
$$
\eqalign{
\Xi_{4{\rm{A}}} &=
\sum_{d=1}^{n-1} \sum_{y=1}^d \sum_{z=d+1}^n
\negthinspace
{
{ \braket{q'}{y}\braket{q}{z} }
\over
{ \braket{q}{y} \braket{y}{z}  \braket{q}{q'}}
}
\cr & \quad\times
\Biggl[
{
{ u^{\alpha}(k_y)
[p+\kappa(1,y{-}1)]_{\alpha\dot\alpha}
[\bar\kappa(z,n)+\bar p']^{\dot\alpha\beta}
u_{\beta}(q) }
\over
{[q+q'+\kappa(y,z{-}1)]^2}
}
\cr & \qquad
-
{
{ u^{\alpha}(k_y)
[p+\kappa(1,y{-}1)]_{\alpha\dot\alpha}
[\bar\kappa(z{+}1,n)+\bar p']^{\dot\alpha\beta}
u_{\beta}(q) }
\over
{[q+q'+\kappa(y,z)]^2}
}
\cr & \qquad
+
[p+\kappa(1,y{-}1)]^2
{
{ u^{\alpha}(k_y)
[q+q'+\kappa(y,z{-}1)]_{\alpha\dot\gamma}
\bar k_z^{\dot\gamma\gamma}
u_{\gamma}(q)}
\over
{[q+q'+\kappa(y,z{-}1)]^2 [q+q'+\kappa(y,z)]^2}
}
\Biggr].
}
\eqlabel\XIfourAnofierz
$$
Finally, we write (\fierz) in the form
$$
 u^{\alpha}(k_y)\thinspace\braket{q}{z}=
 u^{\alpha}(k_z)\thinspace\braket{q}{y}
+ u^{\alpha}(q)\thinspace\braket{y}{z}.
\eqlabel\FIERZZ
$$
to obtain
$$
\eqalign{
\Xi_{4{\rm{A}}} &=
\sum_{d=1}^{n-1} \sum_{y=1}^d \sum_{z=d+1}^n
\negthinspace
{
{ \braket{q'}{y} }
\over
{  \braket{y}{z}  \braket{q}{q'}}
}
\cr & \quad\times
\Biggl[
{
{ u^{\alpha}(k_z)
[p+\kappa(1,y{-}1)]_{\alpha\dot\alpha}
[\bar\kappa(z,n)+\bar p']^{\dot\alpha\beta}
u_{\beta}(q) }
\over
{[q+q'+\kappa(y,z{-}1)]^2}
}
\cr & \qquad
-
{
{ u^{\alpha}(k_z)
[p+\kappa(1,y{-}1)]_{\alpha\dot\alpha}
[\bar\kappa(z{+}1,n)+\bar p']^{\dot\alpha\beta}
u_{\beta}(q) }
\over
{[q+q'+\kappa(y,z)]^2}
}
\cr & \qquad
+
[p+\kappa(1,y{-}1)]^2
{
{ u^{\alpha}(k_z)
[q+q'+\kappa(y,z{-}1)]_{\alpha\dot\gamma}
\bar k_z^{\dot\gamma\gamma}
u_{\gamma}(q)}
\over
{[q+q'+\kappa(y,z{-}1)]^2 [q+q'+\kappa(y,z)]^2}
}
\Biggr]
\cr & +
\sum_{d=1}^{n-1} \sum_{y=1}^d \sum_{z=d+1}^n
\negthinspace
{
{ \braket{q'}{y} }
\over
{ \braket{q}{y}   \braket{q}{q'}}
}
\cr & \quad\times
\Biggl[
{
{ u^{\alpha}(q)
[p+\kappa(1,y{-}1)]_{\alpha\dot\alpha}
[\bar\kappa(z,n)+\bar p']^{\dot\alpha\beta}
u_{\beta}(q) }
\over
{[q+q'+\kappa(y,z{-}1)]^2}
}
\cr & \qquad
-
{
{ u^{\alpha}(q)
[p+\kappa(1,y{-}1)]_{\alpha\dot\alpha}
[\bar\kappa(z{+}1,n)+\bar p']^{\dot\alpha\beta}
u_{\beta}(q) }
\over
{[q+q'+\kappa(y,z)]^2}
}
\cr & \qquad
+
[p+\kappa(1,y{-}1)]^2
{
{ u^{\alpha}(q)
[q+q'+\kappa(y,z{-}1)]_{\alpha\dot\gamma}
\bar k_z^{\dot\gamma\gamma}
u_{\gamma}(q)}
\over
{[q+q'+\kappa(y,z{-}1)]^2 [q+q'+\kappa(y,z)]^2}
}
\Biggr].
}
\eqlabel\XIfourAsplit
$$
We may use  the same procedure to decompose $\Xi_{4{\rm{B}}}$.
The result is
$$
\eqalign{
\Xi_{4{\rm{B}}} &=
\sum_{d=1}^{n-1} \sum_{y=1}^d \sum_{z=d+1}^n
\negthinspace
{
{ \braket{q'}{y} }
\over
{  \braket{y}{z}  \braket{q}{q'}}
}
\cr & \quad\times
\Biggl[
{
{ u^{\alpha}(k_z)
[p+\kappa(1,y)]_{\alpha\dot\alpha}
[\bar\kappa(z{+}1,n)+\bar p']^{\dot\alpha\beta}
u_{\beta}(q) }
\over
{[q+q'+\kappa(y{+}1,z)]^2}
}
\cr & \qquad
-
{
{ u^{\alpha}(k_z)
[p+\kappa(1,y)]_{\alpha\dot\alpha}
[\bar\kappa(z,n)+\bar p']^{\dot\alpha\beta}
u_{\beta}(q) }
\over
{[q+q'+\kappa(y{+}1,z{-}1)]^2}
}
\cr & \qquad
-
[p+\kappa(1,y)]^2
{
{ u^{\alpha}(k_z)
[q+q'+\kappa(y{+}1,z{-}1)]_{\alpha\dot\gamma}
\bar k_z^{\dot\gamma\gamma}
u_{\gamma}(q)}
\over
{[q+q'+\kappa(y{+}1,z{-}1)]^2 [q+q'+\kappa(y{+}1,z)]^2}
}
\Biggr]
\cr & +
\sum_{d=1}^{n-1} \sum_{y=1}^d \sum_{z=d+1}^n
\negthinspace
{
{ \braket{q'}{y}}
\over
{ \braket{q}{y}  \braket{q}{q'}}
}
\cr & \quad\times
\Biggl[
{
{ u^{\alpha}(q)
[p+\kappa(1,y)]_{\alpha\dot\alpha}
[\bar\kappa(z{+}1,n)+\bar p']^{\dot\alpha\beta}
u_{\beta}(q) }
\over
{[q+q'+\kappa(y{+}1,z)]^2}
}
\cr & \qquad
-
{
{ u^{\alpha}(q)
[p+\kappa(1,y)]_{\alpha\dot\alpha}
[\bar\kappa(z,n)+\bar p']^{\dot\alpha\beta}
u_{\beta}(q) }
\over
{[q+q'+\kappa(y{+}1,z{-}1)]^2}
}
\cr & \qquad
-
[p+\kappa(1,y)]^2
{
{ u^{\alpha}(q)
[q+q'+\kappa(y{+}1,z{-}1)]_{\alpha\dot\gamma}
\bar k_z^{\dot\gamma\gamma}
u_{\gamma}(q)}
\over
{[q+q'+\kappa(y{+}1,z{-}1)]^2 [q+q'+\kappa(y{+}1,z)]^2}
}
\Biggr].
}
\eqlabel\XIfourBsplit
$$
The remaining two contributions to $\Xi_4$ are processed
in the same manner, but without the step corresponding
to (\FIERZZ).  Thus, we have
$$
\eqalign{
\Xi_{4{\rm{C}}} &=
\sum_{d=1}^{n-1} \sum_{y=1}^d \sum_{z=d+1}^n
\negthinspace
{
{ \braket{q'}{y} }
\over
{  \braket{y}{z}  \braket{q}{q'}}
}
\cr & \quad\times
\Biggl[
{
{ u^{\alpha}(q)
[p+\kappa(1,y{-}1)]_{\alpha\dot\alpha}
[\bar\kappa(z{+}1,n)+\bar p']^{\dot\alpha\beta}
u_{\beta}(k_z) }
\over
{[q+q'+\kappa(y,z)]^2}
}
\cr & \qquad
-
{
{ u^{\alpha}(q)
[p+\kappa(1,y)]_{\alpha\dot\alpha}
[\bar\kappa(z{+}1,n)+\bar p']^{\dot\alpha\beta}
u_{\beta}(k_z) }
\over
{[q+q'+\kappa(y{+}1,z)]^2}
}
\cr & \qquad
-
[\kappa(z{+}1,n)+p']^2
{
{ u^{\alpha}(q)
k_{y\alpha\dot\gamma}
[\bar q+\bar q'+ \bar\kappa(y{+}1,z)]^{\dot\gamma\gamma}
u_{\gamma}(k_z)}
\over
{[q+q'+\kappa(y{+}1,z)]^2 [q+q'+\kappa(y,z)]^2}
}
\Biggr],
}
\eqlabel\XIfourCsplit
$$
$$
\eqalign{
\Xi_{4{\rm{D}}} &=
\sum_{d=1}^{n-1} \sum_{y=1}^d \sum_{z=d+1}^n
\negthinspace
{
{ \braket{q'}{y} }
\over
{  \braket{y}{z}  \braket{q}{q'}}
}
\cr & \quad\times
\Biggl[
{
{ u^{\alpha}(q)
[p+\kappa(1,y)]_{\alpha\dot\alpha}
[\bar\kappa(z,n)+\bar p']^{\dot\alpha\beta}
u_{\beta}(k_z) }
\over
{[q+q'+\kappa(y{+}1,z{-}1)]^2}
}
\cr & \qquad
-
{
{ u^{\alpha}(q)
[p+\kappa(1,y{-}1)]_{\alpha\dot\alpha}
[\bar\kappa(z,n)+\bar p']^{\dot\alpha\beta}
u_{\beta}(k_z) }
\over
{[q+q'+\kappa(y,z{-}1)]^2}
}
\cr & \qquad
+
[\kappa(z,n)+p']^2
{
{ u^{\alpha}(q)
k_{y\alpha\dot\gamma}
[\bar q+\bar q'+ \bar\kappa(y{+}1,z{-}1)]^{\dot\gamma\gamma}
u_{\gamma}(k_z)}
\over
{[q+q'+\kappa(y{+}1,z{-}1)]^2 [q+q'+\kappa(y,z{-}1)]^2}
}
\Biggr].
}
\eqlabel\XIfourDsplit
$$

The contributions tabulated in equations
(\XIfourAsplit)--(\XIfourDsplit) consist of terms
containing either one  or two propagator factors.
Let us begin with the latter type
of terms.  Denote by $\Xi_{4a}$ the two double
propagator terms containing ${
{ \braket{q'}{y}}
\over
{ \braket{q}{y}  \braket{q}{q'}}
}$
as a prefactor.  That is,
$$
\eqalign{
\Xi_{4a} &\equiv
\sum_{d=1}^{n-1} \sum_{y=1}^d \sum_{z=d+1}^n
\negthinspace
{
{ \braket{q'}{y} }
\over
{ \braket{q}{y}   \braket{q}{q'}}
}
\cr & \quad\times
\Biggl[
[p+\kappa(1,y{-}1)]^2
{
{ u^{\alpha}(q)
[q+q'+\kappa(y,z{-}1)]_{\alpha\dot\gamma}
\bar k_z^{\dot\gamma\gamma}
u_{\gamma}(q)}
\over
{[q+q'+\kappa(y,z{-}1)]^2 [q+q'+\kappa(y,z)]^2}
}
\cr & \qquad
-[p+\kappa(1,y)]^2
{
{ u^{\alpha}(q)
[q+q'+\kappa(y{+}1,z{-}1)]_{\alpha\dot\gamma}
\bar k_z^{\dot\gamma\gamma}
u_{\gamma}(q)}
\over
{[q+q'+\kappa(y{+}1,z{-}1)]^2 [q+q'+\kappa(y{+}1,z)]^2}
}
\Biggr].
}
\eq
$$
Comparison with (\cdef) shows that this is really
$$
\eqalign{
\Xi_{4a} &=
\sum_{d=1}^{n-1} \sum_{y=1}^d \sum_{z=d+1}^n
\negthinspace
\link{y}{q}{q'}
\cr & \quad\times
\biggl[
[p+\kappa(1,y{-}1)]^2
\coef(q{+}q',y,y{+}1,\ldots,z)
\cr & \qquad
-[p+\kappa(1,y)]^2
\coef(q{+}q',y{+}1,y{+}2,\ldots,z)
\biggr].
}
\eqlabel\soclose
$$
If we were to shift the sum over $y$ in
the first term of (\soclose) by one,
both terms appearing in square brackets  would be the same.  Since
the factor outside the square brackets also depends on $y$, we obtain
instead
$$
\eqalign{
\Xi_{4a} &=
\sum_{d=1}^{n-1} \sum_{y=1}^{d-1} \sum_{z=d+1}^n
\negthinspace
\Biggl[
\link{y{+}1}{q}{q'}
-
\link{y}{q}{q'}
\Biggr]\negthinspace
[p+\kappa(1,y)]^2
\coef(q{+}q',y{+}1,y{+}2,\ldots,z)
\cr & \quad
+\sum_{d=1}^{n-1}  \sum_{z=d+1}^n
 \link{1}{q}{q'} \thinspace
p^2 \thinspace \coef(q{+}q',1,\ldots,z)
\cr & \quad
-\sum_{d=1}^{n-1}  \sum_{z=d+1}^n
 \link{d}{q}{q'} \thinspace
[p+\kappa(1,d)]^2 \thinspace \coef(q{+}q',d{+}1,\ldots,z).
}
\eqlabel\recomb
$$
The square brackets in the first contribution to (\recomb)
combine to form $-\link{y}{q}{y{+}1}$ using (\linkidnosum).
The second term vanishes, since $p^2=0$.  The final term
is precisely what is required to extend the first term
to include $y=d$, the proper interpretation of $k_{d+1}$
in this context being $q'$.  Thus,
$$
\eqalign{
\Xi_{4a} &=
-\sum_{d=1}^{n-1} \sum_{y=1}^{d} \sum_{z=d+1}^n
\negthinspace
\link{y}{q}{y{+}1}
[p+\kappa(1,y)]^2
\coef(q{+}q',y{+}1,y{+}2,\ldots,z).
}
\eqlabel\XIfoura
$$
Note that we may extend the sums to include  $d=y=0$ since
such a term is  proportional to $p^2=0$.
Thus, the contribution to the amplitude from (\XIfoura)
may be written
$$
\eqalign{
\amp_{4a} &=
{{i}\over{2}}
(-g\sqrt2)^{n+2} \negthinspace\negthinspace
\permsum{1}{n}
\sum_{d=0}^{n-1} \sum_{y=0}^d \sum_{z=d+1}^n
\negthinspace
{(\Omega[1,d])^i}_{\ell}
{(\Omega[d{+}1,n])_{k}}^{j}
\cr & \qquad\times
{
{ {\braket{p}{q}}^2 \thinspace
[p+\kappa(1,y)]^2 \thinspace
\coef(q{+}q',y{+}1,\ldots,z)}
\over
{ \bra{p}1,\ldots,y\ket{q}
\thinspace \bra{q}y{+}1,\ldots,d\ket{q'}
\thinspace \bra{q}d{+}1,\ldots,n\ket{p'}}
}.
}
\eqlabel\Mfoura
$$
The contribution represented by (\Mfoura) identically cancels
the last term appearing in $\amp_2$ (see equation (\Mtwodone)).

Next, we turn to the double propagator contributions in
(\XIfourAsplit)--(\XIfourBsplit) which have
${
{ \braket{q'}{y}}
\over
{ \braket{y}{z}\braket{q}{q'}  }
}$ as a prefactor.  These two terms are
$$
\eqalign{
\Xi_{4b} & \equiv
\sum_{d=1}^{n-1} \sum_{y=1}^d \sum_{z=d+1}^n
\negthinspace
{
{ \braket{q'}{y} }
\over
{  \braket{y}{z}  \braket{q}{q'}}
}
\cr & \quad\times
\Biggl[
[p+\kappa(1,y{-}1)]^2
{
{ u^{\alpha}(k_z)
[q+q'+\kappa(y,z{-}1)]_{\alpha\dot\gamma}
\bar k_z^{\dot\gamma\gamma}
u_{\gamma}(q)}
\over
{[q+q'+\kappa(y,z{-}1)]^2 [q+q'+\kappa(y,z)]^2}
}
\cr & \qquad
-[p+\kappa(1,y)]^2
{
{ u^{\alpha}(k_z)
[q+q'+\kappa(y{+}1,z{-}1)]_{\alpha\dot\gamma}
\bar k_z^{\dot\gamma\gamma}
u_{\gamma}(q)}
\over
{[q+q'+\kappa(y{+}1,z{-}1)]^2 [q+q'+\kappa(y{+}1,z)]^2}
}
\Biggr].
}
\eqlabel\XIfourbdef
$$
Because of the relations (\spinnorm) and
(\msreplacedotprod a), we may write
$$
\eqalign{
u^{\alpha}(k_z)&
[q+q'+\kappa(y,z{-}1)]_{\alpha\dot\gamma}
\bar k_z^{\dot\gamma\gamma}
u_{\gamma}(q)  =
\cr & =
2k_z \cdot [q+q'+\kappa(y,z{-}1)] \thinspace \braket{z}{q}
\cr & =
\biggl\{
[q+q'+\kappa(y,z)]^2 -[q+q'+\kappa(y,z{-}1)]^2
\biggr\}
\braket{z}{q}.
}
\eq
$$
This produces
$$
\eqalign{
\Xi_{4b} & =
\sum_{d=1}^{n-1} \sum_{y=1}^d \sum_{z=d+1}^n
\negthinspace
{
{ \braket{q'}{y}  \braket{z}{q}}
\over
{  \braket{y}{z}  \braket{q}{q'}}
}
\cr & \quad\times
\Biggl[
-{
{ [p+\kappa(1,y{-}1)]^2 }
\over
{ [q+q'+\kappa(y,z)]^2 }
}
+{
{ [p+\kappa(1,y{-}1)]^2 }
\over
{ [q+q'+\kappa(y,z{-}1)]^2 }
}
\cr & \qquad
+{
{ [p+\kappa(1,y)]^2 }
\over
{ [q+q'+\kappa(y{+}1,z)]^2 }
}
-{
{ [p+\kappa(1,y)]^2 }
\over
{ [q+q'+\kappa(y{+}1,z{-}1)]^2 }
}
\Biggr].
}
\eqlabel\XIfourb
$$
We set (\XIfourb) aside for later cancellation.

There are two double propagator terms not yet accounted for.
These appear in equations (\XIfourCsplit) and (\XIfourDsplit).
They are
$$
\eqalign{
\Xi_{4cd} &\equiv
\sum_{d=1}^{n-1} \sum_{y=1}^d \sum_{z=d+1}^n
\negthinspace
{
{ \braket{q'}{y} }
\over
{  \braket{y}{z}  \braket{q}{q'}}
}
\cr & \quad\times
\Biggl[
[\kappa(z,n)+p']^2
{
{\braket{q}{y}
\bar u_{\dot\gamma}(k_y)
[\bar q+\bar q'+ \bar\kappa(y{+}1,z{-}1)]^{\dot\gamma\gamma}
u_{\gamma}(k_z)}
\over
{[q+q'+\kappa(y{+}1,z{-}1)]^2 [q+q'+\kappa(y,z{-}1)]^2}
}
\cr & \qquad -
[\kappa(z{+}1,n)+p']^2
{
{\braket{q}{y}
\bar u_{\dot\gamma}(k_y)
[\bar q+\bar q'+ \bar\kappa(y{+}1,z)]^{\dot\gamma\gamma}
u_{\gamma}(k_z)}
\over
{[q+q'+\kappa(y{+}1,z)]^2 [q+q'+\kappa(y,z)]^2}
}
\Biggr].
}
\eqlabel\XIfourcddef
$$
We break (\XIfourcddef) into two pieces by applying (\fierz), written
as
$$
u_{\gamma}(k_z) \braket{q}{y} =
-u_{\gamma}(k_y) \braket{z}{q}
-u_{\gamma}(q) \braket{y}{z}.
\eqlabel\onelastfierz
$$
Call the piece of (\XIfourcddef) corresponding to the first
term of (\onelastfierz) $\Xi_{4c}$ and the remainder
$\Xi_{4d}$.

For $\Xi_{4c}$ we have
$$
\eqalign{
\Xi_{4c} &=
\sum_{d=1}^{n-1} \sum_{y=1}^d \sum_{z=d+1}^n
\negthinspace
{
{ \braket{q'}{y} }
\over
{  \braket{y}{z}  \braket{q}{q'}}
}
\cr & \quad\times
\Biggl[
-[\kappa(z,n)+p']^2
{
{\braket{z}{q}
\bar u_{\dot\gamma}(k_y)
[\bar q+\bar q'+ \bar\kappa(y{+}1,z{-}1)]^{\dot\gamma\gamma}
u_{\gamma}(k_y)}
\over
{[q+q'+\kappa(y{+}1,z{-}1)]^2 [q+q'+\kappa(y,z{-}1)]^2}
}
\cr & \qquad +
[\kappa(z{+}1,n)+p']^2
{
{\braket{z}{q}
\bar u_{\dot\gamma}(k_y)
[\bar q+\bar q'+ \bar\kappa(y{+}1,z)]^{\dot\gamma\gamma}
u_{\gamma}(k_y)}
\over
{[q+q'+\kappa(y{+}1,z)]^2 [q+q'+\kappa(y,z)]^2}
}
\Biggr].
}
\eqlabel\XIfourcdef
$$
But, we may write
$$
\eqalign{
\bar u_{\dot\gamma}(k_y)&
[\bar q+\bar q'+ \bar\kappa(y{+}1,z{-}1)]^{\dot\gamma\gamma}
u_{\gamma}(k_y) =
\cr & =
2k_y\cdot[q+q'+\kappa(y{+}1,z{-}1)]
\cr & =
[q+q'+\kappa(y,z{-}1)]^2 - [q+q'+\kappa(y{+}1,z{-}1)]^2,
}
\eq
$$
giving us
$$
\eqalign{
\Xi_{4c} &=
\sum_{d=1}^{n-1} \sum_{y=1}^d \sum_{z=d+1}^n
\negthinspace
{
{ \braket{q'}{y} \braket{z}{q}}
\over
{  \braket{y}{z}  \braket{q}{q'}}
}
\cr & \quad\times
\Biggl[
-{
{ [\kappa(z,n)+p']^2}
\over
{ [q+q'+\kappa(y{+}1,z{-}1)]^2 }
}
+{
{ [\kappa(z,n)+p']^2}
\over
{ [q+q'+\kappa(y,z{-}1)]^2 }
}
\cr & \qquad\quad
+{
{ [\kappa(z{+}1,n)+p']^2}
\over
{ [q+q'+\kappa(y{+}1,z)]^2 }
}
-{
{ [\kappa(z{+}1,n)+p']^2}
\over
{ [q+q'+\kappa(y,z)]^2 }
}
\Biggr].
}
\eqlabel\XIfourc
$$
This is another contribution which will be cancelled later.

The other piece of (\XIfourcddef) is
$$
\eqalign{
\Xi_{4d} &\equiv
\sum_{d=1}^{n-1} \sum_{y=1}^d \sum_{z=d+1}^n
\negthinspace
{
{ \braket{q'}{y} }
\over
{   \braket{q}{q'}}
}
\cr & \quad\times
\Biggl[
-[\kappa(z,n)+p']^2
{
{\bar u_{\dot\gamma}(k_y)
[\bar q+\bar q'+ \bar\kappa(y{+}1,z{-}1)]^{\dot\gamma\gamma}
u_{\gamma}(q)}
\over
{[q+q'+\kappa(y{+}1,z{-}1)]^2 [q+q'+\kappa(y,z{-}1)]^2}
}
\cr & \qquad\quad +
[\kappa(z{+}1,n)+p']^2
{
{\bar u_{\dot\gamma}(k_y)
[\bar q+\bar q'+ \bar\kappa(y{+}1,z)]^{\dot\gamma\gamma}
u_{\gamma}(q)}
\over
{[q+q'+\kappa(y{+}1,z)]^2 [q+q'+\kappa(y,z)]^2}
}
\Biggr].
}
\eqlabel\XIfourddef
$$
Since the two terms in the square brackets differ by 1 unit in
$z$, and there is no other $z$-dependence, we may do the
sum on $z$, the result being only the endpoints:
$$
\eqalign{
\Xi_{4d} &=
\sum_{d=1}^{n-1} \sum_{y=1}^d
\negthinspace
{
{ \braket{q'}{y} }
\over
{   \braket{q}{q'}}
}
\Biggl[
-[\kappa(d{+}1,n)+p']^2
{
{\bar u_{\dot\gamma}(k_y)
[\bar q+\bar q'+ \bar\kappa(y{+}1,d)]^{\dot\gamma\gamma}
u_{\gamma}(q)}
\over
{[q+q'+\kappa(y{+}1,d)]^2 [q+q'+\kappa(y,d)]^2}
}
\cr & \qquad\quad +
p'^2
{
{\bar u_{\dot\gamma}(k_y)
[\bar q+\bar q'+ \bar\kappa(y{+}1,n)]^{\dot\gamma\gamma}
u_{\gamma}(q)}
\over
{[q+q'+\kappa(y{+}1,n)]^2 [q+q'+\kappa(y,n)]^2}
}
\Biggr].
}
\eqlabel\XIfourdalmostdone
$$
Since $p'^2=0$, the second contribution to (\XIfourdalmostdone)
vanishes.  In addition, note that we may add $d=n$ to the
sum appearing in the first contribution with no penalty since
such a term is also proportional to $p'^2$.  Thus
$$
\eqalign{
\Xi_{4d} &=
\sum_{d=1}^{n} \sum_{y=1}^d
\negthinspace
{
{ \braket{q'}{y} }
\over
{   \braket{q}{q'}\braket{q}{y}}
}
[\kappa(d{+}1,n)+p']^2
\coef(q{+}q',d,d{-}1,\ldots,y),
}
\eqlabel\XIfourd
$$
where we have multiplied by $\braket{q}{y}\over\braket{q}{y}$ and
used (\cdef) to identify the factor of $c$.  The contribution
to $\amp_4$ that corresponds to (\XIfourd) is
$$
\eqalign{
\amp_{4d} &=
{{-i}\over{2}}
(-g\sqrt2)^{n+2} \negthinspace\negthinspace
\permsum{1}{n}
\sum_{d=1}^{n} \sum_{y=1}^d
\negthinspace
{(\Omega[1,d])^i}_{\ell}
{(\Omega[d{+}1,n])_{k}}^{j}
\cr & \qquad\times
{
{ \braket{q'}{y} }
\over
{   \braket{q}{q'}\braket{q}{y}}
}
{
{ {\braket{p}{q}}^2 \thinspace
[\kappa(d{+}1,n)+p']^2
\coef(q{+}q',d,d{-}1,\ldots,y)}
\over
{ \bra{p}1,\ldots,d\ket{q'}
\thinspace \bra{q}d{+}1,\ldots,n\ket{p'}}
},
}
\eqlabel\Mfourd
$$
which, exactly cancels the contribution from $\amp_{3{\rm A}}$
given in equation (\MthreeA)  (recall that the spinor
inner product is antisymmetric, $\braket{y}{q'}=-\braket{q'}{y}$).

Having finished with the terms containing two propagators,
we turn to the remaining contributions
with just one propagator.  Let us
begin with the terms proportional to  ${
{ \braket{q'}{y} }
\over
{ \braket{q}{y}  \braket{q}{q'}}
}$.  These contributions come from (\XIfourAsplit) and (\XIfourBsplit)
and comprise the following four terms:
$$
\eqalign{
\Xi_{4e} &\equiv
\sum_{d=1}^{n-1} \sum_{y=1}^d \sum_{z=d+1}^n
\negthinspace
{
{ \braket{q'}{y} }
\over
{ \braket{q}{y}   \braket{q}{q'}}
}
\cr & \quad\times
\Biggl[
{
{ u^{\alpha}(q)
[p+\kappa(1,y{-}1)]_{\alpha\dot\alpha}
[\bar\kappa(z,n)+\bar p']^{\dot\alpha\beta}
u_{\beta}(q) }
\over
{[q+q'+\kappa(y,z{-}1)]^2}
}
\cr & \qquad\quad
-
{
{ u^{\alpha}(q)
[p+\kappa(1,y{-}1)]_{\alpha\dot\alpha}
[\bar\kappa(z{+}1,n)+\bar p']^{\dot\alpha\beta}
u_{\beta}(q) }
\over
{[q+q'+\kappa(y,z)]^2}
}
\cr & \qquad\quad +
{
{ u^{\alpha}(q)
[p+\kappa(1,y)]_{\alpha\dot\alpha}
[\bar\kappa(z{+}1,n)+\bar p']^{\dot\alpha\beta}
u_{\beta}(q) }
\over
{[q+q'+\kappa(y{+}1,z)]^2}
}
\cr & \qquad\quad
-
{
{ u^{\alpha}(q)
[p+\kappa(1,y)]_{\alpha\dot\alpha}
[\bar\kappa(z,n)+\bar p']^{\dot\alpha\beta}
u_{\beta}(q) }
\over
{[q+q'+\kappa(y{+}1,z{-}1)]^2}
}
\Biggr].
}
\eqlabel\XIfouredef
$$
The sum on $z$ may be performed in (\XIfouredef) with the contributions
from the interior points in the summation cancelling in pairs, leaving
just
$$
\eqalign{
\Xi_{4e} &=
\sum_{d=1}^{n} \sum_{y=1}^d
\negthinspace
{
{ \braket{q'}{y} }
\over
{ \braket{q}{y}   \braket{q}{q'}}
}
\cr & \quad\times
\Biggl[
{
{ u^{\alpha}(q)
[p+\kappa(1,y{-}1)]_{\alpha\dot\alpha}
[\bar\kappa(d{+}1,n)+\bar p']^{\dot\alpha\beta}
u_{\beta}(q) }
\over
{[q+q'+\kappa(y,d)]^2}
}
\cr & \qquad\quad
-
{
{ u^{\alpha}(q)
[p+\kappa(1,y{-}1)]_{\alpha\dot\alpha}
\bar p'^{\dot\alpha\beta}
u_{\beta}(q) }
\over
{[q+q'+\kappa(y,n)]^2}
}
\cr & \qquad\quad +
{
{ u^{\alpha}(q)
[p+\kappa(1,y)]_{\alpha\dot\alpha}
\bar p'^{\dot\alpha\beta}
u_{\beta}(q) }
\over
{[q+q'+\kappa(y{+}1,n)]^2}
}
\cr & \qquad\quad
-
{
{ u^{\alpha}(q)
[p+\kappa(1,y)]_{\alpha\dot\alpha}
[\bar\kappa(d{+}1,n)+\bar p']^{\dot\alpha\beta}
u_{\beta}(q) }
\over
{[q+q'+\kappa(y{+}1,d)]^2}
}
\Biggr].
}
\eqlabel\XIfouresummed
$$
We have added the term $d=n$ to (\XIfouresummed) since
the quantity in brackets vanishes for that value of $d$.

At this point, we see that there are pairs of terms in the
square brackets that differ by one unit in $y$.  However,
the prefactor is not independent of $y$.  Following
the same procedure used in the discussion of $\Xi_{3{\rm B}}$
({\it cf.}{ }equations
(\specialsplitstart)--(\XIthreeBdone)), we obtain
$$
\eqalign{
\Xi_{4e} &=
\sum_{d=1}^{n} \sum_{y=1}^d
\link{y}{q}{y{+}1}
{
{ u^{\alpha}(q)
[p+\kappa(1,y)]_{\alpha\dot\alpha}
\bar p'^{\dot\alpha\beta}
u_{\beta}(q) }
\over
{[p+p'+\kappa(1,y)]^2}
}
\cr & \quad
-\sum_{d=1}^{n}
\link{1}{q}{q'}
{
{ u^{\alpha}(q)
p_{\alpha\dot\alpha}
\bar p'^{\dot\alpha\beta}
u_{\beta}(q) }
\over
{[p+p']^2}
}
\cr & \quad
-\sum_{d=1}^{n} \sum_{y=1}^d
\link{y}{q}{y{+}1}
{
{ u^{\alpha}(q)
[p+\kappa(1,y)]_{\alpha\dot\alpha}
[\bar\kappa(d{+}1,n)+\bar p']^{\dot\alpha\beta}
u_{\beta}(q) }
\over
{[q+q'+\kappa(y{+}1,d)]^2}
}
\cr & \quad
+\sum_{d=1}^{n}
\link{1}{q}{q'}
{
{ u^{\alpha}(q)
p_{\alpha\dot\alpha}
[\bar\kappa(d{+}1,n)+\bar p']^{\dot\alpha\beta}
u_{\beta}(q) }
\over
{[q+q'+\kappa(1,d)]^2}
}.
}
\eqlabel\XIfoure
$$
The contribution to $\amp_4$ from (\XIfoure) is
$$
\eqalign{
\amp_{4e} &=
{{-i}\over{2}}
(-g\sqrt2)^{n+2} \negthinspace\negthinspace
\permsum{1}{n}
\sum_{d=1}^{n} \sum_{y=1}^d
\negthinspace
{(\Omega[1,d])^i}_{\ell}
{(\Omega[d{+}1,n])_{k}}^{j}
\cr & \qquad\qquad\times
{
{ {\braket{p}{q}}^2 \thinspace
u^{\alpha}(q)
[p+\kappa(1,y)]_{\alpha\dot\alpha}
\bar p'^{\dot\alpha\beta}
u_{\beta}(q)}
\over
{ \bra{p}1,\ldots,y\ket{q}
\thinspace \bra{q} y{+}1,\ldots, d\ket{q'}
\thinspace \bra{q}d{+}1,\ldots,n\ket{p'}}
}
\cr & \qquad\qquad\times
{ {1}\over{ [p+p'+\kappa(1,y)]^2 } }
\cr & \quad
+{{i}\over{2}}
(-g\sqrt2)^{n+2} \negthinspace\negthinspace
\permsum{1}{n}
\sum_{d=1}^{n}
\negthinspace
{(\Omega[1,d])^i}_{\ell}
{(\Omega[d{+}1,n])_{k}}^{j}
\cr & \qquad\qquad\times
\link{1}{q}{q'}
{
{ {\braket{p}{q}}^2 \thinspace
u^{\alpha}(q)
p_{\alpha\dot\alpha}
\bar p'^{\dot\alpha\beta}
u_{\beta}(q)}
\over
{ [p+p']^2 \thinspace \bra{p}1,\ldots, d\ket{q'}
\thinspace \bra{q}d{+}1,\ldots,n\ket{p'}}
}
\cr & \quad
+{{i}\over{2}}
(-g\sqrt2)^{n+2} \negthinspace\negthinspace
\permsum{1}{n}
\sum_{d=1}^{n} \sum_{y=1}^d
\negthinspace
{(\Omega[1,d])^i}_{\ell}
{(\Omega[d{+}1,n])_{k}}^{j}
\cr &\qquad \qquad\times
{
{ {\braket{p}{q}}^2 \thinspace
u^{\alpha}(q)
[p+\kappa(1,y)]_{\alpha\dot\alpha}
[\bar\kappa(d{+}1,n) + \bar p']^{\dot\alpha\beta}
u_{\beta}(q)}
\over
{ \bra{p}1,\ldots,y\ket{q}
\thinspace \bra{q} y{+}1,\ldots, d\ket{q'}
\thinspace \bra{q}d{+}1,\ldots,n\ket{p'}}
}
\cr & \qquad\qquad\times
{ {1}\over{ [q+q'+\kappa(y{+}1,d)]^2 } }
\cr & \quad
-{{i}\over{2}}
(-g\sqrt2)^{n+2} \negthinspace\negthinspace
\permsum{1}{n}
\sum_{d=1}^{n}
\negthinspace
{(\Omega[1,d])^i}_{\ell}
{(\Omega[d{+}1,n])_{k}}^{j}
\cr & \qquad\qquad\times
\link{1}{q}{q'}
{
{ {\braket{p}{q}}^2 \thinspace
u^{\alpha}(q)
p_{\alpha\dot\alpha}
[\bar\kappa(d{+}1,n) + \bar p']^{\dot\alpha\beta}
u_{\beta}(q)}
\over
{  \bra{p}1,\ldots, d\ket{q'}
\thinspace \bra{q}d{+}1,\ldots,n\ket{p'}}
}
\cr &\qquad \qquad\times
{ {1}\over{[q+q'+\kappa(1,d)]^2 } }.
}
\eqlabel\Mfoure
$$
The third and fourth terms of (\Mfoure) cancel $\amp_{3{\rm B}}$
in its entirety (see equation (\MthreeB)).  The first two
terms of (\Mfoure) have no compensating contributions.

We collect the remaining eight single propagator contributions
from equations (\XIfourAsplit)--(\XIfourDsplit) and form
$$
\eqalign{
\Xi_{4f} &=
\sum_{d=1}^{n-1} \sum_{y=1}^d \sum_{z=d+1}^n
\negthinspace
{
{ \braket{q'}{y} }
\over
{  \braket{y}{z}  \braket{q}{q'}}
}
\Biggl[
{
{ u^{\alpha}(k_z)
[p+\kappa(1,y{-}1)]_{\alpha\dot\alpha}
[\bar\kappa(z,n)+\bar p']^{\dot\alpha\beta}
u_{\beta}(q) }
\over
{[q+q'+\kappa(y,z{-}1)]^2}
}
\cr & \qquad\qquad
-
{
{ u^{\alpha}(k_z)
[p+\kappa(1,y{-}1)]_{\alpha\dot\alpha}
[\bar\kappa(z{+}1,n)+\bar p']^{\dot\alpha\beta}
u_{\beta}(q) }
\over
{[q+q'+\kappa(y,z)]^2}
}
\cr & \qquad\qquad
+
{
{ u^{\alpha}(k_z)
[p+\kappa(1,y)]_{\alpha\dot\alpha}
[\bar\kappa(z{+}1,n)+\bar p']^{\dot\alpha\beta}
u_{\beta}(q) }
\over
{[q+q'+\kappa(y{+}1,z)]^2}
}
\cr & \qquad\qquad
-
{
{ u^{\alpha}(k_z)
[p+\kappa(1,y)]_{\alpha\dot\alpha}
[\bar\kappa(z,n)+\bar p']^{\dot\alpha\beta}
u_{\beta}(q) }
\over
{[q+q'+\kappa(y{+}1,z{-}1)]^2}
}
\cr & \qquad\qquad
+
{
{ u^{\alpha}(q)
[p+\kappa(1,y)]_{\alpha\dot\alpha}
[\bar\kappa(z,n)+\bar p']^{\dot\alpha\beta}
u_{\beta}(k_z) }
\over
{[q+q'+\kappa(y{+}1,z{-}1)]^2}
}
\cr & \qquad\qquad
-
{
{ u^{\alpha}(q)
[p+\kappa(1,y{-}1)]_{\alpha\dot\alpha}
[\bar\kappa(z,n)+\bar p']^{\dot\alpha\beta}
u_{\beta}(k_z) }
\over
{[q+q'+\kappa(y,z{-}1)]^2}
}
\cr & \qquad\qquad
+
{
{ u^{\alpha}(q)
[p+\kappa(1,y{-}1)]_{\alpha\dot\alpha}
[\bar\kappa(z{+}1,n)+\bar p']^{\dot\alpha\beta}
u_{\beta}(k_z) }
\over
{[q+q'+\kappa(y,z)]^2}
}
\cr & \qquad\qquad
-
{
{ u^{\alpha}(q)
[p+\kappa(1,y)]_{\alpha\dot\alpha}
[\bar\kappa(z{+}1,n)+\bar p']^{\dot\alpha\beta}
u_{\beta}(k_z) }
\over
{[q+q'+\kappa(y{+}1,z)]^2}
}
\Biggr].
}
\eqlabel\XIfourfdef
$$
The terms in (\XIfourfdef) occur in pairs with common denominators.
Note that in every pair, we have the form
$$
\eqalign{
u^{\alpha}(k_z) X_{\alpha\dot\alpha}
\bar Y^{\dot\alpha\beta}& u_{\beta}(q)
-u^{\alpha}(q) X_{\alpha\dot\alpha}
\bar Y^{\dot\alpha\beta} u_{\beta}(k_z)=
\cr & =
u^{\alpha}(k_z) \biggl[ X_{\alpha\dot\alpha}
\bar Y^{\dot\alpha\beta} + Y_{\alpha\dot\alpha}
\bar X^{\dot\alpha\beta} \biggr] u_{\beta}(q)
\cr & =
2X\cdot Y \thinspace \braket{z}{q},
}
\eqlabel\WNGIEwge
$$
where  $X$ and $Y$ are momenta and we have applied  (\anticommutator).
If we use (\WNGIEwge) on each of the four pairs in (\XIfourfdef),
we obtain
$$
\eqalign{
\Xi_{4f} &=
\sum_{d=1}^{n-1} \sum_{y=1}^d  \sum_{z=d+1}^n
\negthinspace
{
{ \braket{q'}{y} } \braket{z}{q}
\over
{  \braket{y}{z}  \braket{q}{q'}}
}
\Biggl[
{
{ 2[p+\kappa(1,y{-}1)]\cdot[\kappa(z,n)+p'] }
\over
{[q+q'+\kappa(y,z{-}1)]^2}
}
\cr & \qquad
-
{
{ 2[p+\kappa(1,y{-}1)]\cdot[\kappa(z{+}1,n)+p'] }
\over
{[q+q'+\kappa(y,z)]^2}
}
+
{
{ 2[p+\kappa(1,y)]\cdot[\kappa(z{+}1,n)+p'] }
\over
{[q+q'+\kappa(y{+}1,z)]^2}
}
\cr & \qquad
-
{
{ 2[p+\kappa(1,y)]\cdot[\kappa(z,n)+p'] }
\over
{[q+q'+\kappa(y{+}1,z{-}1)]^2}
}
\Biggr].
}
\eqlabel\XIfourfnearlydone
$$
Each of the four terms in (\XIfourfnearlydone) may be treated
using momentum conservation.  For example, momentum conservation
tells us that
$$
\eqalign{
[q+q'+\kappa(y,z{-}1)]^2 & =
[p+\kappa(1,y{-}1)+\kappa(z,n)+p']^2
\cr & =
[p+\kappa(1,y{-}1)]^2 + [\kappa(z,n)+p']^2
\cr & \quad
+ 2[p+\kappa(1,y{-}1)]\cdot[\kappa(z,n)+p'].
}
\eqlabel\thelasttransformation
$$
When we apply (\thelasttransformation) to (\XIfourfnearlydone),
the result is
$$
\eqalign{
\Xi_{4f} =
\sum_{d=1}^{n-1} \sum_{y=1}^d & \sum_{z=d+1}^n
\negthinspace
{
{ \braket{q'}{y} } \braket{z}{q}
\over
{  \braket{y}{z}  \braket{q}{q'}}
}
\cr  \quad\times
\Biggl[ \quad &1
-{
{ [p+\kappa(1,y{-}1)]^2 }
\over
{[q+q'+\kappa(y,z{-}1)]^2}
}
-{
{ [\kappa(z,n)+p']^2 }
\over
{[q+q'+\kappa(y,z{-}1)]^2}
}
\cr  \qquad
- &1
+
{
{ [p+\kappa(1,y{-}1)]^2 }
\over
{[q+q'+\kappa(y,z)]^2}
}
+
{
{ [\kappa(z{+}1,n)+p']^2 }
\over
{[q+q'+\kappa(y,z)]^2}
}
\cr  \qquad
+&1
-
{
{ [p+\kappa(1,y)]^2 }
\over
{[q+q'+\kappa(y{+}1,z)]^2}
}
-
{
{ [\kappa(z{+}1,n)+p']^2 }
\over
{[q+q'+\kappa(y{+}1,z)]^2}
}
\cr  \qquad
-&1
+
{
{ [p+\kappa(1,y)]^2 }
\over
{[q+q'+\kappa(y{+}1,z{-}1)]^2}
}
+
{
{ [\kappa(z,n)+p']^2 }
\over
{[q+q'+\kappa(y{+}1,z{-}1)]^2}
}
\Biggr].
}
\eqlabel\XIfourf
$$
The 1's cancel among themselves. The remaining terms dispose
of the contributions from $\Xi_{4b}$ and $\Xi_{4c}$ (equations
(\XIfourb) and (\XIfourc) respectively).
\section{Final considerations}

We now collect the surviving terms and combine them into the
amplitude for the process (\monsteramplitude).  These terms are:
the second contribution to $\amp_2$ (equation (\Mtwodone)),
the first two terms of $\amp_{4e}$ (equation (\Mfoure)), and,
of course, the $1\over N$ contribution $\amp_5$ (equation
(\Mfive)).  Let us start by simplifying the remaining non-$1\over N$
pieces, which read
$$
\eqalign{
\amp_T & \equiv
{{i}\over{2}}
(-g\sqrt2)^{n+2} \negthinspace\negthinspace
\permsum{1}{n}
\sum_{d=0}^{n}
\sum_{b=0}^d
{(\Omega[1,d])^{i}}_{\ell}
{(\Omega[d{+}1,n])_{k}}^{j}
\cr & \qquad\quad\times
{
{ {\braket{p}{q}}^2
u^{\alpha}(q)
[p+\kappa(1,b)]_{\alpha\dot\beta}
\bar p'^{\dot\beta\gamma}
u_{\gamma}(q)}
\over
{ \bra{p}1,\ldots,b\ket{q}
\thinspace\bra{q} b{+}1,\ldots,d \ket{q'}
\thinspace\bra{q}  d{+}1,\ldots,n\ket{p'} }
}
\cr & \qquad\quad\times
{
{ 1 }
\over
{[p+p'+\kappa(1,b)]^2 }
}
\cr & \quad
-{{i}\over{2}}
(-g\sqrt2)^{n+2} \negthinspace\negthinspace
\permsum{1}{n}
\sum_{d=1}^{n} \sum_{y=1}^d
\negthinspace
{(\Omega[1,d])^i}_{\ell}
{(\Omega[d{+}1,n])_{k}}^{j}
\cr & \qquad\times
{
{ {\braket{p}{q}}^2 \thinspace
u^{\alpha}(q)
[p+\kappa(1,y)]_{\alpha\dot\alpha}
\bar p'^{\dot\alpha\beta}
u_{\beta}(q)}
\over
{ \bra{p}1,\ldots,y\ket{q}
\thinspace \bra{q} y{+}1,\ldots, d\ket{q'}
\thinspace \bra{q}d{+}1,\ldots,n\ket{p'}}
}
\cr & \qquad\times
{ {1}\over{ [p+p'+\kappa(1,y)]^2 } }
\cr & \quad
+{{i}\over{2}}
(-g\sqrt2)^{n+2} \negthinspace\negthinspace
\permsum{1}{n}
\sum_{d=1}^{n}
\negthinspace
{(\Omega[1,d])^i}_{\ell}
{(\Omega[d{+}1,n])_{k}}^{j}
\cr & \qquad\times
\link{1}{q}{q'}
{
{ {\braket{p}{q}}^2 \thinspace
u^{\alpha}(q)
p_{\alpha\dot\alpha}
\bar p'^{\dot\alpha\beta}
u_{\beta}(q)}
\over
{ [p+p']^2 \thinspace \bra{p}1,\ldots, d\ket{q'}
\thinspace \bra{q}d{+}1,\ldots,n\ket{p'}}
}.
}
\eqlabel\homestretch
$$
The first two contributions to (\homestretch) nearly cancel:  the
remainder reads
$$
\eqalign{
\amp_T &=
{{i}\over{2}}
(-g\sqrt2)^{n+2} \negthinspace\negthinspace
\permsum{1}{n}
\sum_{d=0}^{n}
\negthinspace
{(\Omega[1,d])^i}_{\ell}
{(\Omega[d{+}1,n])_{k}}^{j}
\cr & \qquad\times
{
{ {\braket{p}{q}}^2 \thinspace
u^{\alpha}(q)
p_{\alpha\dot\alpha}
\bar p'^{\dot\alpha\beta}
u_{\beta}(q)}
\over
{ \braket{p}{q}
\thinspace \bra{q} 1,\ldots, d\ket{q'}
\thinspace \bra{q}d{+}1,\ldots,n\ket{p'}}
}
{ {1}\over{ [p+p']^2 } }
\cr & \quad
+{{i}\over{2}}
(-g\sqrt2)^{n+2} \negthinspace\negthinspace
\permsum{1}{n}
\sum_{d=1}^{n}
\negthinspace
{(\Omega[1,d])^i}_{\ell}
{(\Omega[d{+}1,n])_{k}}^{j}
\cr & \qquad\times
\link{1}{q}{q'}
{
{ {\braket{p}{q}}^2 \thinspace
u^{\alpha}(q)
p_{\alpha\dot\alpha}
\bar p'^{\dot\alpha\beta}
u_{\beta}(q)}
\over
{  \bra{p}1,\ldots, d\ket{q'}
\thinspace \bra{q}d{+}1,\ldots,n\ket{p'}}
}
{ {1}\over{ [p+p']^2 } }.
}
\eqlabel\isitreallyalmostdone
$$
The two terms in (\isitreallyalmostdone) have very similar structure,
although the sum in the first contribution contains an extra term.
Taking this into account and pulling out common factors gives
$$
\eqalign{
\amp_T &=
{{i}\over{2}}
(-g\sqrt2)^{n+2} \negthinspace\negthinspace
\permsum{1}{n}
{\delta^i}_{\ell}
{(\Omega[1,n])_{k}}^{j}
\cr & \qquad\times
{
{ {\braket{p}{q}}^2 \braket{q}{p} {\braket{p'}{p}}^{*} \braket{p'}{q} }
\over
{ \braket{p}{q}
\thinspace \braket{q}{q'}
\thinspace \bra{q}1,\ldots,n\ket{p'}}
}
{
{1}
\over
{ \braket{p'}{p}  {\braket{p'}{p}}^{*} }
}
\cr & \quad
+{{i}\over{2}}
(-g\sqrt2)^{n+2} \negthinspace\negthinspace
\permsum{1}{n}
\sum_{d=1}^{n}
\negthinspace
{(\Omega[1,d])^i}_{\ell}
{(\Omega[d{+}1,n])_{k}}^{j}
\cr & \qquad\quad\times
{
{ {\braket{p}{q}}^2 \braket{q}{p} {\braket{p'}{p}}^{*} \braket{p'}{q} }
\over
{  \bra{p}1,\ldots, d\ket{q'}
\thinspace \bra{q}d{+}1,\ldots,n\ket{p'}}
}
{
{1}
\over
{ \braket{p'}{p}  {\braket{p'}{p}}^{*} }
}
\cr & \qquad\quad\times
\Biggl[
\link{p}{q}{1} +
\link{1}{q}{q'}
\Biggr],
}
\eqlabel\itisreallyalmostdone
$$
where we have used the relations in Appendix A to write
everything in terms of spinor inner products.
The square brackets in the second piece of (\itisreallyalmostdone)
may be summed using (\linkidnosum).  Doing this, plus a little more
rearrangement produces
$$
\eqalign{
\amp_T &=
-{{i}\over{2}}
(-g\sqrt2)^{n+2} \negthinspace\negthinspace
\permsum{1}{n}
{
{{\delta^i}_{\ell} \braket{p}{q'}}
\over
{\braket{p}{q'}}
}
{
{{(\Omega[1,n])_{k}}^{j} \braket{q}{p'} }
\over
{\bra{q}1,\ldots,n\ket{p'}}
}
{
{{\braket{p}{q}}^2}
\over
{ \braket{p}{p'}  \braket{q}{q'} }
}
\cr & \quad
-{{i}\over{2}}
(-g\sqrt2)^{n+2} \negthinspace\negthinspace
\permsum{1}{n}
\sum_{d=1}^{n}
{
{{(\Omega[1,d])^i}_{\ell} \braket{p}{q'}}
\over
{\bra{p}1,\ldots,d\ket{q'}}
}
{
{{(\Omega[d{+}1,n])_{k}}^{j} \braket{q}{p'} }
\over
{\bra{q}d{+}1,\ldots,n\ket{p'}}
}
{
{{\braket{p}{q}}^2}
\over
{ \braket{p}{p'}  \braket{q}{q'} }
}.
}
\eqlabel\taDA
$$
Obviously, the first term of (\taDA) simply extends the sum
in the second piece to include $d=0$ once more.  Combination
of (\taDA) with $\amp_5$ (equation (\Mfive)) produces the final
result:
$$
\eqalign{
&\amp^{ij}_{k\ell}(p^{-},q^{-};1^{+},\ldots,n^{+};p'^{+},q'^{+}) =
\cr &\quad=
-{{i}\over{2}}
(-g\sqrt2)^{n+2} \negthinspace\negthinspace
\permsum{1}{n}
\sum_{d=0}^{n}
\Biggl\{
{
{{(\Omega[1,d])^i}_{\ell} \braket{p}{q'}}
\over
{\bra{p}1,\ldots,d\ket{q'}}
}
{
{{(\Omega[d{+}1,n])_{k}}^{j} \braket{q}{p'} }
\over
{\bra{q}d{+}1,\ldots,n\ket{p'}}
}
\cr & \qquad\qquad\qquad -
{ {1}\over{N} }
{
{(\Omega[1,d])^{ij} \braket{p}{p'}}
\over
{\bra{p}1,\ldots,d\ket{p'}}
}
{
{(\Omega[d{+}1,n])_{k\ell} \braket{q}{q'} }
\over
{\bra{q}d{+}1,\ldots,n\ket{q'}}
}
\Biggr\}
{
{{\braket{p}{q}}^2}
\over
{ \braket{p}{p'}  \braket{q}{q'} }
}.
}
\eqlabel\hallelujah
$$
The expression obtained here agrees with the result of
Mangano  from reference \cite\Mangano.  Note that
(\hallelujah) does not include the crossed-channel contribution
which is
required if both quark lines are the same species and helicity.
This is easily written down from (\hallelujah).

As mentioned earlier, it is straightforward to generate the
amplitudes for the remaining  two quark helicity combinations.
The results read
$$
\amp^{ij}_{k\ell}(p^{+},q^{-};1^{+},\ldots,n^{+};p'^{-},q'^{+}) =
{
{ {\braket{p'}{q}}^2 }
\over
{ {\braket{p}{q}}^2 }
}
\amp^{ij}_{k\ell}(p^{-},q^{-};1^{+},\ldots,n^{+};p'^{+},q'^{+}),
\eq
$$
$$
\amp^{ij}_{k\ell}(p^{+},q^{+};1^{+},\ldots,n^{+};p'^{-},q'^{-}) =
{
{ {\braket{p'}{q'}}^2 }
\over
{ {\braket{p}{q}}^2 }
}
\amp^{ij}_{k\ell}(p^{-},q^{-};1^{+},\ldots,n^{+};p'^{+},q'^{+}).
\eq
$$
Thus, we see that the only effect of changing the quark helicities
is a minor change in one of the factors; the gross structure
of the amplitude remains unaltered.
These two amplitudes also agree with reference \cite\Mangano.
\chapter{CONCLUSION}

The complexity of computing scattering amplitudes in QCD
rapidly increases with the number of particles involved.  It is,
therefore, important to develop efficient techniques for computing
different types of amplitudes.  As an attempt in this dierection, we have
studied a gluon current with two off-shell gluons.  This object
appears in the process (\monsteramplitude).
This modified gluon current
has one off-shell gluon, and one ``special'' gluon.
The ``special'' gluon is off mass shell ($k^2 \ne 0$) but
has a transverse polarization vector ($k\cdot\eps=0$).
We have obtained an expression for the modified gluon
current in the case of like helicity gluons.
We have applied this current to the computation
of the process (\monsteramplitude).  In spite of the intermediate
expressions for this amplitude being  quite complicated, the final result
assumes a relatively simple form.

\bigskip
\noindent
{\chapterfont Acknowledgements:}

\noindent
We would like to thank
Dr.~Michelangelo Mangano
and
Dr.~Stephen Parke
for their useful discussions
about the result in reference \cite\Mangano.
\medskip
\noindent
This work was supported in part by the National
Science Foundation.
\loneappendix{MULTISPINOR CONVENTIONS}

Below we list the important results of application of
Weyl-van der Waerden spinor calculus to gauge theories.
Readers interested in the details should refer to
references \cite\QCDrecursions\ and
[\ref{For a brief introduction to properties of two-component
Weyl-van der Waerden spinors, see, for example, M. F. Sohnius,
Phys. Reports {\bf 128} (1985) 39}].

We use the Weyl basis
$$
\gamma^\mu =
\pmatrix{  0      & \sigma^\mu \cr
         \bar\sigma^\mu &      0   \cr},
\eqlabel\weylbasis
$$
for the Dirac matrices.  In (\weylbasis), $\sigma^\mu$
and $\bar\sigma^\mu$ refer to the convenient Lorentz-covariant
grouping of the $2\times2$ Pauli matrices plus the unit matrix:
$$
\sigma^{\mu} \equiv (1, \vec\sigma),
\newlettlabel\sigmamus
$$
$$
\bar\sigma^{\mu} \equiv (1, -\vec\sigma),
\lett
$$
and satisfy the anticommutators
$$
(\bar\sigma^{\mu})^{\dot\alpha\beta}
(\sigma^{\nu})_{\beta\dot\beta}
+
(\bar\sigma^{\nu})^{\dot\alpha\beta}
(\sigma^{\mu})_{\beta\dot\beta}
=
2 g^{\mu\nu} \delta_{\dot\beta}^{\dot\alpha},
\newlettlabel\sigmaanticommutator
$$
$$
(\sigma^{\mu})_{\alpha\dot\beta}
(\bar\sigma^{\nu})^{\dot\beta\beta}
+
(\sigma^{\nu})_{\alpha\dot\beta}
(\bar\sigma^{\mu})^{\dot\beta\beta}
=
2 g^{\mu\nu} \delta_{\alpha}^{\beta}.
\lett
$$

To each Lorentz 4-vector there corresponds a rank two multispinor,
formed from the contraction of the 4-vector with $\sigma^{\mu}$:
$$
\Jms_{\alpha\dot\beta} =
{1\over{\sqrt2}}
\sigma^{\mu}_{\alpha\dot\beta} J_{\mu},
\newlettlabel\msvector
$$
$$
{\Jbar}^{\dot\alpha\beta} =
{1\over{\sqrt2}}
\bar\sigma_{\mu}^{\dot\alpha\beta} J^{\mu}.
\lett
$$
For the purposes of normalization, it is convenient to use a
different convention when converting momenta:
$$
k_{\alpha\dot\beta} =
\sigma^{\mu}_{\alpha\dot\beta} k_{\mu},
\newlettlabel\msmomenta
$$
$$
{\bar k}^{\dot\alpha\beta} =
\bar\sigma_{\mu}^{\dot\alpha\beta} k^{\mu}.
\lett
$$
Useful consequences of (\msmomenta) and (\sigmaanticommutator) are
$$
\bar k^{\dot\alpha\beta} k_{\beta\dot\beta}
= k^2 \delta_{\dot\beta}^{\dot\alpha},
\newlettlabel\slashsquaredidentity
$$
$$
k_{\alpha\dot\beta} \bar k^{\dot\beta\beta}
= k^2 \delta_{\alpha}^{\beta}.
\lett
$$

The spinor indices may be raised and lowered using the
2-component antisymmetric tensor:
$$
u^{\alpha} = \vareps^{\alpha\beta}u_{\beta},
\newlettlabel\raiseandlower
$$
$$
\bar v^{\dot\alpha} = \vareps^{\dot\alpha\dot\beta}\bar v_{\dot\beta},
\lett
$$
$$
\vareps^{\alpha\beta}=\vareps_{\alpha\beta},
\lett
$$
$$
\vareps^{\dot\alpha\dot\beta}=\vareps_{\dot\alpha\dot\beta},
\lett
$$
$$
\vareps_{12}=\vareps_{\dot1\dot2}=1.
\lett
$$
Many useful relations may be easily proven from the Schouten identity
$$
\delta_{\gamma}^{\alpha} \delta_{\delta}^{\beta}
-\delta_{\delta}^{\alpha} \delta_{\gamma}^{\beta}
+\vareps^{\alpha\beta} \vareps_{\gamma\delta}=0,
\eqlabel\SCHOUTEN
$$
the generator of 2-component Fierz transformations.

We denote by $u(k)$ and $\bar u(k)$ the solutions to the
2-component Weyl equations:
$$
\bar k^{\dot\alpha\beta} u_{\beta}(k) = 0,
\newlettlabel\Weylequations
$$
$$
\bar u_{\dot\beta}(k) \bar k^{\dot\beta\alpha} = 0.
\lett
$$
These two spinors are related by complex conjugation
$$
\bar u_{\dot\alpha}(k) = \bigl[u_{\alpha}(k)\bigr]^{*},
\eqlabel\complexconj
$$
and have the normalization
$$
u_{\alpha}(k) \bar u_{\dot\alpha}(k) = k_{\alpha\dot\alpha}.
\eqlabel\spinornormalization
$$
It is useful to define a scalar product
$$
\braket{1}{2} \equiv u^{\alpha}(k_1) u_{\alpha}(k_2),
\eqlabel\scalarproduct
$$
which has two elementary properties
$$
\braket{1}{2} = - \braket{2}{1},
\newlettlabel\antisymmetry
$$
$$
\braket{1}{2} {\braket{1}{2}}^{*} =2k_1\cdot k_2.
\lett
$$
Contraction of $u_{\alpha}(k_1)u_{\beta}(k_2)u^{\gamma}(k_3)
u^{\delta}(k_4)$ into (\SCHOUTEN) produces the extremely
useful relation
$$
\braket{1}{2}\braket{3}{4}
+\braket{1}{3}\braket{4}{2}
+\braket{1}{4}\braket{2}{3}
=0.
\eqlabel\veryhelpful
$$
A second relation of great utility may be derived from (\veryhelpful):
$$
{
{ \braket{1}{2} }
\over
{ \braket{1}{q} \braket{q}{2} }
}
+
{
{ \braket{2}{3} }
\over
{ \braket{2}{q} \braket{q}{3} }
}
=
{
{ \braket{1}{3} }
\over
{ \braket{1}{q} \braket{q}{3} }
}.
\eqlabel\LINK
$$
Equation (\LINK) may be used to demonstrate that
$$
\sum_{j=\ell}^{m-1}
{
{ \braket{j}{j{+}1} }
\over
{ \braket{j}{q} \braket{q}{j{+}1} }
}
=
{
{ \braket{\ell}{m} }
\over
{ \braket{\ell}{q} \braket{q}{m} }
}.
\eqlabel\SUMLINK
$$
A recurring structure is
$$
\bra{p} 1,2,\ldots,n \ket{q} \equiv
\braket{p}{1} \braket{1}{2} \cdots \braket{n}{q}.
\eqlabel\stringdef
$$
We note the following basic properties
of  $\bra{p} 1,2,\ldots,n \ket{q}$:
$$
\bra{p}\thinspace\ket{q} \equiv \braket{p}{q}
\newlett
$$
$$
\bra{p} 1,2,\ldots,j{-}1 \ket{j} \bra{j} j{+}1,j{+}2,\ldots,n \ket{q}
= \bra{p} 1,2,\ldots,n \ket{q}
\lett
$$
$$
\bra{q} n,n{-}1,\ldots,1 \ket{p} =
(-1)^{n-1} \bra{p} 1,2,\ldots,n \ket{q}.
\lett
$$

Helicities $\pm1$ for massless vector bosons may be described by
$$
\eps_{\alpha\dot\alpha}(k^{+}) \equiv
{
{ u_{\alpha}(q) \bar u_{\dot\alpha}(k) }
\over
{ \braket{k}{q} }
},
\newlettlabel\polarizations
$$
$$
\eps_{\alpha\dot\alpha}(k^{-}) \equiv
{
{ u_{\alpha}(k) \bar u_{\dot\alpha}(q) }
\over
{ {\braket{k}{q}}^{*} }
},
\lett
$$
where $q$ is any null-vector such that $k\cdot q\ne0$.  As the
choice of $q$ does not affect any physics result, we will refer
to $u(q)$ and $\bar u(q)$ as gauge spinors.  The  corresponding
polarization vectors $\eps^{\mu}(k)$ defined through (\msvector)
differ from the ``standard'' polarization vectors
$$
\vareps_{0}^{\mu}(k^{\pm}) =
\left(
0, \mp{1\over{\sqrt2}},{{-i}\over{\sqrt2}},0
\right),
\newlettlabel\standardpolarizations
$$
$$
k^{\mu}=(k,0,0,k),
\lett
$$
by a phase and a gauge transformation depending
on $q$.\cite\QCDrecursions

To save accounting for a large number of indices, an efficient
method is to initially write quantities in the usual formalism
and then convert to multispinor notation at a later stage using
the substitutions
$$
k\cdot k' =
{1\over2} \bar k^{\dot\alpha\alpha}{k'}_{\alpha\dot\alpha} =
{1\over2} k_{\alpha\dot\alpha} {\bar k}^{\prime\dot\alpha\alpha},
\newlettlabel\conversiondot
$$
$$
k\cdot\eps(k') =
{1\over{\sqrt2}} \bar k^{\dot\alpha\alpha}\eps_{\alpha\dot\alpha}(k') =
{1\over{\sqrt2}} k_{\alpha\dot\alpha}\bar\eps^{\dot\alpha\alpha}(k'),
\lett
$$
$$
\eps(k)\cdot\eps(k') =
\bar\eps^{\dot\alpha\alpha}(k)\eps_{\alpha\dot\alpha}(k') =
\eps_{\alpha\dot\alpha}(k)\bar\eps^{\dot\alpha\alpha}(k'),
\lett
$$
for Lorentz dot products and
$$
{1\over2}(1-\gamma_5)\psi \longrightarrow \psi_{\alpha},
\newlettlabel\conversionfermion
$$
$$
{1\over2}(1+\gamma_5)\psi \longrightarrow \psi^{\dot\alpha},
\lett
$$
$$
{1\over2}(1-\gamma_5)\Jslash
{1\over2}(1+\gamma_5)
\longrightarrow \sqrt2 \thinspace \Jms_{\alpha\dot\alpha},
\lett
$$
$$
{1\over2}(1+\gamma_5)\Jslash
{1\over2}(1-\gamma_5)
\longrightarrow \sqrt2 \thinspace \Jbar^{\dot\alpha\alpha},
\lett
$$
$$
{1\over2}(1-\gamma_5)\slash{k}{1\over2}(1+\gamma_5)
\longrightarrow k_{\alpha\dot\alpha},
\lett
$$
$$
{1\over2}(1+\gamma_5)\slash{k}{1\over2}(1-\gamma_5)
\longrightarrow \bar k^{\dot\alpha\alpha},
\lett
$$
in strings of Dirac matrices.  Note the unequal treatments of momenta
versus other 4-vectors caused by the conventions (\msvector)
and (\msmomenta).

\vfill\eject

\global \chap =1
\vfill \eject \vglue .2in
\centerline {\headingfont REFERENCES}
\vglue .5in\baselineskip =\single
\parindent =16pt \parskip =\single
\item {1.}F. A. Berends and W. T. Giele, Nucl. Phys.
{\fam \bffam \twelvbf B306} (1988) 759; C. Dunn and T.--M. Yan,
Nucl. Phys. {\fam \bffam \twelvbf B352} (1991) 402;
C. Dunn, thesis, Cornell University (1990) \hfill \par
\item {2.}J. Schwinger, {\fam \itfam \twelvit Particles,
Sources and Fields,}{ }Vol. I, Addison Wesley, 1970;
Ann. Phys. {\fam \bffam \twelvbf 119}{ }(1979) 192 \hfill \par
\item {3.} The spinor technique was first introduced by the
CALCUL collaboration, in the context of massless Abelian gauge
theory: P. De Causmaecker, R. Gastmans, W. Troost, and T.T. Wu,
Phys. Lett. {\fam \bffam \twelvbf 105B}{ }(1981) 215; P. De Causmaecker,
R. Gastmans, W. Troost and T.T. Wu, Nucl. Phys.
{\fam \bffam \twelvbf B206}{ }(1982) 53; F. A. Berends,
R. Kleiss, P. De Causmaecker, R. Gastmans, W. Troost and
T.T. Wu, Nucl. Phys. {\fam \bffam \twelvbf B206}{ }(1982) 61;
F.A. Berends, P. De Causmaecker, R. Gastmans, R. Kleiss, W. Troost
and T.T. Wu, Nucl. Phys. {\fam \bffam \twelvbf B239}{ }(1984) 382;
{\fam \bffam \twelvbf B239}{ }(1984) 395;
{\fam \bffam \twelvbf B264}{ }(1986) 243;
{\fam \bffam \twelvbf B264}{ }(1986) 265
\vskip \smallskipamount \item { } By now, many
papers have been published
on the subject. A partial list of references follows.
\vskip \smallskipamount \item { } P. De Causmaecker, thesis,
Leuven University (1983); R. Farrar and F. Neri, Phys. Lett.
{\fam \bffam \twelvbf 130B}{ }(1983) 109; R. Kleiss, Nucl. Phys.
{\fam \bffam \twelvbf B241}{ }(1984) 61; Z. Xu,
D.H. Zhang and Z. Chang,
Tsingua University preprint TUTP-84/3, 84/4,
and 84/5a (1984), and Nucl.
Phys. {\fam \bffam \twelvbf B291}{ }(1987) 392; J.F. Gunion and
Z. Kunszt, Phys. Lett. {\fam \bffam \twelvbf 161B}{ }(1985) 333;
F.A. Berends, P.H. Davereldt and R. Kleiss, Nucl. Phys.
{\fam \bffam \twelvbf B253}{ } (1985) 441; R. Kleiss and
W.J. Stirling, Nucl. Phys. {\fam \bffam \twelvbf B262}{ }(1985) 235;
J.F. Gunion and Z. Kunszt,
Phys. Lett. {\fam \bffam \twelvbf 159B}{ }(1985)
167; {\fam \bffam \twelvbf 161B}{ }(1985) 333; S.J. Parke and
T.R. Taylor, Phys. Rev. Lett. {\fam \bffam \twelvbf 56}{ }(1986) 2459;
Z. Kunszt, Nucl. Phys. {\fam \bffam \twelvbf B271}{ }(1986) 333;
J.F. Gunion and J. Kalinowski, Phys. Rev.
{\fam \bffam \twelvbf D34}{ }(1986) 2119; R. Kleiss and
W.J. Stirling, Phys. Lett. {\fam \bffam \twelvbf 179B}{ }(1986) 159;
M. Mangano and S.J. Parke,
Nucl. Phys. {\fam \bffam \twelvbf B299}{ }(1988)
673; M. Mangano, S.J. Parke and Z. Xu, Nucl. Phys.
{\fam \bffam \twelvbf B298}{ } (1988) 653; D.A. Kosower, B.--H. Lee
and V.P. Nair, Phys. Lett. {\fam \bffam \twelvbf 201B}{ } (1988) 85;
M. Mangano and S.J. Parke,
Nucl. Phys. {\fam \bffam \twelvbf B299}{ }(1988)
673; F.A. Berends and W.T. Giele, Nucl. Phys.
{\fam \bffam \twelvbf B313}{ }(1989) 595; M. Mangano, Nucl.
Phys. {\fam \bffam \twelvbf B315}{ }(1989) 391; D.A. Kosower,
Nucl. Phys. {\fam \bffam \twelvbf B335}{ }(1990) 23; D.A. Kosower,
Phys. Lett. {\fam \bffam \twelvbf B254}{ }(1991) 439; Z. Bern and
D.A. Kosower, Nucl. Phys. {\fam \bffam \twelvbf B379}{ }(1992) 451;
C.S. Lam, McGill preprint McGill/92-32 (1992) \hfill \par
\item {4.}Many of the results for processes containing six or fewer
particles are collected in R. Gastmans and T.T. Wu, The Ubiquitous
Photon: Helicity Method for QED and QCD (Oxford University Press,
New York, 1990) \hfill \par
\item {5.}M. Mangano and S. Parke, Phys. Reports
{\fam \bffam \twelvbf 200}{ }(1991) 301 \hfill \par
\item {6.}M. Mangano, Nucl. Phys. {\fam \bffam \twelvbf B309}{ }(1988)
461 \hfill \par
\item {7.}For a brief introduction to properties of two-component
Weyl-van der Waerden spinors, see, for example, M. F. Sohnius,
Phys. Reports {\fam \bffam \twelvbf 128} (1985) 39 \hfill \par

\vfill\eject\bye

%
%

\input pictex

%
%
\font\twelvrm=cmr12
\font\ninerm=cmr9
\font\twelvi=cmmi12
\font\ninei=cmmi9
\font\twelvex=cmex10 scaled\magstep1
\font\twelvbf=cmbx12
\font\ninebf=cmbx9
\font\twelvit=cmti12
\font\twelvsy=cmsy10 scaled\magstep1
\font\ninesy=cmsy9
\font\twelvtt=cmtt12

\font\twelvsl=cmsl12

\font\abstractfont=cmr10
\font\abstractitalfont=cmti10

\def\twelvepoint{\def\rm{\fam0\twelvrm}
   \textfont0=\twelvrm \scriptfont0=\ninerm \scriptscriptfont0=\sevenrm
   \textfont1=\twelvi \scriptfont1=\ninei \scriptscriptfont1=\seveni
   \textfont2=\twelvsy \scriptfont2=\ninesy \scriptscriptfont2=\sevensy
   \textfont3=\twelvex \scriptfont3=\tenex \scriptscriptfont3=\tenex
        \textfont\itfam=\twelvit \def\it{\fam\itfam\twelvit}
        \textfont\slfam=\twelvsl \def\sl{\fam\slfam\twelvsl}
        \textfont\ttfam=\twelvtt \def\tt{\fam\ttfam\twelvtt}
        \textfont\bffam=\twelvbf \def\bf{\fam\bffam\twelvbf}
        \scriptfont\bffam=\ninebf  \scriptscriptfont\bffam=\sevenbf
        \skewchar\ninei='177
        \skewchar\twelvi='177
        \skewchar\seveni='177
}

\newdimen\normalwidth
\newdimen\double
\newdimen\single
\newdimen\indentlength          \indentlength=.5in

\newif\ifdrafton

\def\galley{
 \draftonfalse
 \twelvepoint
 \rm
 \font\chapterfont=cmbx10 scaled\magstep1
 \font\sectionfont=cmbx12
 \font\subsectionfont=cmbx12
 \font\headingfont=cmr10 scaled\magstep2
 \font\titlefont=cmbx10 scaled\magstep2
 \normalwidth=5.7in
 \double=.34in
 \single=.17in
 \hsize=\normalwidth
 \vsize=8.7in
 \hoffset=0.48in
 \voffset=0.1in
 \hfuzz=0.5pt
 \baselineskip=\double plus 2pt minus 2pt }

\parindent=\indentlength
\clubpenalty=10000
\widowpenalty=10000
\displaywidowpenalty=500
\overfullrule=2pt
\tolerance=100

\newcount\chapterno     \chapterno=0
\newcount\sectionno     \sectionno=0
\newcount\appno         \appno=0
\newcount\subsectionno  \subsectionno=0
\newcount\eqnum \eqnum=0
\newcount\refno \refno=0
\newcount\chap
\newcount\figno \figno=0
\newcount\tableno \tableno=0
\newcount\lettno \lettno=0

\def\bodypaging{
 \headline={\ifodd\chap \hfil \else \tenrm\hfil\twelvrm\folio \fi}
 \footline={\rm \ifodd\chap \global\chap=0 \tenrm\hfil\twelvrm\folio\hfil
 \else \hfil \fi}}

\galley             

\pageno=81 \bodypaging 	 

\vfil
$$
\beginpicture

\setcoordinatesystem units <1in,1in> point at 3.5 9.0


\ellipticalarc axes ratio 1:2 360 degrees from 3.675 5.5 center at 3.5 5.5

\ellipticalarc axes ratio 2:1 360 degrees from 5.20 7.25 center at 4.85 7.25

\ellipticalarc axes ratio 2:1 360 degrees from 2.5 7.25 center at 2.15 7.25

\ellipticalarc axes ratio 2:1 360 degrees from 5.20 3.75 center at 4.85 3.75


\setlinear

\plot 1.00 7.25  1.80 7.25 /
\plot 2.50 7.25  4.50 7.25 /
\plot 5.20 7.25  6.00 7.25 /

\plot 1.00 3.75  4.50 3.75 /
\plot 5.20 3.75  6.00 3.75 /

\setsolid


\setquadratic

\plot 3.50 7.25  3.55 7.20  3.50 7.15  3.45 7.10  3.50 7.05
                 3.55 7.00  3.50 6.95  3.45 6.90  3.50 6.85
                 3.55 6.80  3.50 6.75  3.45 6.70  3.50 6.65
                 3.55 6.60  3.50 6.55  3.45 6.50  3.50 6.45
                 3.55 6.40  3.50 6.35  3.45 6.30  3.50 6.25
                 3.55 6.20  3.50 6.15  3.45 6.10  3.50 6.05
                 3.55 6.00  3.50 5.95  3.45 5.90  3.50 5.85 /

\plot 3.50 5.15  3.55 5.10  3.50 5.05  3.45 5.00  3.50 4.95
                 3.55 4.90  3.50 4.85  3.45 4.80  3.50 4.75
                 3.55 4.70  3.50 4.65  3.45 4.60  3.50 4.55
                 3.55 4.50  3.50 4.45  3.45 4.40  3.50 4.35
                 3.55 4.30  3.50 4.25  3.45 4.20  3.50 4.15
                 3.55 4.10  3.50 4.05  3.45 4.00  3.50 3.95
                 3.55 3.90  3.50 3.85  3.45 3.80  3.50 3.75 /


\setshadesymbol <z,z,z,z> ({\fiverm .})

\setshadegrid span <.2pt>

\setlinear

\vshade 1.35 7.25 7.25
        1.40 7.20 7.30 /

\vshade 2.95 7.25 7.25
        3.00 7.20 7.30 /

\vshade 4.00 7.25 7.25
        4.05 7.20 7.30 /

\vshade 5.60 7.25 7.25
        5.65 7.20 7.30 /

\vshade 2.15 3.75 3.75
        2.20 3.70 3.80 /

\vshade 4.00 3.75 3.75
        4.05 3.70 3.80 /

\vshade 5.60 3.75 3.75
        5.65 3.70 3.80 /


\setshadesymbol <z,z,z,z> ({\fiverm .})

\setshadegrid span <1.75pt>

\setquadratic

\vshade 1.8000 7.2500000 7.2500000
        1.8875 7.1342484 7.3657516
        1.9750 7.0984456 7.4015544
        2.0625 7.0805570 7.4194430
        2.1500 7.0750000 7.4250000
        2.2375 7.0805570 7.4194430
        2.3250 7.0984456 7.4015544
        2.4125 7.1342484 7.3657516
        2.5000 7.2500000 7.2500000 /

\vshade 4.5000 7.2500000 7.2500000
        4.5875 7.1342484 7.3657516
        4.6750 7.0984456 7.4015544
        4.7625 7.0805570 7.4194430
        4.8500 7.0750000 7.4250000
        4.9375 7.0805570 7.4194430
        5.0250 7.0984456 7.4015544
        5.1125 7.1342484 7.3657516
        5.2000 7.2500000 7.2500000 /

\vshade 4.5000 3.7500000 3.7500000
        4.5875 3.6342484 3.8657516
        4.6750 3.5984456 3.9015544
        4.7625 3.5805570 3.9194430
        4.8500 3.5750000 3.9250000
        4.9375 3.5805570 3.9194430
        5.0250 3.5984456 3.9015544
        5.1125 3.6342484 3.8657516
        5.2000 3.7500000 3.7500000 /

\hshade 5.1500 3.5000000 3.5000000
        5.2375 3.3842484 3.6157516
        5.3250 3.3484456 3.6515544
        5.4125 3.3305570 3.6694430
        5.5000 3.3250000 3.6750000
        5.5875 3.3305570 3.6694430
        5.6750 3.3484456 3.6515544
        5.7625 3.3842484 3.6157516
        5.8500 3.5000000 3.5000000 /


 \startrotation by  0.965926   0.258819 about      3.669443       5.587500
 \setquadratic
 \plot
     3.669443       5.587500
     3.719443       5.637500
     3.769443       5.587500
     3.819443       5.537500
     3.869443       5.587500
     3.919443       5.637500
     3.969443       5.587500
     4.019443       5.537500
     4.069443       5.587500
     4.119443       5.637500
     4.169443       5.587500
 /
 \stoprotation

 \startrotation by  0.965926  -0.258819 about      3.669443       5.412500
 \setquadratic
 \plot
     3.669443       5.412500
     3.719443       5.462500
     3.769443       5.412500
     3.819443       5.362500
     3.869443       5.412500
     3.919443       5.462500
     3.969443       5.412500
     4.019443       5.362500
     4.069443       5.412500
     4.119443       5.462500
     4.169443       5.412500
 /
 \stoprotation

 \startrotation by -1.000000   0.000000 about      3.325000       5.500000
 \setquadratic
 \plot
     3.325000       5.500000
     3.375000       5.550000
     3.425000       5.500000
     3.475000       5.450000
     3.525000       5.500000
     3.575000       5.550000
     3.625000       5.500000
     3.675000       5.450000
     3.725000       5.500000
     3.775000       5.550000
     3.825000       5.500000
 /
 \stoprotation

 \startrotation by -0.906308  -0.422618 about      3.348446       5.325000
 \setquadratic
 \plot
     3.348446       5.325000
     3.398446       5.375000
     3.448446       5.325000
     3.498446       5.275000
     3.548446       5.325000
     3.598446       5.375000
     3.648446       5.325000
     3.698446       5.275000
     3.748446       5.325000
     3.798446       5.375000
     3.848446       5.325000
 /
 \stoprotation

 \startrotation by -0.906308   0.422618 about      3.348446       5.675000
 \setquadratic
 \plot
     3.348446       5.675000
     3.398446       5.725000
     3.448446       5.675000
     3.498446       5.625000
     3.548446       5.675000
     3.598446       5.725000
     3.648446       5.675000
     3.698446       5.625000
     3.748446       5.675000
     3.798446       5.725000
     3.848446       5.675000
 /
 \stoprotation


 \startrotation by  0.000000  -1.000000 about      2.150000       7.075000
 \setquadratic
 \plot
     2.150000       7.075000
     2.200000       7.125000
     2.250000       7.075000
     2.300000       7.025000
     2.350000       7.075000
     2.400000       7.125000
     2.450000       7.075000
     2.500000       7.025000
     2.550000       7.075000
     2.600000       7.125000
     2.650000       7.075000
 /
 \stoprotation

 \startrotation by -0.422618  -0.906308 about      1.975000       7.094844
 \setquadratic
 \plot
     1.975000       7.094844
     2.025000       7.144845
     2.075000       7.094844
     2.125000       7.044844
     2.175000       7.094844
     2.225000       7.144845
     2.275000       7.094844
     2.325000       7.044844
     2.375000       7.094844
     2.425000       7.144845
     2.475000       7.094844
 /
 \stoprotation

 \startrotation by  0.422618  -0.906308 about      2.325000       7.094844
 \setquadratic
 \plot
     2.325000       7.094844
     2.375000       7.144845
     2.425000       7.094844
     2.475000       7.044844
     2.525000       7.094844
     2.575000       7.144845
     2.625000       7.094844
     2.675000       7.044844
     2.725000       7.094844
     2.775000       7.144845
     2.825000       7.094844
 /
 \stoprotation

 \startrotation by  0.000000   1.000000 about      2.150000       7.425000
 \setquadratic
 \plot
     2.150000       7.425000
     2.200000       7.475000
     2.250000       7.425000
     2.300000       7.375000
     2.350000       7.425000
     2.400000       7.475000
     2.450000       7.425000
     2.500000       7.375000
     2.550000       7.425000
     2.600000       7.475000
     2.650000       7.425000
 /
 \stoprotation

 \startrotation by -0.422618   0.906308 about      1.975000       7.401555
 \setquadratic
 \plot
     1.975000       7.401555
     2.025000       7.451555
     2.075000       7.401555
     2.125000       7.351554
     2.175000       7.401555
     2.225000       7.451555
     2.275000       7.401555
     2.325000       7.351554
     2.375000       7.401555
     2.425000       7.451555
     2.475000       7.401555
 /
 \stoprotation

 \startrotation by  0.422618   0.906308 about      2.325000       7.401555
 \setquadratic
 \plot
     2.325000       7.401555
     2.375000       7.451555
     2.425000       7.401555
     2.475000       7.351554
     2.525000       7.401555
     2.575000       7.451555
     2.625000       7.401555
     2.675000       7.351554
     2.725000       7.401555
     2.775000       7.451555
     2.825000       7.401555
 /
 \stoprotation


 \startrotation by -0.766044  -0.642788 about      4.587500       3.634248
 \setquadratic
 \plot
     4.587500       3.634248
     4.637500       3.684248
     4.687500       3.634248
     4.737500       3.584249
     4.787500       3.634248
     4.837500       3.684248
     4.887500       3.634248
     4.937500       3.584249
     4.987500       3.634248
     5.037500       3.684248
     5.087500       3.634248
 /
 \stoprotation

 \startrotation by -0.224951  -0.974370 about      4.762500       3.580557
 \setquadratic
 \plot
     4.762500       3.580557
     4.812500       3.630557
     4.862500       3.580557
     4.912500       3.530557
     4.962500       3.580557
     5.012500       3.630557
     5.062500       3.580557
     5.112500       3.530557
     5.162500       3.580557
     5.212500       3.630557
     5.262500       3.580557
 /
 \stoprotation

 \startrotation by  0.224951  -0.974370 about      4.937500       3.580557
 \setquadratic
 \plot
     4.937500       3.580557
     4.987500       3.630557
     5.037500       3.580557
     5.087500       3.530557
     5.137500       3.580557
     5.187500       3.630557
     5.237500       3.580557
     5.287500       3.530557
     5.337500       3.580557
     5.387500       3.630557
     5.437500       3.580557
 /
 \stoprotation

 \startrotation by  0.766045  -0.642787 about      5.112500       3.634248
 \setquadratic
 \plot
     5.112500       3.634248
     5.162500       3.684248
     5.212500       3.634248
     5.262500       3.584249
     5.312500       3.634248
     5.362500       3.684248
     5.412500       3.634248
     5.462500       3.584249
     5.512500       3.634248
     5.562500       3.684248
     5.612500       3.634248
 /
 \stoprotation


 \startrotation by  0.000000   1.000000 about      4.850000       7.425000
 \setquadratic
 \plot
     4.850000       7.425000
     4.900000       7.475000
     4.950000       7.425000
     5.000000       7.375000
     5.050000       7.425000
     5.100000       7.475000
     5.150000       7.425000
     5.200000       7.375000
     5.250000       7.425000
     5.300000       7.475000
     5.350000       7.425000
 /
 \stoprotation

 \startrotation by -0.241922  -0.970296 about      4.762500       7.080557
 \setquadratic
 \plot
     4.762500       7.080557
     4.812500       7.130557
     4.862500       7.080557
     4.912500       7.030557
     4.962500       7.080557
     5.012500       7.130557
     5.062500       7.080557
     5.112500       7.030557
     5.162500       7.080557
     5.212500       7.130557
     5.262500       7.080557
 /
 \stoprotation

 \startrotation by  0.241922  -0.970296 about      4.937500       7.080557
 \setquadratic
 \plot
     4.937500       7.080557
     4.987500       7.130557
     5.037500       7.080557
     5.087500       7.030557
     5.137500       7.080557
     5.187500       7.130557
     5.237500       7.080557
     5.287500       7.030557
     5.337500       7.080557
     5.387500       7.130557
     5.437500       7.080557
 /
 \stoprotation


\put {$\bar q$} at 0.84 7.25

\put {$\bar q$} at 0.84 3.75

\put {$q$} at 6.16 7.25

\put {$q$} at 6.16 3.75

\put {$\longrightarrow$} at 1.25 7.45

\put {$p$} at 1.25 7.6

\put {$\longleftarrow$} at 5.75 7.45

\put {$p'$} at 5.75 7.6

\put {$\longrightarrow$} at 1.25 3.55

\put {$q$} at 1.25 3.4

\put {$\longleftarrow$} at 5.75 3.55

\put {$q'$} at 5.75 3.4

\put {$\{v{+}1,\ldots,t\}$} at 2.15 8.10

\put {$\{t{+}1,\ldots,w\}$} at 4.85 8.10

\put {$\{w{+}1,\ldots,n\}$} at 4.85 2.90

\put {$\{1,\ldots,u\}$} at 2.35 5.5

\put {$\{u{+}1,\ldots,v\}$} at 4.75 5.5

\put {$\uparrow\enspace k_0 = q+\kappa(w{+}1,n)+q'$} at 4.6 4.5

\endpicture
$$
\bigskip
\noindent
Figure 1.  Contributions to
$q \thinspace \bar q \longrightarrow q \thinspace \bar q
\thinspace g \thinspace g \cdots g$.  The antiquark line
of momentum
$q$ has no gluons attached to it in accordance with  the
gauge choice made in the text.

\vfil\eject\bye
%
%

\input pictex

%
%
\font\twelvrm=cmr12
\font\ninerm=cmr9
\font\twelvi=cmmi12
\font\ninei=cmmi9
\font\twelvex=cmex10 scaled\magstep1
\font\twelvbf=cmbx12
\font\ninebf=cmbx9
\font\twelvit=cmti12
\font\twelvsy=cmsy10 scaled\magstep1
\font\ninesy=cmsy9
\font\twelvtt=cmtt12

\font\twelvsl=cmsl12

\font\abstractfont=cmr10
\font\abstractitalfont=cmti10

\def\twelvepoint{\def\rm{\fam0\twelvrm}
   \textfont0=\twelvrm \scriptfont0=\ninerm \scriptscriptfont0=\sevenrm
   \textfont1=\twelvi \scriptfont1=\ninei \scriptscriptfont1=\seveni
   \textfont2=\twelvsy \scriptfont2=\ninesy \scriptscriptfont2=\sevensy
   \textfont3=\twelvex \scriptfont3=\tenex \scriptscriptfont3=\tenex
        \textfont\itfam=\twelvit \def\it{\fam\itfam\twelvit}
        \textfont\slfam=\twelvsl \def\sl{\fam\slfam\twelvsl}
        \textfont\ttfam=\twelvtt \def\tt{\fam\ttfam\twelvtt}
        \textfont\bffam=\twelvbf \def\bf{\fam\bffam\twelvbf}
        \scriptfont\bffam=\ninebf  \scriptscriptfont\bffam=\sevenbf
        \skewchar\ninei='177
        \skewchar\twelvi='177
        \skewchar\seveni='177
}

\newdimen\normalwidth
\newdimen\double
\newdimen\single
\newdimen\indentlength          \indentlength=.5in

\newif\ifdrafton

\def\galley{
 \draftonfalse
 \twelvepoint
 \rm
 \font\chapterfont=cmbx10 scaled\magstep1
 \font\sectionfont=cmbx12
 \font\subsectionfont=cmbx12
 \font\headingfont=cmr10 scaled\magstep2
 \font\titlefont=cmbx10 scaled\magstep2
 \normalwidth=5.7in
 \double=.34in
 \single=.17in
 \hsize=\normalwidth
 \vsize=8.7in
 \hoffset=0.48in
 \voffset=0.1in
 \hfuzz=0.5pt
 \baselineskip=\double plus 2pt minus 2pt }

\parindent=\indentlength
\clubpenalty=10000
\widowpenalty=10000
\displaywidowpenalty=500
\overfullrule=2pt
\tolerance=100

\newcount\chapterno     \chapterno=0
\newcount\sectionno     \sectionno=0
\newcount\appno         \appno=0
\newcount\subsectionno  \subsectionno=0
\newcount\eqnum \eqnum=0
\newcount\refno \refno=0
\newcount\chap
\newcount\figno \figno=0
\newcount\tableno \tableno=0
\newcount\lettno \lettno=0

\def\bodypaging{
 \headline={\ifodd\chap \hfil \else \tenrm\hfil\twelvrm\folio \fi}
 \footline={\rm \ifodd\chap \global\chap=0 \tenrm\hfil\twelvrm\folio\hfil
 \else \hfil \fi}}

\galley             

\pageno=82 \bodypaging 	 

\vfil
$$
\beginpicture

\setcoordinatesystem units <1in,1in> point at 3.5 9.0


\ellipticalarc axes ratio 1:2 360 degrees from 3.675 5.5 center at 3.5 5.5

\ellipticalarc axes ratio 2:1 360 degrees from 5.20 7.25 center at 4.85 7.25

\ellipticalarc axes ratio 2:1 360 degrees from 2.5 7.25 center at 2.15 7.25

\ellipticalarc axes ratio 2:1 360 degrees from 5.20 3.75 center at 4.85 3.75


\setlinear

\plot 1.00 7.25  1.80 7.25 /
\plot 2.50 7.25  4.50 7.25 /
\plot 5.20 7.25  6.00 7.25 /

\plot 1.00 3.75  4.50 3.75 /
\plot 5.20 3.75  6.00 3.75 /

\setsolid


\setquadratic

\plot 3.50 7.25  3.55 7.20  3.50 7.15  3.45 7.10  3.50 7.05
                 3.55 7.00  3.50 6.95  3.45 6.90  3.50 6.85
                 3.55 6.80  3.50 6.75  3.45 6.70  3.50 6.65
                 3.55 6.60  3.50 6.55  3.45 6.50  3.50 6.45
                 3.55 6.40  3.50 6.35  3.45 6.30  3.50 6.25
                 3.55 6.20  3.50 6.15  3.45 6.10  3.50 6.05
                 3.55 6.00  3.50 5.95  3.45 5.90  3.50 5.85 /

\plot 3.50 5.15  3.55 5.10  3.50 5.05  3.45 5.00  3.50 4.95
                 3.55 4.90  3.50 4.85  3.45 4.80  3.50 4.75
                 3.55 4.70  3.50 4.65  3.45 4.60  3.50 4.55
                 3.55 4.50  3.50 4.45  3.45 4.40  3.50 4.35
                 3.55 4.30  3.50 4.25  3.45 4.20  3.50 4.15
                 3.55 4.10  3.50 4.05  3.45 4.00  3.50 3.95
                 3.55 3.90  3.50 3.85  3.45 3.80  3.50 3.75 /


\setshadesymbol <z,z,z,z> ({\fiverm .})

\setshadegrid span <.2pt>

\setlinear

\vshade 1.35 7.25 7.25
        1.40 7.20 7.30 /

\vshade 2.95 7.25 7.25
        3.00 7.20 7.30 /

\vshade 4.00 7.25 7.25
        4.05 7.20 7.30 /

\vshade 5.60 7.25 7.25
        5.65 7.20 7.30 /

\vshade 2.15 3.75 3.75
        2.20 3.70 3.80 /

\vshade 4.00 3.75 3.75
        4.05 3.70 3.80 /

\vshade 5.60 3.75 3.75
        5.65 3.70 3.80 /


\setshadesymbol <z,z,z,z> ({\fiverm .})

\setshadegrid span <1.75pt>

\setquadratic

\vshade 1.8000 7.2500000 7.2500000
        1.8875 7.1342484 7.3657516
        1.9750 7.0984456 7.4015544
        2.0625 7.0805570 7.4194430
        2.1500 7.0750000 7.4250000
        2.2375 7.0805570 7.4194430
        2.3250 7.0984456 7.4015544
        2.4125 7.1342484 7.3657516
        2.5000 7.2500000 7.2500000 /

\vshade 4.5000 7.2500000 7.2500000
        4.5875 7.1342484 7.3657516
        4.6750 7.0984456 7.4015544
        4.7625 7.0805570 7.4194430
        4.8500 7.0750000 7.4250000
        4.9375 7.0805570 7.4194430
        5.0250 7.0984456 7.4015544
        5.1125 7.1342484 7.3657516
        5.2000 7.2500000 7.2500000 /

\vshade 4.5000 3.7500000 3.7500000
        4.5875 3.6342484 3.8657516
        4.6750 3.5984456 3.9015544
        4.7625 3.5805570 3.9194430
        4.8500 3.5750000 3.9250000
        4.9375 3.5805570 3.9194430
        5.0250 3.5984456 3.9015544
        5.1125 3.6342484 3.8657516
        5.2000 3.7500000 3.7500000 /

\hshade 5.1500 3.5000000 3.5000000
        5.2375 3.3842484 3.6157516
        5.3250 3.3484456 3.6515544
        5.4125 3.3305570 3.6694430
        5.5000 3.3250000 3.6750000
        5.5875 3.3305570 3.6694430
        5.6750 3.3484456 3.6515544
        5.7625 3.3842484 3.6157516
        5.8500 3.5000000 3.5000000 /


 \startrotation by  0.965926   0.258819 about      3.669443       5.587500
 \setquadratic
 \plot
     3.669443       5.587500
     3.719443       5.637500
     3.769443       5.587500
     3.819443       5.537500
     3.869443       5.587500
     3.919443       5.637500
     3.969443       5.587500
     4.019443       5.537500
     4.069443       5.587500
     4.119443       5.637500
     4.169443       5.587500
 /
 \stoprotation

 \startrotation by  0.965926  -0.258819 about      3.669443       5.412500
 \setquadratic
 \plot
     3.669443       5.412500
     3.719443       5.462500
     3.769443       5.412500
     3.819443       5.362500
     3.869443       5.412500
     3.919443       5.462500
     3.969443       5.412500
     4.019443       5.362500
     4.069443       5.412500
     4.119443       5.462500
     4.169443       5.412500
 /
 \stoprotation

 \startrotation by -1.000000   0.000000 about      3.325000       5.500000
 \setquadratic
 \plot
     3.325000       5.500000
     3.375000       5.550000
     3.425000       5.500000
     3.475000       5.450000
     3.525000       5.500000
     3.575000       5.550000
     3.625000       5.500000
     3.675000       5.450000
     3.725000       5.500000
     3.775000       5.550000
     3.825000       5.500000
 /
 \stoprotation

 \startrotation by -0.906308  -0.422618 about      3.348446       5.325000
 \setquadratic
 \plot
     3.348446       5.325000
     3.398446       5.375000
     3.448446       5.325000
     3.498446       5.275000
     3.548446       5.325000
     3.598446       5.375000
     3.648446       5.325000
     3.698446       5.275000
     3.748446       5.325000
     3.798446       5.375000
     3.848446       5.325000
 /
 \stoprotation

 \startrotation by -0.906308   0.422618 about      3.348446       5.675000
 \setquadratic
 \plot
     3.348446       5.675000
     3.398446       5.725000
     3.448446       5.675000
     3.498446       5.625000
     3.548446       5.675000
     3.598446       5.725000
     3.648446       5.675000
     3.698446       5.625000
     3.748446       5.675000
     3.798446       5.725000
     3.848446       5.675000
 /
 \stoprotation


 \startrotation by  0.000000  -1.000000 about      2.150000       7.075000
 \setquadratic
 \plot
     2.150000       7.075000
     2.200000       7.125000
     2.250000       7.075000
     2.300000       7.025000
     2.350000       7.075000
     2.400000       7.125000
     2.450000       7.075000
     2.500000       7.025000
     2.550000       7.075000
     2.600000       7.125000
     2.650000       7.075000
 /
 \stoprotation

 \startrotation by -0.422618  -0.906308 about      1.975000       7.094844
 \setquadratic
 \plot
     1.975000       7.094844
     2.025000       7.144845
     2.075000       7.094844
     2.125000       7.044844
     2.175000       7.094844
     2.225000       7.144845
     2.275000       7.094844
     2.325000       7.044844
     2.375000       7.094844
     2.425000       7.144845
     2.475000       7.094844
 /
 \stoprotation

 \startrotation by  0.422618  -0.906308 about      2.325000       7.094844
 \setquadratic
 \plot
     2.325000       7.094844
     2.375000       7.144845
     2.425000       7.094844
     2.475000       7.044844
     2.525000       7.094844
     2.575000       7.144845
     2.625000       7.094844
     2.675000       7.044844
     2.725000       7.094844
     2.775000       7.144845
     2.825000       7.094844
 /
 \stoprotation

 \startrotation by  0.000000   1.000000 about      2.150000       7.425000
 \setquadratic
 \plot
     2.150000       7.425000
     2.200000       7.475000
     2.250000       7.425000
     2.300000       7.375000
     2.350000       7.425000
     2.400000       7.475000
     2.450000       7.425000
     2.500000       7.375000
     2.550000       7.425000
     2.600000       7.475000
     2.650000       7.425000
 /
 \stoprotation

 \startrotation by -0.422618   0.906308 about      1.975000       7.401555
 \setquadratic
 \plot
     1.975000       7.401555
     2.025000       7.451555
     2.075000       7.401555
     2.125000       7.351554
     2.175000       7.401555
     2.225000       7.451555
     2.275000       7.401555
     2.325000       7.351554
     2.375000       7.401555
     2.425000       7.451555
     2.475000       7.401555
 /
 \stoprotation

 \startrotation by  0.422618   0.906308 about      2.325000       7.401555
 \setquadratic
 \plot
     2.325000       7.401555
     2.375000       7.451555
     2.425000       7.401555
     2.475000       7.351554
     2.525000       7.401555
     2.575000       7.451555
     2.625000       7.401555
     2.675000       7.351554
     2.725000       7.401555
     2.775000       7.451555
     2.825000       7.401555
 /
 \stoprotation


 \startrotation by -0.766044  -0.642788 about      4.587500       3.634248
 \setquadratic
 \plot
     4.587500       3.634248
     4.637500       3.684248
     4.687500       3.634248
     4.737500       3.584249
     4.787500       3.634248
     4.837500       3.684248
     4.887500       3.634248
     4.937500       3.584249
     4.987500       3.634248
     5.037500       3.684248
     5.087500       3.634248
 /
 \stoprotation

 \startrotation by -0.224951  -0.974370 about      4.762500       3.580557
 \setquadratic
 \plot
     4.762500       3.580557
     4.812500       3.630557
     4.862500       3.580557
     4.912500       3.530557
     4.962500       3.580557
     5.012500       3.630557
     5.062500       3.580557
     5.112500       3.530557
     5.162500       3.580557
     5.212500       3.630557
     5.262500       3.580557
 /
 \stoprotation

 \startrotation by  0.224951  -0.974370 about      4.937500       3.580557
 \setquadratic
 \plot
     4.937500       3.580557
     4.987500       3.630557
     5.037500       3.580557
     5.087500       3.530557
     5.137500       3.580557
     5.187500       3.630557
     5.237500       3.580557
     5.287500       3.530557
     5.337500       3.580557
     5.387500       3.630557
     5.437500       3.580557
 /
 \stoprotation

 \startrotation by  0.766045  -0.642787 about      5.112500       3.634248
 \setquadratic
 \plot
     5.112500       3.634248
     5.162500       3.684248
     5.212500       3.634248
     5.262500       3.584249
     5.312500       3.634248
     5.362500       3.684248
     5.412500       3.634248
     5.462500       3.584249
     5.512500       3.634248
     5.562500       3.684248
     5.612500       3.634248
 /
 \stoprotation


 \startrotation by  0.000000   1.000000 about      4.850000       7.425000
 \setquadratic
 \plot
     4.850000       7.425000
     4.900000       7.475000
     4.950000       7.425000
     5.000000       7.375000
     5.050000       7.425000
     5.100000       7.475000
     5.150000       7.425000
     5.200000       7.375000
     5.250000       7.425000
     5.300000       7.475000
     5.350000       7.425000
 /
 \stoprotation

 \startrotation by -0.241922  -0.970296 about      4.762500       7.080557
 \setquadratic
 \plot
     4.762500       7.080557
     4.812500       7.130557
     4.862500       7.080557
     4.912500       7.030557
     4.962500       7.080557
     5.012500       7.130557
     5.062500       7.080557
     5.112500       7.030557
     5.162500       7.080557
     5.212500       7.130557
     5.262500       7.080557
 /
 \stoprotation

 \startrotation by  0.241922  -0.970296 about      4.937500       7.080557
 \setquadratic
 \plot
     4.937500       7.080557
     4.987500       7.130557
     5.037500       7.080557
     5.087500       7.030557
     5.137500       7.080557
     5.187500       7.130557
     5.237500       7.080557
     5.287500       7.030557
     5.337500       7.080557
     5.387500       7.130557
     5.437500       7.080557
 /
 \stoprotation


\put {$\bar q$} at 0.84 7.25

\put {$\bar q$} at 0.84 3.75

\put {$q$} at 6.16 7.25

\put {$q$} at 6.16 3.75

\put {$\longrightarrow$} at 1.25 7.45

\put {$p$} at 1.25 7.6

\put {$\longleftarrow$} at 5.75 7.45

\put {$p'$} at 5.75 7.6

\put {$\longrightarrow$} at 1.25 3.55

\put {$q$} at 1.25 3.4

\put {$\longleftarrow$} at 5.75 3.55

\put {$q'$} at 5.75 3.4

\put {$\{1,\ldots,b\}$} at 2.15 8.10

\put {$\{e{+}1,\ldots,n\}$} at 4.85 8.10

\put {$\{c{+}1,\ldots,d\}$} at 4.85 2.90

\put {$\{b{+}1,\ldots,c\}$} at 2.25 5.5

\put {$\{d{+}1,\ldots,e\}$} at 4.75 5.5

\put {$\uparrow\enspace k_0 = q+\kappa(c{+}1,d)+q'$} at 4.6 4.5

\endpicture
$$
\bigskip
\noindent
Figure 2.  Contributions to
$q \thinspace \bar q \longrightarrow q \thinspace \bar q
\thinspace g \thinspace g \cdots g$, showing the natural
way to label the gluon momenta for the purpose of evaluating
the non-$1\over N$ contribution.

\vfil\eject\bye